\documentclass[rmp,aps,amsfonts,amssymb,nofootinbib,floats,twocolumn]{revtex4}
\usepackage{graphics}
\usepackage{epsfig}

\begin{document}
\title{Angle-resolved photoemission spectroscopy of the cuprate superconductors\footnote{\vspace{-1.75cm}{\bf Reviews of Modern Physics, in
press.}}}
\author{Andrea Damascelli}
\affiliation{Stanford Synchrotron Radiation Laboratory, Stanford University, Stanford, CA\,94305, USA}
\author{Zhi-Xun Shen}
\affiliation{Department of Physics, Applied Physics and Stanford Synchrotron
        Radiation Laboratory,
        \\Stanford University, Stanford, CA\,94305, USA}
\author{Zahid Hussain}
\affiliation{Advanced Light Source, Lawrence Berkeley National Laboratory, Berkeley, CA\,94720, USA}

\begin{abstract}
The last decade witnessed significant progress in angle-resolved
photoemission spectroscopy (ARPES) and its applications. Today,
ARPES experiments with 2 meV energy resolution and $0.2^\circ$
angular resolution are a reality even for photoemission on solid
systems. These technological advances and the improved sample
quality have enabled ARPES to emerge as a leading tool in the
investigation of the high-$T_c$ superconductors. This paper
reviews the most recent ARPES results on the cuprate
superconductors and their insulating parent and sister compounds,
with the purpose of providing an updated summary of the extensive
literature in this field. The low energy excitations are discussed
with emphasis on some of the most relevant issues, such as the
Fermi surface and remnant Fermi surface, the superconducting gap,
the pseudogap and $d$-wave-like dispersion, evidence of electronic
inhomogeneity and nano-scale phase separation, the emergence of
coherent quasiparticles through the superconducting transition,
and many-body effects in the one-particle spectral function due to
the interaction of the charge with magnetic and/or lattice degrees
of freedom. Given the dynamic nature of the field, we chose to
focus mainly on reviewing the experimental data, as on the
experimental side a general consensus has been reached whereas
interpretations and related theoretical models can vary
significantly. The first part of the paper introduces
photoemission spectroscopy in the context of strongly interacting
systems, along with an update on the state-of-the-art
instrumentation. The second part provides a brief overview of the
scientific issues relevant to the investigation of the low energy
electronic structure by ARPES. The rest of the paper is devoted to
the review of experimental results from the cuprates and the
discussion is organized along conceptual lines: normal-state
electronic structure, interlayer interaction, superconducting gap,
coherent superconducting peak, pseudogap, electron self energy and
collective modes. Within each topic, ARPES data from the various
copper oxides are presented.
\end{abstract}

\maketitle

\tableofcontents

\section{Introduction}

\label{sec:introduc}

The discovery of superconductivity at 30 K in the LaBaCuO ceramics
by \textcite{BednorzJG:PoshTc} has opened the era of {\it
high-$T_c$ superconductivity}, changing the history of a
phenomenon that had before been confined to very low temperatures
[until 1986 the maximum value of $T_c$ was limited to the 23 K
observed in Nb$_3$Ge \cite{GavalerJR:SupN-G,TestardiLR:Supwoa}].
This unexpected result prompted intense activity in the field of
ceramic oxides and has led to the synthesis of compounds with
increasingly higher $T_c$, all characterized by a layered crystal
structure with one or more CuO$_2$ planes per unit cell, and a
quasi two-dimensional (2D) electronic structure.  By 1987, a $T_c$
of approximately 90 K (i.e., higher than the boiling point of
liquid nitrogen at 77 K) was already observed in
YBa$_2$Cu$_3$O$_{7-\delta}$ \cite{WuMK:Sup9Kn}. The record $T_c$
of 133.5 K (at atmospheric pressure) was later obtained in
HgBa$_2$Ca$_2$Cu$_3$O$_{8+x}$ \cite{SchillingA:Supa1K}.

\begin{figure}[b!]
\centerline{\epsfig{figure=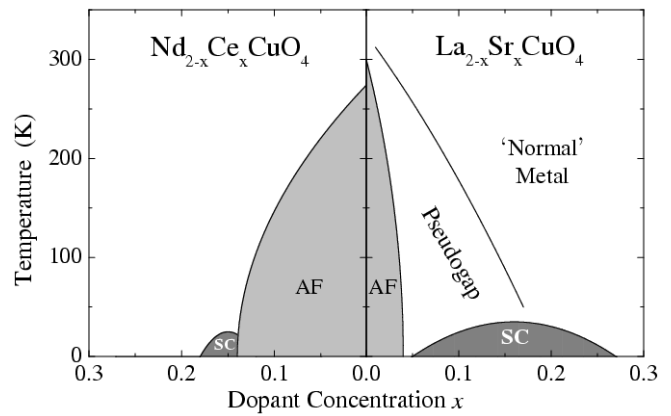,width=1\linewidth,clip=}}
\vspace{0.1cm} \caption{Phase diagram of n and p-type
superconductors.}\label{rmp_pd}\end{figure}

One may wonder whether the impact of the discovery by
\textcite{BednorzJG:PoshTc} would have been somewhat overlooked if
MgB$_2$, with its recently ascertained 39 K $T_c$, had already
been discovered [\textcite{NagamatsuJ:Sup3Km}; for a review see:
\textcite{DayC:1}]. However, independent of the values of $T_c$
the observation of superconductivity in the ceramic copper oxides
was in itself an unexpected and surprising result. In fact,
ceramic materials are typically insulators, and this is also the
case for the undoped copper oxides. However, when doped the latter
can become poor metals in the normal state and high-temperature
superconductors (HTSCs) upon reducing the temperature (see in
Fig.\,\ref{rmp_pd} the phenomenological phase diagram of electron
and hole-doped HTSCs, here represented by Nd$_{2-x}$Ce$_x$CuO$_4$
and La$_{2-x}$Sr$_x$CuO$_4$ respectively).
In addition, the detailed investigation of their phase diagram
revealed that the macroscopic properties of the copper oxides are
profoundly influenced by {\it strong electron-electron
correlations} (i.e., large Coulomb repulsion $U$). Naively, this
is not expected to favor the emergence of superconductivity for
which electrons must be bound together to form Cooper pairs. Even
though the approximate $T^2$ dependence of the resistivity
observed in the overdoped metallic regime was taken as evidence
for {\it Fermi liquid} (FL) behavior, the applicability of FL
theory (which describes electronic excitations in terms of an
interacting gas of renormalized {\it quasiparticles}, see
Sec.\,\ref{sec:spectral}) to the `normal' metallic state of HTSCs
is questionable, because many properties do not follow canonical
FL behavior \cite{OrensteinJ:Advtph}. This breakdown of FL theory
and of the single particle picture becomes most dramatic upon
approaching the undoped line of the phase diagram ($x\!=\!0$ in
Fig.\,\ref{rmp_pd}), where one finds the antiferromagnetic (AF)
Mott insulator (see Sec.\,\ref{sec:MottIns}).

Therefore, the cuprate HTSCs have attracted great interest not
only for the application potential related to their high $T_c$,
but also for their scientific significance. This stems from the
fact that the HTSCs highlight a major intellectual crisis in the
quantum theory of solids which, in the form of one-electron band
theory, has been very successful in describing good metals (like
Cu) but has proven inadequate for strongly correlated electron
systems. In turn, the Bardeen-Cooper-Schrieffer (BCS) theory
[\textcite{Bardeen:1}; see also \textcite{schrieffer:1}], which
was developed for FL-like metals and has been so successful in
describing conventional superconductors, does not seem to have the
appropriate foundation for the description of high-$T_c$
superconductivity. In order to address the scope of the current
approach in the quantum theory of solids and the validity of the
proposed alternative models, a detailed comparison with
experiments that probe the electronic properties and the nature of
the elementary excitations is required.

In this context, angle-resolved photoemission spectroscopy (ARPES)
plays a major role because it is the most direct method of
studying the electronic structure of solids (see
Sec.\,\ref{sec:arpes}). Its large impact on the development of
many-body theories stems from the fact that this technique
provides information on the single particle Green's function,
which can be calculated starting from a microscopic Hamiltonian.
Driven by the demand, a significant improvement in instrumental
resolution and detection efficiency has taken place over the last
decade. Owing also to the continuously improved sample quality,
the stage was set for ARPES to emerge as a leading experimental
probe in the study of the HTSCs. Indeed, many of the ARPES results
have broadly impacted the field
\cite{LEVIBG:DOOSBF,levi:1,levi:2,mitton:1,mitton:2}, such as the
observation of dispersive electronic features
\cite{TakahashiT:Evia-r,OlsonCG:SupgB-}, the discovery of the
$d$-wave superconducting gap \cite{WellsBO:Evik-d,ShenZ-X:Anolga},
and of the normal-state pseudogap
\cite{MarshallDS:Uncese,LoeserAG:Excgtn,DingH:Speept}.

In such a rapidly evolving field, one is always presented with a
dilemma when deciding whether and how to write a review. On the
one hand it may be premature to proceed with an extensive review,
but on the other hand it is helpful for the broader community to
have access to a summary of the current state of the subject. It
is our intention to present a `snapshot' of the investigation of
the HTSCs by ARPES taken at the time of completion of this
article. This will help the readers, especially those who are not
photoemission experts, to sort through the extensive literature,
learn about the outstanding problems, and become aware of the
level of consensus. The downside is that some of the most recent
results we will discuss may be outpaced by the rapid advance of
the field. With this in mind, we limit the scope of our review to
ARPES results only, without making detailed comparison with other
techniques except when necessary.  We will mainly focus on results
reported after 1994, as on the previous period another extensive
review paper is already
available.\footnote{\textcite{ShenZX:Elesps}. For more recent
reviews see
\textcite{randeria:1,lynch:1,TohyamaT:Ang-re,DamascelliA:Motios,GoldenMS:elesch,JohnsonPD:Phoss-,KimC:Spi-ch}.}
In addition, because over the last two years we have witnessed a
dramatic improvement of energy and especially momentum resolution
with the introduction of the Scienta SES200 electron analyzer,
more emphasis will be given to ARPES data obtained with this kind
of spectrometer, whenever available. As this review paper will
deal mostly with experimental results, we will refer only to those
theoretical concepts and models which are explicitly relevant to
the discussion of the ARPES data on the HTSCs.  Furthermore, as
the doping evolution is probably the best comparison between
theory and experiment, we will discuss ARPES data on different
HTSCs and on their insulating parent compounds focusing, in
particular, on systematic changes in the electronic structure
which may be relevant to the development of a comprehensive
picture for the evolution from Mott insulator to overdoped
superconductor.

\section{Angle-resolved photoemission spectroscopy}
\label{sec:arpes}

\subsection{General description}

Photoelectron spectroscopy is a general term which refers to all
techniques based on the photoelectric effect originally observed
by \textcite{Hertz:1}. This was later explained as a manifestation
of the quantum nature of light by \textcite{EinsteinA:1}, who
recognized that when light is incident on a sample an electron can
absorb a photon and escape from the material with a maximum
kinetic energy $h\nu\!-\!\phi$ (where $\nu$ is the photon
frequency and $\phi$, the material work function, is a measure of
the potential barrier at the surface that prevents the valence
electrons from escaping, and is typically around 4-5 eV for
metals). As a description of the spectroscopic techniques based on
the detection of photoemitted electrons is beyond the scope of
this review, we will only summarize those experimental and
theoretical aspects which will be relevant to the discussion of
the ARPES results presented in the course of the paper. For a more
detailed description of ARPES and other photoelectron
spectroscopies, we refer the reader to the extensive literature
available on the subject.\footnote{A very detailed list of review
articles and books dedicated to photoelectron spectroscopy on
solids was recently given by \textcite{lynch:1} in their book
dealing with photoemission spectroscopy on high-temperature
superconductors, and is here reproduced with some additions:
\textcite{SmithNV:Phopm.,EastmanDE:Phosm.,carlsonta:1,FeuerbacherB:Phoesc,brundle:1,brundle:2,wertheim:1,cardona:1,ley:1,mahan:1,feuerbacher:1,nemoshkalenko,WilliamsRH:Phosst,lindau:1,InglesfieldJE:Eles.,wendin:1,Leckey:1,plummer:1,SmithNV:Phos.,margaritondo:1,HimpselFJ:Ang-re,CourthsR:Phoec.,SmithKE:elesss,kevan:1,bachrach:1,ShenZX:Elesps,hufner:1,BraunJ:thea-r,Grioni:1}.}

\begin{figure}[b!]
\centerline{\epsfig{figure=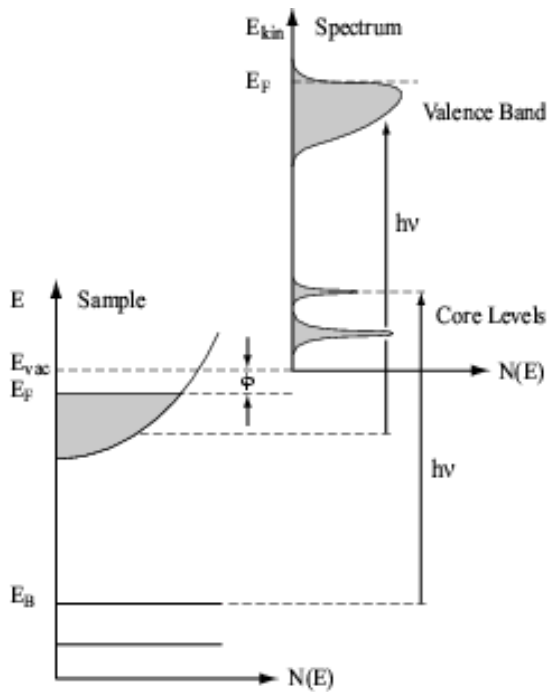,width=7.3cm,clip=}}
\vspace{0cm} \caption{Energetics of the photoemission process
\cite{hufner:1}. The electron energy distribution produced by the
incoming photons, and measured as a function of the kinetic energy
$E_{kin}$ of the photoelectrons (right), is more conveniently
expressed in terms of the binding energy $E_B$ (left) when one
refers to the density of states inside the solid ($E_B\!=\!0$ at
$E_F$).}\label{energetics}\end{figure}

The energetics of the photoemission process and of the geometry of
an ARPES experiment are sketched in Fig.\ref{energetics} and
Fig.\ref{geometry}a, respectively. A beam of monochromatized
radiation supplied either by a gas-discharge lamp or by a
synchrotron beamline is incident on a sample (which has to be a
properly aligned single crystal, in order to perform momentum
resolved measurements). As a result, electrons are emitted by the
photoelectric effect and escape into the vacuum in all directions.
By collecting the so-called photoelectrons with an electron energy
analyzer characterized by a finite acceptance angle, one measures
the kinetic energy $E_{kin}$ of the photoelectrons for a given
emission angle. This way, the photoelectron momentum ${\bf p}$ is
also completely determined: its modulus is given by
$p\!=\!\sqrt{2mE_{kin}}$ and its components parallel and
perpendicular to the sample surface are obtained from the polar
($\vartheta$) and azimuthal ($\varphi$) emission angles.

Within the non-interacting electron picture and by taking
advantage of total energy and momentum conservation laws (note
that the photon momentum can be neglected at the low photon
energies typically used in ARPES experiments), the kinetic energy
and momentum of the photoelectron can be related to the binding
energy $E_B$ and crystal-momentum $\hbar{\bf k}$ inside the solid:
\begin{eqnarray}
E_{kin}&&=h\nu-\phi-|E_B| \label{eq:enegy}
\\
{\bf p}_{\|}&&=\hbar{\bf
k}_{\|}=\sqrt{2mE_{kin}}\cdot\sin\vartheta \label{eq:momentum}
\end{eqnarray}
\noindent
Here $\hbar{\bf k}_{\|}$ is the component parallel to the surface
of the electron crystal momentum in the {\it extended} zone
scheme. Upon going to larger $\vartheta$ angles, one actually
probes electrons with {\bf k} lying in higher-order Brillouin
zones. By subtracting the corresponding reciprocal lattice vector
{\bf G}, the {\it reduced} electron crystal momentum in the first
Brillouin zone is obtained. Note that the perpendicular component
of the wave vector ${\bf k}_{\bot}$ is not conserved across the
sample surface due to the lack of translational symmetry along the
surface normal.
This implies that, in general, even experiments
performed for all ${\bf k}_{\|}$ (i.e., by collecting
photoelectrons at all possible angles) will not allow a complete
determination of the total crystal wave vector ${\bf k}$ [unless
some {\it a priori} assumption is made for the dispersion $E({\bf
k})$ of the electron final states involved in the photoemission
process]. However, several specific experimental methods for
absolute three dimensional band mapping have also been developed
[see, e.g.,
\textcite{hufner:1,StrocovVN:Newmab,StrocovVN:Absbmb}].

A particular case in which the uncertainty in ${\bf k}_{\bot}$ is
less relevant is that of the low-dimensional systems characterized
by an anisotropic electronic structure and, in particular, a
negligible dispersion along the $z$ axis (i.e., along the surface
normal, see Fig.\,\ref{geometry}a). The electronic dispersion is
then almost exclusively determined by ${\bf k}_{\|}$, as in the
case of the 2D copper oxide superconductors which we will focus on
throughout this paper [note, however, that possible complications
arising from a finite three-dimensionality of the initial and/or
final states involved in the photoemission process should always
be carefully considered \cite{lindroos:1}].  As a result, one can
map out in detail the electronic dispersion relations $E({\bf
k}_{\|})$ simply by tracking, as a function of ${\bf p}_{\|}$, the
energy position of the peaks detected in the ARPES spectra for
different take off angles (as in Fig.\,\ref{geometry}b, where both
direct and inverse photoemission spectra for a single band
dispersing through the Fermi energy $E_F$ are shown). Furthermore,
as an additional bonus of the lack of $z$ dispersion, one can
directly identify the width of the photoemission peaks with the
lifetime of the photohole \cite{SMITHNV:PHOLQL}, which contains
information on the intrinsic correlation effects of the system and
is formally described by the imaginary part of the electron self
energy (see Sec.\,\ref{sec:spectral}). On the contrary, in 3D
systems the linewidth contains contributions from both photohole
and photoelectron lifetimes, with the latter reflecting final
state scattering processes and thus the finite probing depth; as a
consequence, isolating the intrinsic many-body effects becomes a
much more complicated problem.

Before moving on to the discussion of some theoretical issues, it
is worth pointing out that most of the ARPES experiments are
performed at photon energies in the ultraviolet (in particular for
$h\nu\!<\!100$ eV). The main reason is that by working at lower
photon energies it is possible to achieve higher energy and
momentum resolution. This is easy to see for the case of the
momentum resolution $\Delta{\bf k}_{\|}$ which, from
Eq.\,\ref{eq:momentum} and neglecting the contribution due to the
finite energy resolution, is given by:
\begin{equation}
\Delta{\bf k}_{\|}\simeq
\sqrt{2mE_{kin}/\hbar^2}\cdot\cos\vartheta\!\cdot\!\Delta\vartheta
\label{eq:res}
\end{equation}
where $\Delta\vartheta$ corresponds to the finite acceptance angle
of the electron analyzer. From Eq.\,\ref{eq:res} it is clear that
the momentum resolution will be better at lower photon energy
(i.e., lower $E_{kin}$), and for larger polar angles $\vartheta$
(note that one can effectively improve the momentum resolution by
extending the measurements to momenta outside the first Brillouin
zone). By working at low photon energies there are also some
additional advantages: first, for a typical beamline it is easier
to achieve high energy resolution (see Sec.\,\ref{sec:stateart});
second, one can completely disregard the photon momentum
$\kappa\!=\!2\pi/\lambda$ in Eq.\,\ref{eq:momentum}, as for 100 eV
photons the momentum is 3\% (0.05 \AA$^{-1}$) of the typical
Brillouin zone size of the cuprates ($2\pi/a\!\simeq\!1.6$
\AA$^{-1}$), and at 21.2 eV (the HeI$\alpha$ line typically used
on ARPES systems equipped with a gas-discharge lamp) it is only
0.5\% (0.008 \AA$^{-1}$). If on the contrary the photon momentum
is not negligible, the photoemission process does not involve
vertical transitions and $\kappa$ must be explicitly taken into
account in Eq.\,\ref{eq:momentum}. For example, for 1487 eV
photons (the Al K$_{\alpha}$ line commonly used in X-ray
photoemission) $\kappa\!\simeq\!0.76$ \AA$^{-1}$ that corresponds
to 50\% of the zone size.

A major drawback of working at low photon energies is the extreme
surface sensitivity. The mean free path for unscattered
photoelectrons is characterized by a minimum of approximately 5
\AA\ at 20-100 eV kinetic energies \cite{SeahMP:Quaess}, which are
typical values in ARPES experiments. This means that a
considerable fraction of the total photoemission intensity will be
representative of the topmost surface layer, especially on systems
characterized by a large structural/electronic anisotropy and, in
particular, by relatively large $c$-axis lattice parameters, such
as the cuprates. Therefore, in order to learn about the bulk
electronic structure ARPES experiments have to be performed on
atomically clean and well-ordered systems, which implies that {\it
fresh} and {\it flat} surfaces have to be `prepared' immediately
prior to the experiment in ultra-high vacuum conditions (typically
at pressures lower than 5$\times\!10^{-11}$ torr). So far, the
best ARPES results on copper oxide superconductors were obtained
on samples cleaved {\it in situ}, which however requires a natural
cleavage plane for the material under investigation and explains
why not all the cuprates are suitable for ARPES experiments.

\subsection{Three-step model and sudden approximation}
\label{sec:sudden}

To develop a formal description of the photoemission process, one
has to calculate the transition probability $w_{fi}$ for an
optical excitation between the $N$-electron ground state
$\Psi_i^N$ and one of the possible final states $\Psi_f^N$. This
can be approximated by Fermi's golden rule:
\begin{equation}
w_{fi}=\frac{2\pi}{\hbar}|\langle\Psi_f^N|H_{int}|\Psi_i^N\rangle|^2\delta(E_f^N-E_i^N-h\nu)
\label{eq:wfi}
\end{equation}
where $E_i^N\!=\!E_i^{N-1}\!-\!E_B^{\bf k}$ and
$E_f^N\!=\!E_f^{N-1}\!+\!E_{kin}$ are the initial and final-state
energies of the $N$-particle system ($E_B^{\bf k}$ is the binding
energy of the photoelectron with kinetic energy $E_{kin}$ and
momentum ${\bf k}$). The interaction with the photon is treated as
a perturbation given by:
\begin{equation}
H_{int}=\frac{e}{2mc}({\bf A}\!\cdot\!{\bf p}+{\bf p}\!\cdot\!{\bf
A})= \frac{e}{mc}\,{\bf A}\!\cdot\!{\bf p} \label{eq:hint}
\end{equation}
where {\bf p} is the electronic momentum operator and {\bf A} is
the electromagnetic vector potential (note that the gauge
$\Phi\!=\!0$ was chosen for the scalar potential $\Phi$, and the
quadratic term in ${\bf A}$ was dropped because in the linear
optical regime it is typically negligible with respect to the
linear terms). In Eq.\,\ref{eq:hint} we also made use of the
commutator relation $[{\bf p},{\bf
A}]\!=\!i\hbar\nabla\!\cdot\!{\bf A}$ and dipole approximation
[i.e., {\bf A} constant over atomic dimensions and therefore
$\nabla\!\cdot\!{\bf A}\!=\!0$, which holds in the ultraviolet].
Although this is a routinely used approximation, it should be
noted that $\nabla\!\cdot\!{\bf A}$ might become important at the
{\it surface} where the electromagnetic fields may have a strong
spatial dependence, giving rise to a significant intensity for
indirect transitions. This surface photoemission contribution,
which is proportional to ($\varepsilon-1$) where $\varepsilon$ is
the medium dielectric function, can interfere with the bulk
contribution resulting in asymmetric lineshapes for the bulk
direct-transition peaks.\footnote{For more details on the surface
photoemission effects see:
\textcite{feuerbacher:1,MillerT:Intbbs,HansenED:SurpA,HansenED:Ovetsp}.}
At this point, a more rigorous approach is to proceed with the
so-called {\it one-step model} in which photon absorption,
electron removal, and electron detection are treated as a single
coherent process.\footnote{See for example
\textcite{mitchell:1,makinson:1,buckingham:1,MahanGD:Thepsm,SchaichWL:Modctt,FeibelmanPJ:Phos-C,PendryJB:Thep.,pendry:1,LiebschA:Thepla,liebsch:1,LindroosM:Sursa-,LindroosM:novdmF,BansilA:Ang-re,BansilA:Mateet,BansilA:Impmet}.}
In this case bulk, surface, and vacuum have to be included in the
Hamiltonian describing the crystal, which implies that not only
bulk states have to be considered, but also surface and evanescent
states, and surface resonances. However, due to the complexity of
the one-step model, photoemission data are usually discussed
within the {\it three-step model} that, although purely
phenomenological, has proven to be rather successful
\cite{FanHY:1,berglund:1,FeibelmanPJ:Phos-C}. Within this
approach, the photoemission process is subdivided into three
independent and sequential steps:
\begin{figure*}[t!]
\centerline{\epsfig{figure=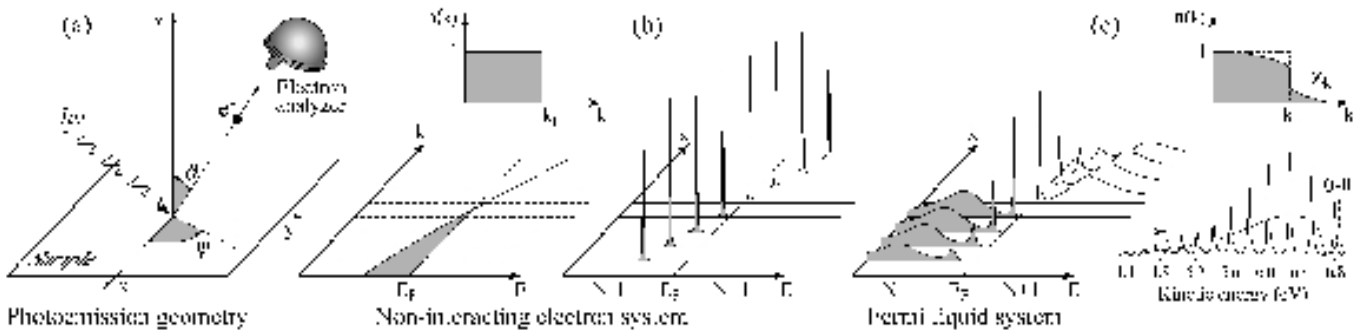,width=18cm,clip=}}
\caption{(a) Geometry of an ARPES experiment; the emission
direction of the photoelectron is specified by the polar
($\vartheta$) and azimuthal ($\varphi$) angles. Momentum resolved
one-electron removal and addition spectra for: (b) a
non-interacting electron system (with a single energy band
dispersing across $E_F$); (c) an interacting Fermi liquid system
\cite{sawatzky:1,meinders:1}. The corresponding ground-state
($T\!=\!0$ K) momentum distribution function $n({\bf k})$ is also
shown. (c) Bottom right: photoelectron spectrum of gaseous
hydrogen and the ARPES spectrum of solid hydrogen developed from
the gaseous one \cite{sawatzky:1}. }\label{geometry}\end{figure*}
\begin{itemize}
\item[(i)]{Optical excitation of the electron in the {\it bulk}.}
\item[(ii)]{Travel of the excited electron to the surface.}
\item[(iii)]{Escape of the photoelectron into vacuum.}
\end{itemize}
 The total photoemission intensity is then given by the product
of three independent terms: the total probability for the optical
transition, the scattering probability for the travelling
electrons, and the transmission probability through the surface
potential barrier. Step (i) contains all the information about the
intrinsic electronic structure of the material and will be
discussed in detail below. Step (ii) can be described in terms of
an effective mean free path, proportional to the probability that
the excited electron will reach the surface without scattering
(i.e, with no change in energy and momentum). The inelastic
scattering processes, which determine the surface sensitivity of
photoemission (as discussed in the previous section), also give
rise to a continuous background in the spectra which is usually
ignored or subtracted. Step (iii) is described by a transmission
probability through the surface, which depends on the energy of
the excited electron and the material work function $\phi$.

In evaluating step (i), and therefore the photoemission intensity
in terms of the transition probability $w_{fi}$, it would be
convenient to factorize the wavefunctions in Eq.\,\ref{eq:wfi}
into photoelectron and $(N\!-\!1)$-electron terms, as we have done
for the corresponding energies. This however is far from trivial
because during the photoemission process itself the system will
relax. The problem simplifies within the {\it sudden
approximation}, which is extensively used in many-body
calculations of the photoemission spectra from interacting
electron systems, and is in principle applicable only to high
kinetic-energy electrons. In this limit, the photoemission process
is assumed to be {\it sudden}, with no post-collisional
interaction between the photoelectron and the system left behind
(in other words, an electron is instantaneously removed and the
effective potential of the system changes discontinuously at that
instant). The final state $\Psi^{N}_{f}$ can then be written as:
\begin{equation}
\Psi_f^N={\mathcal{A}}\, \phi_f^{\bf k}\, \Psi^{N-1}_f
\end{equation}
where $\mathcal{A}$ is an antisymmetric operator that properly
antisymmetrizes the $N$-electron wavefunction so that the Pauli
principle is satisfied, $\phi_f^{\bf k}$ is the wavefunction of
the photoelectron with momentum ${\bf k}$, and $\Psi^{N-1}_{f}$ is
the final state wavefunction of the $(N\!-\!1)$-electron system
left behind, which can be chosen as an excited state with
eigenfunction $\Psi^{N-1}_{m}$ and energy $E^{N-1}_m$. The {\it
total} transition probability is then given by the sum over {\it
all} possible excited states $m$. Note, however, that the sudden
approximation is inappropriate for low kinetic energy
photoelectrons, which may need longer than the system response
time to escape into vacuum. In this case, the so-called {\it
adiabatic limit}, one can no longer factorize $\Psi_f^N$ in two
independent parts and the detailed screening of photoelectron and
photohole has to be taken into account \cite{GadzukJW:Excedc}. It
is important to mention that, however, there is evidence that the
sudden approximation is justified for the cuprate HTSCs even at
photon energies as low as 20\,eV [\textcite{RanderiaM:Momdsr};
Sec.\,\ref{sec:spectral}].

For the initial state, let us first assume for simplicity that
$\Psi^{N}_{i}$ is a single Slater determinant (i.e., Hartree-Fock
formalism), so that we can write it as the product of a
one-electron orbital $\phi_i^{\bf k}$ and an $(N\!-\!1)$-particle
term:
\begin{equation}
\Psi_i^N={\mathcal{A}}\, \phi_i^{\bf k}\, \Psi^{N-1}_i
\end{equation}
More generally, however, $\Psi^{N-1}_{i}$ should be expressed as
$\Psi^{N-1}_i\!=\!c_{\bf k}\Psi^{N}_i$, where $c_{\bf k}$ is the
annihilation operator for an electron with momentum $\bf k$. This
also shows that $\Psi^{N-1}_{i}$ is {\it not} an eigenstate of the
$(N\!-\!1)$ particle Hamiltonian, but is just what remains of the
$N$-particle wavefunction after having pulled out one electron. At
this point, we can write the matrix elements in Eq.\,\ref{eq:wfi}
as:
\begin{equation}
\langle\Psi_f^N|H_{int}|\Psi_i^N\rangle\!=\!\langle\phi_f^{\bf
k}|H_{int}|\phi_i^{\bf k}\rangle
\langle\Psi^{N-1}_{m}|\Psi^{N-1}_{i}\rangle
\end{equation}
where $M_{f,i}^{\bf k}\!\equiv\!\langle\phi_f^{\bf
k}|H_{int}|\phi_i^{\bf k}\rangle$ is the one-electron dipole
matrix element, and the second term is the $(N\!-\!1)$-electron
overlap integral. Here, we replaced $\Psi_f^{N-1}$ with an
eigenstate $\Psi_m^{N-1}$, as discussed above. The total
photoemission intensity measured as a function of $E_{kin}$ at a
momentum ${\bf k}$, namely $I({\bf
k},E_{kin})\!=\!\sum_{f,i}w_{f,i}$, is then proportional to:
\begin{equation}
\sum_{f,i}|M_{f,i}^{\bf k}|^2 \sum_{m}
|c_{m,i}|^2\delta(E_{kin}\!+\!E_{m}^{N-1}\!-E_i^N\!-\!h\nu)
\label{eq:photocurrent}
\end{equation}
where
$|c_{m,i}|^2\!=\!|\langle\Psi^{N-1}_{m}|\Psi^{N-1}_{i}\rangle|^2$
is the probability that the removal of an electron from state $i$
will leave the $(N\!-\!1)$-particle system in the excited state
$m$. From here we see that, if
$\Psi^{N-1}_{i}\!=\!\Psi^{N-1}_{m_0}$ for one particular
$m\!=\!m_0$, the corresponding $|c_{m_0,i}|^2$ will be unity and
all the others $c_{m,i}$ zero; in this case, if also $M_{f,i}^{\bf
k}\!\neq\!0$, the ARPES spectra will be given by a delta function
at the Hartree-Fock orbital energy $E_B^{\bf k}\!=\!-\epsilon_{\bf
k}$, as shown in Fig.\,\ref{geometry}b (i.e., non-interacting
particle picture). In the strongly correlated systems, however,
many of the $|c_{m,i}|^2$ will be different from zero because the
removal of the photoelectron results in a strong change of the
system effective potential and, in turn, $\Psi^{N-1}_{i}$ will
have an overlap with many of the eigenstates $\Psi^{N-1}_{m}$.
Therefore, the ARPES spectra will not consist of single delta
functions but will show a main line and several satellites
according to the number of excited states $m$ created in the
process (Fig.\,\ref{geometry}c).

This is very similar to the situation encountered in photoemission
from molecular hydrogen \cite{siegbahn:1} in which not simply a
single peak but many lines separated by few tenths of eV from each
other are observed (solid line in Fig.\,\ref{geometry}c, bottom
right). These so-called `shake-up' peaks correspond to the
excitations of the different vibrational states of the H$_2^+$
molecule. In the case of solid hydrogen (dashed line in
Fig.\,\ref{geometry}c, bottom right), as discussed by
\textcite{sawatzky:1}, the vibrational excitations would develop
in a broad continuum while a sharp peak would be observed for the
fundamental transition (from the ground state of the H$_2$ to the
one of the H$_2^+$ molecule). Note that the fundamental line would
also be the only one detected in the adiabatic limit, in which
case the $(N\!-\!1)$-particle system is left in its ground state.

\subsection{One-particle spectral function}
\label{sec:spectral}

In the discussion of photoemission on solids, and in particular on
the correlated electron systems in which many $|c_{m,i}|^2$ in
Eq.\,\ref{eq:photocurrent} are different from zero, the most
powerful and commonly used approach is based on the Green's
function formalism.\footnote{See for example:
\textcite{Abrikosov:1,hedin:1,FetterAL:1,mahan:2,economou:1,rickayzen:1}.}
 In this context, the propagation of a single electron in a many-body
system is described by the {\it time-ordered} one-electron Green's
function ${\mathcal{G}}(t-t^\prime)$, which can be interpreted as
the probability amplitude that an electron added to the system in
a Bloch state with momentum ${\bf k}$ at a time zero will still be
in the same state after a time $|t\!-\!t^\prime|$. By taking the
Fourier transform, ${\mathcal{G}}(t\!-\!t^\prime)$ can be
expressed in energy-momentum representation resulting in
${\mathcal{G}}({\bf k},\omega)\!=\!G^+({\bf
k},\omega)\!+\!G^-({\bf k},\omega)$, where $G^+({\bf k},\omega)$
and $G^-({\bf k},\omega)$ are the one-electron addition and
removal Green's function, respectively. At $T\!=\!0$:
\begin{equation}
G^\pm({\bf
k},\omega)=\sum_{m}\frac{|\langle\Psi^{N\pm1}_m|c^\pm_{\bf k}
|\Psi^N_i\rangle|^2}{\omega- E_m^{N\pm 1}+ E^N_i\pm i\eta}
\end{equation}
where the operator $c_{\bf k}^+\!=\!c_{{\bf k}\sigma}^\dag$
($c_{\bf k}^-\!=\!c_{{\bf k}\sigma}^{ }$) creates (annihilates) an
electron with energy $\omega$, momentum ${\bf k}$, and spin
$\sigma$ in the $N$-particle initial state $\Psi^N_i$, the
summation runs over all possible $({N\!\pm\!1})$-particle
eigenstates $\Psi_m^{N\pm1}$ with eigenvalues $E_m^{N\pm1}$, and
$\eta$ is a positive infinitesimal (note also that from here on we
will take $\hbar\!=\!1$). In the limit $\eta\!\rightarrow\!0^+$
one can make use of the identity
$(x\!\pm\!i\eta)^{-1}\!=\!{\mathcal{P}}(1/x)\!\mp\!i\pi\delta(x)$,
where $\mathcal{P}$ denotes the principle value, to obtain the
{\it one-particle spectral function}  $A({\bf
k},\omega)\!=\!A^+({\bf k},\omega)\!+\!A^-({\bf
k},\omega)\!=\!-(1/\pi){\rm Im}\,G({\bf k},\omega)$, with:
\begin{equation}
A^\pm({\bf
k},\omega)\!=\!\!\!\sum_{m}|\langle\Psi^{N\pm1}_m\!|c^\pm_{\bf k}
|\Psi^N_i\rangle|^2 \delta(\omega\!-\!E_m^{N\pm 1}\!\!+\!E^N_i)
\label{eq:akw0}
\end{equation}
and $G({\bf k},\omega)\!=\!G^+({\bf k},\omega)\!+\![G^{-}({\bf
k},\omega)]^*$, which defines the {\it retarded} Green's function.
Note that $A^-({\bf k},\omega)$ and $A^+({\bf k},\omega)$ define
the one-electron removal and addition spectra which one can probe
with direct and inverse photoemission, respectively. This is
evidenced, for the direct case, by the comparison between the
expression for $A^-({\bf k},\omega)$ and
Eq.\,\ref{eq:photocurrent} for the photoemission intensity (note
that in the latter $\Psi^{N-1}_i\!=\!c_{\bf k}\Psi^{N}_i$ and the
energetics of the photoemission process has been explicitly
accounted for). Finite temperatures effect can be taken into
account by extending the Green's function formalism just
introduced to $T\!\neq\!0$ [see, e.g., \textcite{mahan:2}]. In the
latter case, by invoking once again the sudden approximation the
intensity measured in an ARPES experiment on a 2D  single-band
system can be conveniently written as:
\begin{equation}
I({\bf k},\omega)=I_0({\bf k},\nu,{\bf A})f(\omega)A({\bf
k},\omega) \label{eq:ikw} \label{eq:intone}
\end{equation}
where ${\bf k}\!=\!{\bf k}_\parallel$ is the in-plane electron
momentum, $\omega$ is the electron energy with respect to the
Fermi level, and $I_0({\bf k},\nu,{\bf A})$ is proportional to the
squared one-electron matrix element $|M_{f,i}^{\bf k}|^2$ and
therefore depends on the electron momentum, and on the energy and
polarization of the incoming photon. We also introduced the Fermi
function $f(\omega)\!=\!(e^{\omega/k_BT}\!+\!1)^{-1}$ which
accounts for the fact that direct photoemission probes only the
occupied electronic states. Note that in Eq.\,\ref{eq:ikw} we
neglected the presence of any extrinsic background and the
broadening due to finite energy and momentum resolution, which
however have to be carefully considered when performing a
quantitative analysis of the ARPES spectra (Sec.\,\ref{sec:matele}
and Eq.\,\ref{eq:ikwreal}).

The corrections to the Green's function originating from
electron-electron correlations can be conveniently expressed in
terms of the electron {\it proper self energy} $\Sigma({\bf
k},\omega)\!=\!\Sigma^{\prime}({\bf
k},\omega)\!+\!i\Sigma^{\prime\prime}({\bf k},\omega)$. Its real
and imaginary part contain all the information on the energy
renormalization and lifetime, respectively, of an electron with
band energy $\epsilon_k$ and momentum ${\bf k}$ propagating in a
many-body system. The Green's and spectral functions expressed in
terms of the self energy are then given by:
\begin{equation}
G({\bf k},\omega)=\frac{1}{\omega-\epsilon_{\bf k}-\Sigma({\bf
k},\omega)} \label{eq:gwk}
\end{equation}
\vspace{-.55cm}
\begin{equation} A({\bf
k},\omega)=-\frac{1}{\pi}\frac{\Sigma^{\prime\prime}({\bf
k},\omega)}{[\omega-\epsilon_{\bf k}-\Sigma^{\prime}({\bf
k},\omega)]^2+[\Sigma^{\prime\prime}({\bf k},\omega)]^2}
\label{eq:akw1}
\end{equation}
It is worth emphasizing that because $G(t,t^\prime)$ is a linear
response function to an external perturbation, the real and
imaginary parts of its Fourier transform $G({\bf k},\omega)$ have
to satisfy causality and, therefore, are related by Kramers-Kronig
relations. This implies that if the full $A({\bf
k},\omega)\!=\!-(1/\pi){\rm Im}\,G({\bf k},\omega)$ is available
from photoemission and inverse photoemission, one can calculate
${\rm Re}\,G({\bf k},\omega)$ and then obtain both the real and
imaginary parts of the self energy directly from
Eq.\,\ref{eq:gwk}. However, due to the lack of high-quality
inverse photoemission data, this analysis is usually performed
only using ARPES spectra by taking advantage of certain
approximations \cite{NormanMR:Exttes}. This will be discussed in
more detail in Sec.\,\ref{sec:selfmodes} together with other
practical methods used to estimate the self-energy corrections.

In general, the exact calculation of $\Sigma({\bf k},\omega)$ and,
in turn, of $A({\bf k},\omega)$ is an extremely difficult task. In
the following, as an example we will briefly consider the
interacting FL case \cite{landau:1,landau:2,landau:3}. Let us
start from the trivial $\Sigma({\bf k},\omega)\!=\!0$
non-interacting case.  The $N$-particle eigenfunction $\Psi^N$ is
a single Slater determinant and we always end up in a single
eigenstate when removing or adding an electron with momentum ${\bf
k}$. Therefore, $G({\bf
k},\omega)\!=\!1/(\omega\!-\!\epsilon_k\!\pm\!i\eta)$ has only one
pole for each ${\bf k}$, and $A({\bf
k},\omega)\!=\!\delta(\omega\!-\!\epsilon_k)$ consists of a single
line at the band energy $\epsilon_k$ (as in
Fig.\,\ref{geometry}b). In this case, the occupation numbers
$n_{{\bf k}\sigma}^{ }\!=\!c_{{\bf k}\sigma}^\dag c_{{\bf
k}\sigma}^{ }$ are good quantum numbers and for a metallic system
the {\it momentum distribution} [i.e., the expectation value
$n({\bf k})\!\equiv\!\langle n_{{\bf k}\sigma}\rangle$, quite
generally independent of the spin $\sigma$ for nonmagnetic
systems], is characterized by a sudden drop from 1 to 0 at ${\bf
k}\!=\!{\bf k}_F$ (Fig.\,\ref{geometry}b, top), which defines a
sharp Fermi surface (FS). If we now switch on the
electron-electron correlation adiabatically, (so that the system
remains at equilibrium), any particle added into a Bloch state has
a certain probability of being scattered out of it by a collision
with another electron, leaving the system in an excited state in
which additional electron-hole pairs have been created. The
momentum distribution $n({\bf k})$ will now show a discontinuity
smaller than 1 at ${\bf k}_F$ and a finite occupation probability
for ${\bf k}\!>\!{\bf k}_F$ even at $T\!=\!0$
(Fig.\,\ref{geometry}c, top). As long as $n({\bf k})$ shows a
finite discontinuity $Z_{\bf k}\!>\!0$ at ${\bf k}\!=\!{\bf k}_F$,
we can describe the correlated Fermi sea in terms of well defined
{\it quasiparticles}, i.e. electrons {\it dressed} with a manifold
of excited states, which are characterized by a pole structure
similar to the one of the non-interacting system but with
renormalized energy $\varepsilon_{\bf k}$ and mass $m^*$, and a
finite lifetime $\tau_{\bf k}\!=\!1/\Gamma_{\bf k}$. In other
words, the properties of a FL are similar to those of a free
electron gas with damped quasiparticles. As the bare-electron
character of the quasiparticle or pole strength (also called
coherence factor) is $Z_{\bf k}\!<\!1$ and the total spectral
weight must be conserved (see Eq.\,\ref{eq:intakw}), we can
separate $G({\bf k},\omega)$ and $A({\bf k},\omega)$ into a {\it
coherent} pole part and an {\it incoherent} smooth part without
poles \cite{pines:1}:
\begin{equation}
G({\bf k},\omega)=\frac{Z_{\bf k}}{\omega-\varepsilon_{\bf
k}+i\Gamma_{\bf k}}+G_{inch} \label{eq:gfl}
\end{equation}
\vspace{-.55cm}
\begin{equation} A({\bf
k},\omega)=Z_{\bf k}\frac{\Gamma_{\bf k}/\pi}
{(\omega-\varepsilon_{\bf k})^2+\Gamma_{\bf k}^2}+A_{inch}
\end{equation}
where $Z_{\bf
k}\!=\!(1\!-\!\frac{\partial\Sigma^{\prime}}{\partial\omega})^{-1}$,
$\varepsilon_{\bf k}\!=\!Z_{\bf k}(\epsilon_{\bf
k}\!+\!\Sigma^{\prime})$, $\Gamma_{\bf k}\!=\!Z_{\bf
k}|\Sigma^{\prime\prime}|$, and the self energy and its
derivatives are evaluated at $\omega\!=\!\varepsilon_{\bf k}$. It
should be emphasized that the FL description is valid only in
proximity to the FS and rests on the condition $\varepsilon_{\bf
k}\!-\!\mu\!\gg\!|\Sigma^{\prime\prime}|$ for small
$(\omega\!-\!\mu)$ and $({\bf k}\!-\!{\bf k}_F)$. Furthermore,
$\Gamma_{\bf k}\!\propto\![(\pi k_BT)^2\!+\!(\varepsilon_{\bf
k}\!-\!\mu)^2]$ for a FL system in two or more dimensions
\cite{luttinger:2,pines:1}, although additional logarithmic
corrections should be included in the two-dimensional case
\cite{HodgesC:EffFsg}. By comparing the electron removal and
addition spectra for a FL of quasiparticles with those of a
non-interacting electron system (in the lattice periodic
potential), the effect of the self-energy correction becomes
evident (see Fig.\,\ref{geometry}c and \,\ref{geometry}b,
respectively). The quasiparticle peak has now a finite lifetime
(due to $\Sigma^{\prime\prime}$), and it sharpens up rapidly thus
emerging from the broad incoherent component upon approaching the
Fermi level, where the lifetime is infinite corresponding to a
well defined quasiparticle [note that the coherent and incoherent
part of $A({\bf k},\omega)$ represent the main line and satellite
structure discussed in the previous section and shown in
Fig.\,\ref{geometry}c, bottom right]. Furthermore, the peak
position is shifted with respect to the bare band energy
$\epsilon_{\bf k}$ (due to $\Sigma^{\prime}$): as the
quasiparticle mass is larger than the band mass because of the
dressing ($m^*\!>\!m$), the total dispersion (or bandwidth) will
be smaller ($|\varepsilon_{\bf k}|\!<\!|\epsilon_{\bf k}|$).

Given that many of the normal state properties of the cuprate
HTSCs do not follow the canonical FL behavior, it is worth
illustrating the {\it marginal Fermi liquid} (MFL) model, which
was specifically proposed as a phenomenological characterization
of the HTSCs and will be referred to in Sec.\,\ref{sec:selfnodal}
\cite{VarmaCM:Phetns,VarmaCM:erratum,AbrahamsE:Whaape}. In
particular, the motivation of the MFL description was to account
for the anomalous response observed at optimal doping in, e.g.,
electrical resistivity, Raman scattering intensity, and nuclear
spin relaxation rate. The MFL assumptions are as follows: (i)
there are momentum independent excitations over most of the
Brillouin zone that contribute to spin and charge polarizability
$\chi({\bf q},\omega,T)$; (ii) the latter has a scale invariant
form as a function of frequency and temperature, namely ${\rm
Im}\chi\!\propto\!f(\omega/T)$. To visualize the implications of
this approach, we compare the MFL and FL self energy, neglecting
for simplicity any momentum dependence:
\begin{equation}
\Sigma_{FL}(\omega)\!=\!\alpha\omega + i\beta\left[\omega^2+(\pi
k_B T)^2 \right]\\
\label{eq:spectrMFL}
\label{eq:seFL}
\end{equation}
\vspace{-.55cm}
\begin{equation}
\Sigma_{MFL}(\omega)\!=\!\lambda \left[
\omega\ln\frac{x}{\omega_c}-i\frac{\pi}{2}x \right]
\label{eq:spectrMFL}
\label{eq:seMFL}
\end{equation}
Here $x\!\approx\!{\rm max}(|\omega|,T)$, $\mu\!=\!0$, $\omega_c$
is an ultraviolet cutoff, and $\lambda$ is a coupling constant
(which could in principle be momentum dependent). From
Eq.\,\ref{eq:seFL} and\,\ref{eq:seMFL} at $T\!=\!0$ we see that,
while in a FL the quasiparticles are well defined because
$\Sigma^{\prime\prime}(\omega)/\Sigma^{\prime}(\omega)$ vanishes
as $\omega$ for $\omega\!\rightarrow\!0$ and $Z_{\bf k}$ is finite
at ${\bf k}\!=\!{\bf k}_F$, in a MFL
$\Sigma^{\prime\prime}(\omega)/\Sigma^{\prime}(\omega)\!\propto\!1/\ln\omega$
is only {\it marginally} singular for $\omega\!\rightarrow\!0$,
and there are no FL-like quasiparticles because $Z_{\bf k}$
vanishes as $1/\ln\omega$ at the FS (in turn, for ${\bf
k}\!=\!{\bf k}_F$ the corresponding Green's and spectral functions
are entirely incoherent). As for a MFL description of the HTSCs,
note that from Eq.\,\ref{eq:seMFL} one obtains a contribution
linear in $T$ to the electrical resistivity (i.e.,
$\omega\!=\!0$), consistent with experiments at optimal doping.
Furthermore, the MFL self energy has been used for the lineshape
analysis of the ARPES spectra from
Bi$_2$Sr$_2$CaCu$_2$O$_{8+\delta}$ [Bi2212; see
Sec.\,\ref{sec:selfnodal} and \textcite{AbrahamsE:Whaape}]. As a
last remark, it should be emphasized that the scale invariant
low-energy excitation spectrum assumed in the MFL model is
characteristic of fluctuations associated with a $T\!=\!0$ quantum
critical point, as also discussed for the HTSCs at optimal doping
\cite{VarmaCM:Phetns,VarmaCM:Psepq-}.

Among the general properties of the spectral function there are
also several sum rules. A fundamental one, which in discussing the
FL model was implicitly used to state that $\int\!d\omega
A_{ch}\!=\!Z_{\bf k}$ and $\int\!d\omega A_{inch}\!=\!1\!-\!Z_{\bf
k}$ (where $A_{ch}$ and $A_{inch}$ refer to coherent and
incoherent parts of the spectral function, respectively), is:
\begin{equation}
\int_{-\infty}^{+\infty}d\omega A({\bf k},\omega)=1
\label{eq:intakw}
\end{equation}
which reminds us that $A({\bf k},\omega)$ describes the
probability of removing/adding an electron with  momentum {\bf k}
and energy $\omega$ to a many-body system. However, as it also
requires the knowledge of the electron addition part of the
spectral function, it is not so useful in the analysis of ARPES
data. A sum rule more relevant to this task is:
\begin{equation}
\int_{-\infty}^{+\infty}d\omega f(\omega)A({\bf k},\omega)=n({\bf
k}) \label{eq:nk}
\end{equation}
which solely relates the one-electron removal spectrum to the
momentum distribution $n({\bf k})$. When electronic correlations
are important and the occupation numbers are no longer good
quantum numbers, the discontinuity at ${\bf k}_F$ is reduced (as
discussed for the FL case) but a drop in $n({\bf k})$ is usually
still observable even for strong correlations \cite{nozieres:1}.
\begin{figure}[t!]
\centerline{\epsfig{figure=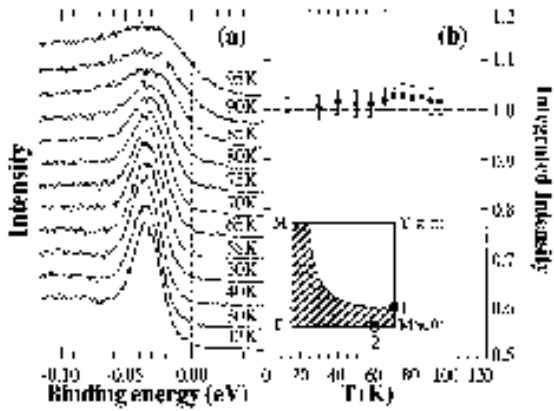,width=8cm,clip=}}
\caption{(a) Temperature dependence of the ARPES spectra from
Bi$_2$Sr$_2$CaCu$_2$O$_{8+\delta}$ ($T_c\!=\!87$\,K) measured at
${\bf k}\!=\!{\bf k}_F$ (point 1 in the Brillouin zone sketch).
(b) Temperature dependence of the integrated intensity
\cite{RanderiaM:Momdsr}.}\label{sudden}\end{figure}
Note, however, that great care is necessary in making use of
Eq.\,\ref{eq:nk} because the integral of Eq.\,\ref{eq:ikw} does
not give simply $n({\bf k})$ but rather $I_0({\bf k},\nu,{\bf
A})n({\bf k})$. Nevertheless, by tracking in momentum space the
{\it loci} of steepest descent of the experimentally determined
$n({\bf k})$, i.e. the maxima in $|\nabla\!_{\bf k}\,n({\bf k})|$,
it has been possible to identify not only the FS in Bi2212
[\textcite{CampuzanoJC:Dirop-}; Sec.\,\ref{sec:ns_Bi2212}] but
also the {\it remnant}-FS in the insulating parent compound
Ca$_2$CuO$_2$Cl$_2$ [\textcite{RonningF:PhoerF};
Sec.\,\ref{sec:_ccocRFS}].\footnote{For a more extended discussion
on the different methods typically used to experimentally
determine the FS, see:
\textcite{StraubT:Man-bo,KippL:HowdFv,RossnagelK:Thr-di}.}

An approximate sum rule was proposed by
\textcite{RanderiaM:Momdsr} under the assumptions that at low
energies $A({\bf k}_F,-\omega)\!=\!A({\bf k}_F,\omega)$, and
$I_0({\bf k},\nu,{\bf A})$ does not have significant dependence on
$\nu$ and $T$ (the dependence on ${\bf k}$ or {\bf A} in this case
will appear to be irrelevant). It states that $n({\bf k}_F)$ is
{\it independent of temperature} (note that here ${\bf k}\!=\!{\bf
k}_F$ necessarily, as for {\bf k} near but not equal to ${\bf
k}_F$, different $T$ dependence is expected for states below and
above the chemical potential). This approximate sum rule was
tested on temperature dependent ARPES spectra taken on Bi2212 at
${\bf k}_F$ near ($\pi$,0) which, as we will discuss in detail in
Sec.\,\ref{sec:speak_Bi2212}, are characterized by a remarkable
change in lineshape below the superconducting phase transition
(Fig.\,\ref{sudden}a). Indeed, as shown in Fig.\,\ref{sudden}b,
the integrated intensity of the ARPES spectra is temperature
independent, which not only satisfies the sum rule but also
suggests the validity of the sudden approximation for Bi2212 even
at 19 eV photon energy (since it is on this approximation that
Eq.\,\ref{eq:ikw} and the above analysis rest). It was also noted
that, as the low energy spectral weight in all the quasi 2D copper
oxides is mostly representative of the electronic states belonging
to the CuO$_2$ planes (Cu\,3$d$ and O\,2$p$), the validity of the
sudden approximation at this relatively low photon energy and, in
turn, of the $A({\bf k},\omega)$ interpretation of the ARPES
spectra (Eq.\,\ref{eq:ikw}), should be regarded as a general
property of this class of compounds \cite{RanderiaM:Momdsr}.

\subsection{Matrix elements and finite resolution effects}
\label{sec:matele}

As discussed in the previous section and summarized by
Eq.\,\ref{eq:ikw}, ARPES directly probes the one-particle spectral
function $A({\bf k},\omega)$. However, in extracting quantitative
information from the experiment, not only the effect of the matrix
element term $I_0({\bf k},\nu,{\bf A})$ has to be taken into
account, but also the finite experimental resolution and the
extrinsic continuous background due to the secondaries (those
electrons which escape from the solid after having suffered
inelastic scattering events and, therefore, with a reduced
$E_{kin}$). The latter two effects may be explicitly accounted for
by considering a more realistic expression for the photocurrent
$I({\bf k},\omega)$:
\begin{equation}
\int\!\!d\tilde{\omega}d\tilde{{\bf k}}\,I_0(\tilde{{\bf
k}},\!\nu,\!{\bf A})f(\tilde{\omega})A(\tilde{{\bf
k}},\!\tilde{\omega})R(\omega\!-\!\tilde{\omega})Q({\bf
k}\!-\!\tilde{{\bf k}})\!+\!B \label{eq:ikwreal}
\end{equation}
which consists of the convolution of Eq.\,\ref{eq:ikw} with energy
($R$) and momentum ($Q$) resolution functions [$R$ is typically a
Gaussian, $Q$ may be more complicated], and of the background
correction B.
\begin{figure}[t!]
\centerline{\epsfig{figure=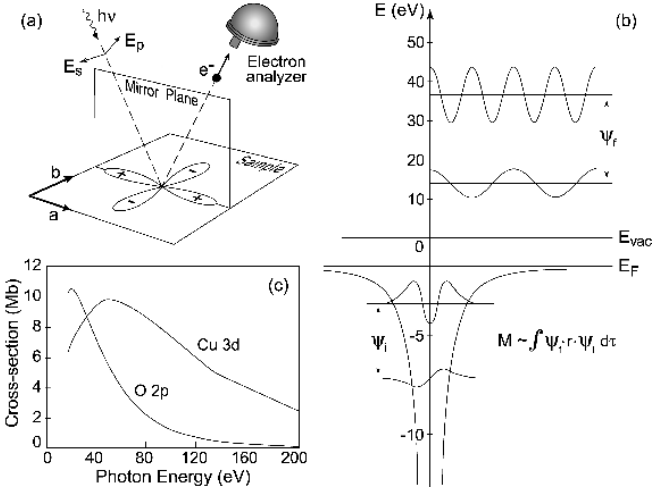,width=1\linewidth,clip=}}
\caption{(a) Mirror plane emission from a $d_{x^2-y^2}$ orbital.
(b) Sketch of the optical transition between atomic orbitals with
different angular momenta (the wavefunctions of the harmonic
oscillator are here used for simplicity) and free electron
wavefunctions with different kinetic energies [after
\textcite{hufner:1}]. (c) Calculated photoionization
cross-sections for Cu 3$d$ and O 2$p$ atomic levels [after
\textcite{YehJJ:Atospc}].} \label{matrelem}
\end{figure}
Of the several possible forms for the background function B
\cite{hufner:1}, two are more frequently used: (i) the step-edge
background (with three parameters for height, energy position, and
width of the step-edge), which reproduces the background observed
all the way to $E_F$ in an unoccupied region of momentum space;
(ii) the Shirley background
$B_{Sh}(\omega)\!\propto\!\int^\mu_\omega d\omega^\prime
P(\omega^\prime)$, which allows one to extract from the measured
photocurrent $I(\omega)\!=\!P(\omega)\!+\!c_{Sh} B_{Sh}(\omega)$
the contribution $P(\omega)$ due to the unscattered electrons
[here the only parameter is the constant $c_{Sh}$;
\textcite{ShirleyDA:Hig-re}].

Let us now very briefly illustrate the effect of the matrix
element term $I_0({\bf k},\nu,{\bf A})\!\propto\!|M_{f,i}^{\bf
k}|^2$, which is responsible for the dependence of the
photoemission data on photon energy and experimental geometry, and
may even result in complete suppression of the intensity
\cite{gobeligw:1,DietzE:Polda-,HermansonJ:Fin-st,EberhardtW:Dipsro}.
By using the commutation relation $\hbar{\bf p}\!=\!-i[{\bf
x},H]$, we can write $|M^{\bf k}_{f,i}|^2\!\propto\!|\langle
\phi_f^{\bf k}|\mbox{\boldmath$\varepsilon$}\!\cdot\!{\bf
x}|\phi_i^{\bf k}\rangle |^2$, where {\boldmath$\varepsilon$} is a
unit vector along the polarization direction of the vector
potential {\bf A}.  As in Fig\,\ref{matrelem}a, let us consider
photoemission from a $d_{x^2-y^2}$ orbital, with the detector
located in the mirror plane (when the detector is out of the
mirror plane, the problem is more complicated because of the lack
of an overall well defined even/odd symmetry). In order to have
non vanishing photoemission intensity, the whole integrand in the
overlap integral must be an even function under reflection with
respect to the mirror plane. Because odd parity final states would
be zero everywhere on the mirror plane and therefore also at the
detector, the final state wavefunction $\phi_f^{\bf k}$ itself
must be even. In particular, at the detector the photoelectron is
described by an even parity plane-wave state $e^{i{\bf kr}}$ with
momentum in the mirror plane and fronts orthogonal to it
\cite{HermansonJ:Fin-st}. In turn, this implies that
$(\mbox{\boldmath$\varepsilon$}\!\cdot\!{\bf x})|\phi_i^{\bf
k}\rangle$ must be even. In the case depicted in
Fig\,\ref{matrelem}a where $|\phi_i^{\bf k}\rangle$ is also even,
the photoemission process is symmetry allowed for {\bf A} even or
in-plane (i.e., $\mbox{\boldmath$\varepsilon$}_p\!\cdot\!{\bf x}$
depends only on in-plane coordinates and is therefore even under
reflection with respect to the plane) and forbidden for {\bf A}
odd or normal to the mirror plane (i.e.,
$\mbox{\boldmath$\varepsilon$}_s\!\cdot\!{\bf x}$ is odd as it
depends on normal-to-the-plane coordinates). Following the
notations of \textcite{MesotJ:DettFs}, for a generic initial state
of either even or odd symmetry with respect to the mirror plane,
the polarization conditions resulting in an overall even matrix
element can be summarized as:
\begin{equation}
\left\langle{\phi_f^{\bf k}}\right|{\bf A}\!\cdot\!{\bf
p}\left|{\phi_i^{\bf k}}\right\rangle \left\{{\matrix{{\phi_i^{\bf
k}\,\,even\,\,\left\langle + \right|+\left| + \right\rangle
\,\,\Rightarrow \,\,{\bf A}\,\,even} \vspace{0.1cm}\cr
{\!\!\!\phi_i^{\bf k}\,\,odd\,\,\,\,\,\left\langle +
\right|-\left| - \right\rangle \,\,\Rightarrow \,\,{\bf
A}\,\,odd}\cr }}\right.
\end{equation}

In order to discuss the photon energy dependence, from
Eq.\,\ref{eq:hint} and by considering a plane wave $e^{i{\bf kr}}$
for the photoelectron at the detector, one may more conveniently
write $|M^{\bf
k}_{f,i}|^2\!\propto\!|(\mbox{\boldmath$\varepsilon$}\!\cdot\!{\bf
k})\langle\phi_i^{\bf k}|e^{i{\bf kr}}\rangle |^2$. The overlap
integral, as sketched in Fig\,\ref{matrelem}b, strongly depends on
the details of the initial state wavefunction (peak position of
the radial part and oscillating character of it), and on the
wavelength of the outgoing plane wave. Upon increasing the photon
energy, both $E_{kin}$ and {\bf k} increase, and $M^{\bf k}_{f,i}$
changes in a non-necessarily monotonic fashion (see
Fig\,\ref{matrelem}c, for the Cu 3$d$ and the O 2$p$ atomic case).
In fact, the photoionization cross section is usually
characterized by one minimum in free atoms, the so-called Cooper
minimum \cite{Cooperjw:1}, and a series of them in solids
\cite{MolodtsovSL:Coomtp}.

Before concluding this section, it has to be emphasized that the
description of photoemission based on the sudden approximation and
the three-step model, although artificial and oversimplified,
allows an intuitive understanding of the process. However, for a
quantitative analysis of the ARPES spectra, calculations based on
the one-step model are generally required. In this case, surface
discontinuity, multiple scattering, finite lifetime effects, and
also matrix elements for initial and final state crystal
wavefunctions are included and accounted for by first principle
calculations, as we will discuss in Sec.\,\ref{sec:ns_Bi2212} for
the case of Bi2212 \cite{BansilA:Impmet}.

\subsection{State-of-the-art photoemission}
\label{sec:stateart}

In the early stages of the investigation of the HTSCs, ARPES
proved to be very successful in detecting dispersive electronic
features \cite{TakahashiT:Evia-r,OlsonCG:SupgB-,OlsonCG:Hig-re},
the $d$-wave superconducting gap \cite{ShenZ-X:Anolga}, and the
normal state pseudogap
\cite{MarshallDS:Uncese,LoeserAG:Excgtn,DingH:Speept}. Over the
past decade, a great deal of effort has been invested in further
improving this technique.
\begin{figure}[b!]
\centerline{\epsfig{figure=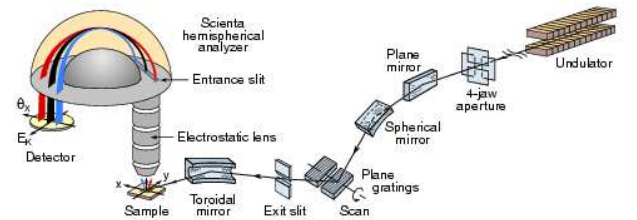,width=1\linewidth,clip=}}
\caption{Generic beamline equipped with a plane grating
monochromator and a Scienta spectrometer [Color].} \label{scienta}
\end{figure}
This resulted in an order of magnitude improvement in both energy
and momentum resolution, thus ushering in a new era in electron
spectroscopy and allowing a detailed comparison between theory and
experiment. The reasons for this progress are twofold: the
availability of dedicated photoemission beamlines on high-flux
second and third generation synchrotron facilities [for a
description of synchrotron radiation technology and applications
see \textcite{Koch:1}], and the development of the Scienta
electron spectrometers \cite{BeamsonG:PeratS,MartenssonN:verhre}.

The configuration of a generic angle-resolved photoemission
beamline is shown in Fig.\,\ref{scienta}.
\begin{figure}[t!]
\centerline{\epsfig{file=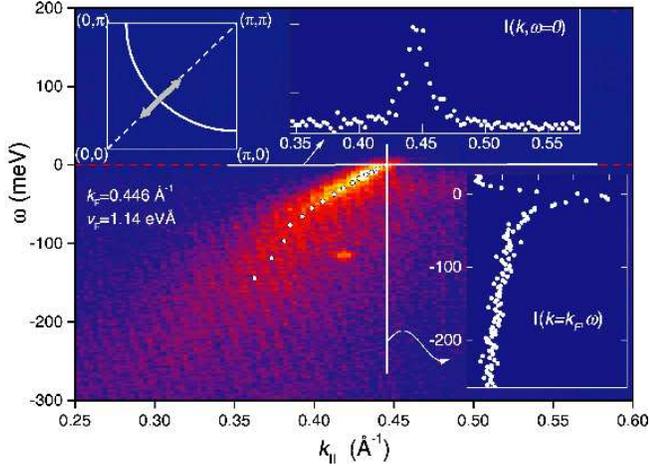,clip=,angle=0,width=0.99\linewidth}}
\caption{Energy ($\omega$) versus momentum (${\bf k}_\parallel$)
image plot of the photoemission intensity from
Bi$_2$Sr$_2$CaCu$_2$O$_{8+\delta}$ along (0,0)-($\pi$,$\pi$). This
$k$-space cut was taken across the FS (see sketch of the 2D
Brillouin zone) and allows a direct visualization of the photohole
spectral function $A({\bf k},\omega)$ (weighted by Fermi
distribution and matrix elements): the quasiparticle dispersion
can be clearly followed up to $E_F$, as emphasized by the white
circles. Energy scans at constant momentum (right) and momentum
scans at constant energy (top) define {\it energy distribution
curves} (EDCs) and {\it momentum distribution curves} (MDCs),
respectively. After \textcite{VallaT:Eviqcb} [Color].}
\label{valla1}
\end{figure}
A beam of white radiation is produced in a wiggler or an undulator
(so-called `insertion devices', these are the straight sections of
the electron storage ring where radiation is produced), is
monochromatized at the desired photon energy by a grating
monochromator, and is focused on the sample. Alternatively, a
gas-discharge lamp can be used as a radiation source (once
properly monochromatized, to avoid complications due to the
presence of different satellites and refocused to a small spot
size, essential for high angular resolution). However, synchrotron
radiation offers important advantages: it covers a wide spectral
range (from the visible to the X-ray region) with an intense and
highly polarized continuous spectrum, while a discharge lamp
provides only a few unpolarized resonance lines at discrete
energies. Photoemitted electrons are then collected by the
analyzer, where kinetic energy and emission angle are determined
(the whole system is in high vacuum at pressures lower than
5$\times\!10^{-11}$ torr).

A conventional hemispherical analyzer consists of a multi-element
electrostatic input lens, a hemispherical deflector with entrance
and exit slits, and an electron detector (i.e., a channeltron or a
multi-channel detector). The heart of the analyzer is the
deflector which consists of two concentric hemispheres (of radius
$R_1$ and $R_2$). These are kept at a potential difference $\Delta
V$, so that only those electrons reaching the entrance slit with
kinetic energy within a narrow range centered at
$E_{pass}\!=\!e\Delta V/(R_1/R_2\!-\!R_2/R_1)$ will pass through
this hemispherical capacitor, thus reaching the exit slit and then
the detector. This way it is possible to measure the kinetic
energy of the photoelectrons with an energy resolution given by
$\Delta E_a\!=\!E_{pass}(w/R_0\!+\!\alpha^2\!/4)$, where
$R_0\!=\!(R_1\!+\!R_2)/2$, $w$ is the width of the entrance slit,
and $\alpha$ is the acceptance angle. The role of the
electrostatic lens is that of decelerating and focusing the
photoelectrons onto the entrance slit. By scanning the lens
retarding potential one can effectively record the photoemission
intensity versus the photoelectron kinetic energy. One of the
innovative characteristics of the Scienta analyzer is the
two-dimensional position-sensitive detector consisting of two
micro-channel plates and a phosphor plate in series, followed by a
CCD camera.
\begin{figure}[b!]
\centerline{\epsfig{figure=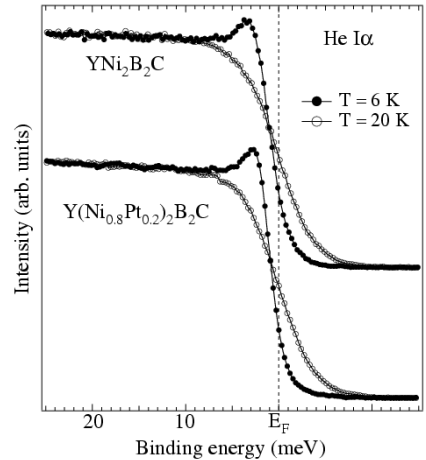,width=6cm,clip=}}
\caption{Angle-integrated photoemission data from Ni borocarbides
taken with 2.0 meV resolution
\cite{YokoyaT:Ult-re}.}\label{yokoya}\end{figure}
\begin{figure*}[t!]
\centerline{\epsfig{figure=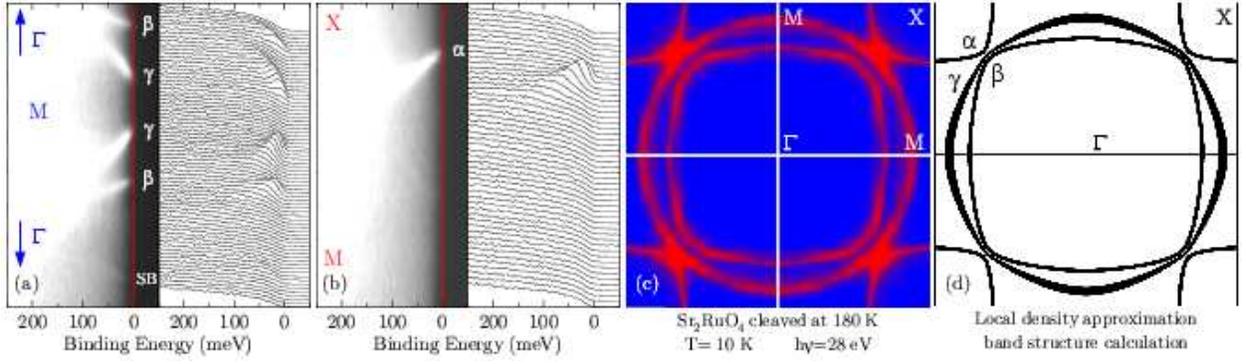,width=16.5cm,clip=}}
\vspace{0cm} \caption{ARPES spectra and corresponding intensity
plot along $\Gamma$-M (a), and M-X (b). Measured (c) and
calculated (d) FSs \cite{MazinII:Fersfi}. All data were taken at
10\,K on Sr$_2$RuO$_4$ crystals cleaved at 180\,K
\cite{DamascelliA:Fersss} [Color].} \label{srofs}
\end{figure*}
In this case, no exit slit is required: the electrons, spread
apart along the $Y$ axis of the detector (Fig.\,\ref{scienta}) as
a function of their kinetic energy due to the travel through the
hemispherical capacitor, are detected simultaneously [in other
words, a range of electron energies is dispersed over one
dimension of the detector and can be measured in parallel;
scanning the lens voltage is in principle no longer necessary, at
least for narrow energy windows (a few percent of $E_{pass}$)].
Furthermore, contrary to a conventional electron analyzer in which
the momentum information is averaged over all the photoelectrons
within the acceptance angle (typically $\pm1^\circ$), the Scienta
system can be operated in angle-resolved mode, which provides
energy-momentum information not only at a single $k$-point but
along an extended cut in $k$-space. In particular, the
photoelectrons within an angular window of $\sim\!14^\circ$ along
the direction defined by the analyzer entrance slit are focused on
different $X$ positions on the detector (Fig.\,\ref{scienta}). It
is thus possible to measure multiple energy distribution curves
simultaneously for different photoelectron angles, obtaining a 2D
snapshot of energy versus momentum (Fig.\,\ref{valla1}).

The Scienta SES200 analyzer ($R_0\!=\!200$ mm) typically allows
energy and angular resolutions of approximately a few meV and
$0.2^\circ$, respectively (for the 21.2 eV photons of the
HeI$\alpha$ line, as one can obtain from Eq.\,\ref{eq:momentum},
0.2$^\circ$ corresponds to $\sim$1\% of the cuprates'
Brillouin-zone edge $\pi/a$). Note, however, that in estimating
the total energy resolution achievable on a beamline one has to
take into account also $\Delta E_{m}$ of the monochromator, which
can be adjusted with entrance and exit slits (the ultimate
resolution a monochromator can deliver is given by its resolving
power $R\!=\!E/\Delta E_{m}$; it can be as good as 1-2 meV for 20
eV photons but worsens upon increasing photon energy). To maximize
the signal intensity at the desired total $\Delta E$,
monochromator and analyzer should be operated at comparable
resolutions.

To date, one of the most impressive examples of high energy
resolution for a photoemission experiment on a solid sample was
reported by \textcite{YokoyaT:Ult-re}, who performed photoemission
measurements on the Ni borocarbides in a system equipped with a
Scienta SES2002 electron analyzer (newer version of the SES200),
and a Gammadata high-flux discharge lamp combined with a toroidal
grating monochromator. With this system, capable of an energy
resolution of 1.5 meV, \textcite{YokoyaT:Ult-re} measured
angle-integrated photoemission spectra with 2.0 meV resolution on
YNi$_2$B$_2$C and Y(Ni$_{0.8}$Pt$_{0.2}$)$_2$B$_2$C, which are
characterized by a superconducting transition at 15.4 and 12.1 K,
respectively (see Fig.\,\ref{yokoya}). Due to the extremely high
resolution, they successfully detected the opening of the small
superconducting gap, as evidenced by the shift to high binding
energies of the 6 K spectra leading-edge midpoint (which is
instead located at $E_F$ at 20 K, as expected for a metal), and by
the appearance of a peak below $E_F$ which directly reflects the
piling up of the density of states due to the gap opening. By a
detailed analysis, the authors could conclude in favor of an
anisotropic $s$-wave superconducting gap with
$\Delta_{max}\!\simeq\!2.2$ and 1.5 meV in the pure and Pt-doped
samples, respectively.

It has to be emphasized that, when angle resolved experiments are
performed, one has to compromise the energy resolution to improve
the angular resolution. Therefore, in angle resolved experiments,
$\Delta E$ is typically set in the range of 5-20 meV. To
illustrate the capability of state-of-the-art ARPES and how
critical the improvement in angle resolution has been, the novel
superconductor Sr$_2$RuO$_4$ is a particularly good example. This
system is isostructural to the archetypal cuprate parent compound
La$_2$CuO$_4$ (see Fig.\,\ref{bz_lsco}), but RuO$_{2}$ planes
replace the CuO$_{2}$ planes. Its low-energy electronic structure,
as predicted by band-structure calculations is characterized by
three bands crossing the chemical potential
\cite{SinghDJ:RelS24,OguchiT:Elebst}. These define a complex FS
comprised of two electron pockets and one hole pocket
(Fig.\,\ref{srofs}d), which have been clearly observed in de
Haas-van Alphen experiments
\cite{MackenzieAP:Quaotl,BergemannC:DetttF}. On the other hand,
early photoemission measurements suggested a different topology
\cite{YokoyaT:Ang-re,YokayaT:ExtVHs,LuDH:Fersev}, which generated
a certain degree of controversy in the field
\cite{PuchkovAV:ARPrS2}. This issue was conclusively resolved only
by taking advantage of the high energy and momentum resolution of
the `new generation' of ARPES data: it was then recognized that a
surface reconstruction \cite{MatzdorfR:Fersbl} and, in turn, the
detection of several direct and folded surface bands were
responsible for the conflicting interpretations
\cite{DamascelliA:Fersss,DamascelliA:FersS2,DamascelliA:1,shenkm:1}.
Fig.\,\ref{srofs}a,b show high resolution ($\Delta E\!=\!14$ meV,
$\Delta k\!=\!1.5$\% of the zone edge) ARPES data taken at 10\,K
with 28\,eV photons on a Sr$_2$RuO$_4$ single crystal cleaved at
180 K [for Sr$_2$RuO$_4$, as recently discovered, the
high-temperature cleaving suppresses surface contributions to the
photoemission signal and allows one to isolate the bulk electronic
structure; \textcite{DamascelliA:Fersss}]. Many well defined
quasiparticle peaks disperse towards the Fermi energy and
disappear upon crossing $E_F$.
\begin{figure}[t!]
\centerline{\epsfig{figure=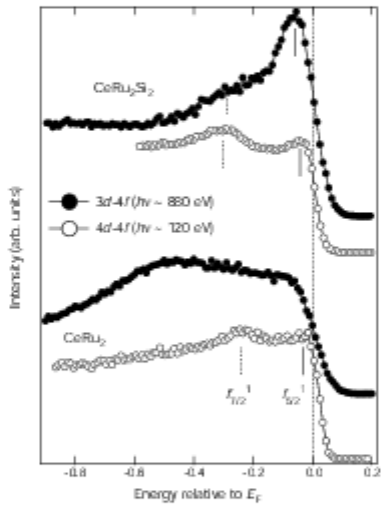,width=5.5cm,clip=}}
\vspace{0cm} \caption{High energy angle-integrated photoemission
data from Ce compounds at $T\!=\!20$ K
\cite{SekiyamaA:Probsc}.}\label{suga}\end{figure}
A Fermi energy intensity map (Fig.\,\ref{srofs}c) can then be
obtained by integrating the spectra over a narrow energy window
about $E_F$ ($\pm10$ meV). As the spectral function (multiplied by
the Fermi function) reaches its maximum at $E_F$ when a band
crosses the Fermi energy, the FS is identified by the local maxima
of the intensity map. Following this method, the three sheets of
FS are clearly resolved and are in excellent agreement with the
theoretical calculations (Fig.\,\ref{srofs}d).

The improvement in synchrotron radiation technology results in the
availability of high resolution beamlines operating at
increasingly higher photon energies. The significance of this
progress is well exemplified by the resonance photoemission
experiments on Ce compounds performed by
\textcite{SekiyamaA:Probsc} on a high resolution soft X-ray
(500-1500 eV) photoemission system [this consists of a Scienta
SES200 spectrometer, combined with a varied line-spacing plane
grating monochromator on a high brilliance beamline at the
SPring-8 synchrotron facility \cite{SaitohY:Twihub}]. Ce compounds
are characterized by a very different degree of hybridization
between the 4$f$ electronic states and other valence bands; the
strength of the hybridization is stronger the larger the Kondo
temperature $T_K$. However, although CeRu$_2$Si$_2$ and CeRu$_2$
are characterized by very different $T_K$ (approximately 22 and
1000 K, respectively), earlier photoemission studies reported
similar spectra for the Ce 4$f$ electronic
states.\footnote{\textcite{PattheyF:Hig-re,WeschkeE:Surbes,KaindlG:Staph-,JoyceJJ:Tem-in,DuoL:Ce4sst,GarnierM:Apptsi}.}
By performing angle-integrated high resolution photoemission
experiments at the 3$d$-4$f$ ($h\nu\!\simeq\!880$ eV, $\Delta
E\!\simeq\!100$ meV) and 4$d$-4$f$ ($h\nu\!\simeq\!120$ eV,
$\Delta E\!\simeq\!50$ meV) resonances (see Fig.\,\ref{suga}),
\textcite{SekiyamaA:Probsc} observed that, while the spectra for
the two compounds are indeed qualitatively similar at 120 eV
photon energy, they are remarkably different at 880 eV. As the
photoelectron mean free path increases from approximately 5 to
almost 20 \AA\ upon increasing the photon energy from 120 to 880
eV \cite{SeahMP:Quaess}, it was concluded that the 4$d$-4$f$
spectra mainly reflect the surface 4$f$ electronic states. These
are different from those of the bulk and are not representative of
the intrinsic electronic properties of the two compounds, which
are more directly probed at 880 eV: the 3$d$-4$f$ spectra show a
prominent structure corresponding to the tail of a Kondo peak in
CeRu$_2$Si$_2$, and a broader feature reflecting the more
itinerant character of the 4$f$ electrons in CeRu$_2$
\cite{SekiyamaA:Probsc}.

At this point, it is worth emphasizing that while the examples
discussed in this section underline certain shortcomings of
photoemission and may raise some doubts concerning the general
validity of the ARPES results, on the other hand they demonstrate
that, by taking full advantage of the momentum and energy
resolution as well as of the photon energy range now available,
state-of-the-art ARPES is not only a reliable technique but is
also a unique tool for {\it momentum space microscopy}.

\section{From Mott insulator to high-{\boldmath$T_c$} superconductor}
\label{sec:MottIns}

\begin{figure}[b!]
\centerline{\epsfig{figure=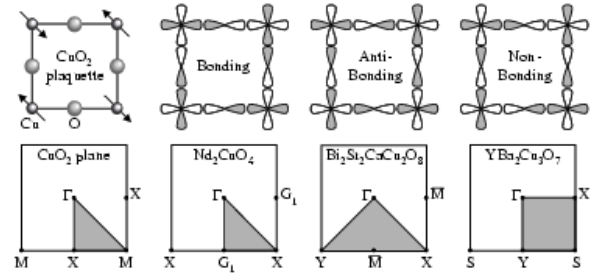,width=8cm,clip=}}
\vspace{0cm} \caption{Cu-O$_2$ plaquette, phase at ($\pi$,$\pi$)
of Cu $d_{x^2-y^2}$ and O 2$p$ orbitals for bonding, antibonding
and nonbonding hybridized wavefunctions for the bare CuO$_2$ plane
(i.e., square lattice; see also Fig.\,\ref{band_mott}a), and 2D
projected Brillouin zones with conventional notations for
different copper oxides (shaded areas represent the irreducible
symmetry units).} \label{bz_tot}
\end{figure}

\begin{figure}[b!]
\centerline{\epsfig{figure=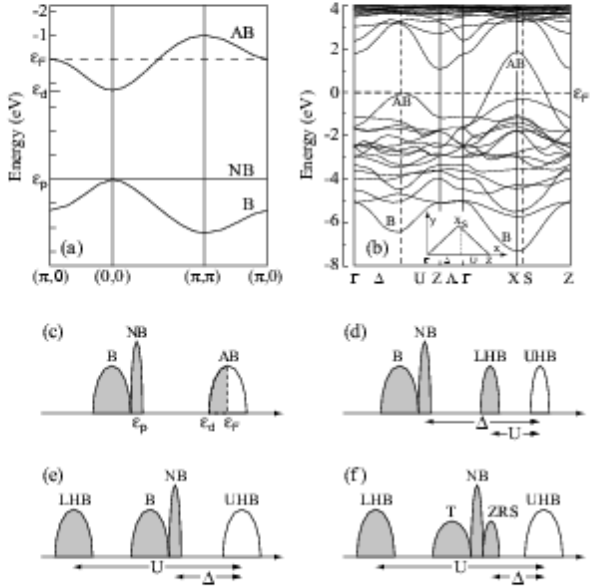,width=8.5cm,clip=}}
\caption{(a) Bonding, antibonding, and nonbonding bands for the
square lattice \cite{fuldeP:1} at half filling, and (c)
corresponding density of states [(b) shows band structure
calculations for La$_2$CuO$_4$ \cite{MattheissLF:Elebps}]. The
system is metallic in absence of electronic correlations, and
becomes a Mott (d) or a charge-transfer (e) insulator,
respectively, for $\Delta\!>\!U\!>\!W$ and $U\!>\Delta\!>\!W$. In
the latter case, due to the hybridization with the upper Hubbard
band, the nonbonding band further splits in triplet and Zhang-Rice
singlet states (f).} \label{band_mott}
\end{figure}

In the following we summarize the basic characteristics of the
crystal and electronic structures of the copper
oxides.\footnote{For a more detailed description see:
\textcite{PickettWE:Elesth,DagottoE:Coreh-,auerbach:1,fuldeP:1,rao:1,MarkiewiczRS:surtVH,MARKIEWICZRS:VANSHS,ImadaM:Met-in,TokuraY:Orbpt-,OrensteinJ:Advtph,SachdevS:Quaccg}.}
\begin{figure}[t!]
\centerline{\epsfig{figure=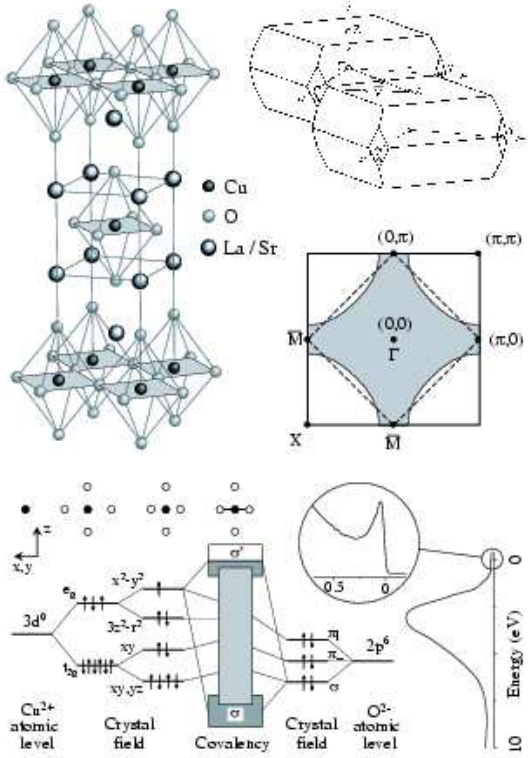,width=7.5cm,clip=}}
\caption{(top) Crystal structure, 3D Brillouin zone (body-centered
tetragonal) and its 2D projection for LSCO (diamond: FS at half
filling calculated with only nearest neighbor hopping; grey area:
FS obtained including also next-nearest neighbor hopping). Note
that $\overline{\!M}$ is the midpoint along $\Gamma$-$Z$ and not a
true symmetry point. (bottom) Crystal field splitting and
hybridization giving rise to the Cu-O bands \cite{FinkJ:Elessh},
and generic LSCO ARPES spectrum (the circle shows the low-energy
scale we will focus on in this review).} \label{bz_lsco}
\end{figure}
We will use, as an example, the archetypical cuprate
superconductor La$_{2-x}$Sr$_2$CuO$_4$ (LSCO) and its parent
compound La$_2$CuO$_4$ (see Fig.\,\ref{rmp_pd}), whose undistorted
high-temperature tetragonal (HTT) structure is sketched in
Fig.\,\ref{bz_lsco}. Upon lowering the temperature, several
structural phase transitions occur, characterized by coherent
rotations of the CuO$_6$ octahedra [see, e.g.,
\textcite{KimuraH:Striaw}]. However, no discernible effect has
been found in photoemission and ARPES data are usually discussed
within HTT notations. The corresponding three-dimensional
Brillouin zone, which is most relevant to the study of the
momentum-resolved electronic properties, is also sketched in
Fig.\,\ref{bz_lsco}. However, as the cuprate HTSCs have a quasi 2D
electronic structure with weak dispersion along the $z$ axis, in
the discussion of the ARPES data we will refer to the 2D projected
zones as the one presented in Fig.\,\ref{bz_lsco} for LSCO or in
Fig.\,\ref{bz_tot} for other systems. As emphasized in
Fig.\,\ref{bz_lsco}, the most important structural element is
represented by the CuO$_2$ planes which form single (as in LSCO)
or multi-layer blocks separated from each other by the so-called
{\it charge reservoir layers} (La/Sr in Fig.\,\ref{bz_lsco}).
Depending on the number $N$ of CuO$_2$ planes contained within the
characteristic blocks ($N$ is also the number of Cu ions per
formula unit), the cuprates are classified into, e.g.,
single-layer [LSCO, Bi$_2$Sr$_2$CuO$_{6+\delta}$,
Nd$_{2-x}$Ce$_x$CuO$_4$, and (Sr,Ca)$_2$CuO$_2$Cl$_2$], bilayer
[Bi$_2$Sr$_2$CaCu$_2$O$_{8+\delta}$ and
YBa$_2$Cu$_3$O$_{7-\delta}$], and trilayer materials
[Bi$_2$Sr$_2$Ca$_2$Cu$_3$O$_{10+\delta}$], {\it et cetera}. This
structural characteristic profoundly affects the superconducting
properties: within each family of cuprates $T_c$ increases with
$N$, at least for $N\!\leq\!3$
\cite{DiStasioM:Noncdl,TarasconJM:Prespt}. For instance, within
the Bi-based cuprate HTSCs, a maximum $T_c$ of 34, 96, and 110\,K
is found for $N\!=\!1$,\,2, and 3, respectively \cite{eisaki:1}.
By substituting different elements in the reservoir layers or by
varying their oxygen content (other methods are also possible,
depending on the system) one can dope charge carriers into the
CuO$_2$ planes. The latter are believed to be responsible for
high-temperature superconductivity as the Cu-O bands are the
lowest energy electronic states and therefore directly determine
the macroscopic electronic properties. For LSCO this is clearly
indicated by the local-density approximation (LDA) band structure
calculations presented in Fig.\,\ref{band_mott}b, where all the
bands between $E_F$ and 8 eV binding energy appear to be of Cu
3$d$ or O 2$p$ character [a schematic picture of the origin of the
Cu-O bands in the cuprates is given at the bottom of
Fig.\,\ref{bz_lsco}, where the effect of crystal field splitting
and, in particular, Jahn-Teller \cite{jahn:1} distortion of the
octahedron on the Cu $e_g$ and $t_{2g}$ levels is also shown].
Analogous results are obtained for a square lattice with three
orbitals (Cu $d_{x^2-y^2}$, and O 2$p_x$ and 2$p_y$) at half
filling (i.e., one electron per Cu $d_{x^2-y^2}$ orbital
corresponding to $x\!=\!0$ in the phase diagram of
Fig.\,\ref{rmp_pd}), which emphasize the presence of one
antibonding band at the Fermi level, and of nonbonding and bonding
bands at higher binding energy [see Fig.\,\ref{band_mott}a, and
also Fig.\,\ref{band_mott}c and Fig.\,\ref{bz_tot} where the
corresponding metallic density of states and the symmetry of the
hybridized wavefunctions at ($\pi$,$\pi$) are shown].

The band structures of Fig.\,\ref{band_mott}a,b imply metallic
behavior and a FS with volume equal to half of the Brillouin zone.
In particular, for the calculations of Fig.\,\ref{band_mott}a in
which only nearest neighbor hopping (Cu-O, $t_{pd}$) was
considered, a diamond-like FS is obtained (Fig.\,\ref{bz_lsco}). A
distortion of this FS takes place with the more realistic
inclusion of also next-nearest neighbor hopping (O-O, $t_{pp}$),
which results in the FS given by the grey area in
Fig.\,\ref{bz_lsco} \cite{DICKINSONPH:CALAPT}. These results seem
to well correspond to the photoemission spectrum shown in
Fig.\,\ref{bz_lsco}, where a several eV broad valence band and a
low-energy quasiparticle peak are observed (note that the
$\sim\!500$\,meV region shown in the enlargement is the energy
range that we will be mostly dealing with throughout this
article).
However, the experimental spectrum of Fig.\,\ref{bz_lsco} was
obtained on optimally doped LSCO ($x\!=\!0.15$ point in
Fig.\,\ref{rmp_pd}), whereas on the basis of the band structure
results of Fig.\,\ref{band_mott}b metallic behavior is expected
also for undoped La$_2$CuO$_4$ ($x\!=\!0$ point in
Fig.\,\ref{rmp_pd}), which on the contrary is an AF insulator.
This contradiction reflects the failure of the independent
particle picture (assumed in band calculations like those of
Fig.\,\ref{band_mott}a,b) and suggests that the undoped parent
compounds of the cuprate superconductors may belong to the class
of the {\it Mott-Hubbard insulators}
\cite{mott:1,mott:2,mott:3,hubbard:1,hubbard:2,ANDERSONPW:3}.
These systems, because of the odd number of electrons per unit
cell are erroneously predicted by band theory to be paramagnetic
metals, with a partially filled {\it d}-band in the case of
transition metal oxides such as, e.g., CoO. The reason for this
failure lies in the on-site electron-electron repulsion $U$, which
is much larger than the bandwidth $W$. As a consequence, the
conduction band splits in upper and lower Hubbard bands and these
compounds are rather good insulators with an optical gap $U$ of a
few eV between the two Hubbard bands. Similarly in the case of the
copper oxides, as the on-site electron-electron repulsion $U$ for
the Cu 3$d$ electrons is comparable to the bandwidth $W\!=\!8t$
(which is the tight binding result for the square lattice), the
antibonding band splits in upper and lower Hubbard bands and
charge fluctuations are suppressed (Fig.\,\ref{band_mott}d). It
has to be emphasized, however, that in the cuprates the Cu-O
charge-transfer energy $\Delta$ is smaller than the on-site
Coulomb repulsion $U$ (Fig.\,\ref{band_mott}e), which
characterizes these compounds more precisely as {\it
charge-transfer insulators} \cite{ZaanenJ:Banges}.

Therefore the cuprates should be described in terms of the
three-band extended Hubbard model
\cite{VarmaCM:Chates,varma:1,EmeryVJ:Theh-T}, where Cu
3$d_{x^2-y^2}$, as well as O 2$p_x$ and 2$p_y$ orbitals are
explicitly considered. However, because of the finite
hybridization between the correlated Cu and the
non-interacting-like O orbitals, the first electron-removal states
correspond to the O-derived Zhang-Rice singlet band
\cite{ZhangFC:EffHsC}. It was then suggested that the cuprates may
be considered equivalent to an effective single-band Mott-Hubbard
system with the Zhang-Rice singlet band playing the role of the
lower Hubbard band, and an in-plane Cu-derived band as the upper
Hubbard band. These states are separated by an effective Mott gap
of the order of $\Delta$ (Fig.\,\ref{band_mott}f). Note that
although not universally agreed
upon,\footnote{\textcite{VarmaCM:Chates,varma:1,EmeryVJ:Theh-T}.
It should be emphasized however that the existence and stability
of Zhang-Rice singlets in the cuprates, and in turn the
description in terms of an effective one-band Hubbard model, have
recently received further support from spin-resolved photoemission
studies on Bi$_2$Sr$_2$CaCu$_2$O$_{8+\delta}$, which suggest the
pure singlet character of the first ionization states
\cite{BrookesNB:DetZ-R,TjengLH:Spi-re}.} this line of thinking is
widely used in the literature and supports the early proposal by
\textcite{ANDERSONPW:RESVBS} that the essential physics of the
cuprates would be captured by the one-band Hubbard model; this
contains a single kinetic-energy term proportional to the
nearest-neighbor hopping amplitude $t$, in addition to the Hubbard
$U$ term that favors electron localization and results in
`frustration' of the kinetic energy:
\begin{equation}
H=-t\sum_{\langle ij\rangle,\sigma}
(c^\dag_{i\sigma}c_{j\sigma}+{\rm H.c.})+U\sum_i
n_{i\uparrow}n_{i\downarrow}
\end{equation}
Here $c^\dag_{i\sigma}$ ($c_{i\sigma}$) creates (annihilates) an
electron or hole on site $i$ with spin $\sigma$, $\langle ij
\rangle$ identifies nearest-neighbor pairs, and
$n_{i\sigma}\!=\!c^\dag_{i\sigma}c_{i\sigma}$ is the number
operator. In the strong coupling limit ($U\!\gg\!t$) at half
filling ($x\!=\!0$, i.e., one electron per Cu site in a
3$d_{x^2-y^2}$ orbital), the AF state \cite{ANDERSONPW:2} results
from the fact that, when nearest-neighbor spins are antiparallel
to each other, the electrons gain kinetic energy by undergoing
virtual hopping to neighboring sites (because of the Pauli
principle, hopping is forbidden for parallel spins). Note that by
projecting out the doubly occupied states at large $U$
\cite{DagottoE:Coreh-}, the Hubbard Hamiltonian simplifies into
the $t$-$J$ Hamiltonian, which is more commonly used in studying
the low-lying excitations of the 1/2-filled AF insulator:
\begin{equation}
H=-t\!\sum_{\langle ij\rangle,\sigma}
(\tilde{c}^\dag_{i\sigma}\tilde{c}_{j\sigma}+{\rm
H.c.})+J\!\sum_{\langle ij\rangle} ({\bf S}_i\!\cdot\!{\bf
S}_j-\frac{n_{i}n_{j}}{4})
\end{equation}
where the operator
$\tilde{c}_{i\sigma}\!=\!c_{i\sigma}(1\!-n_{i\,-\sigma})$ excludes
double occupancy, $J\!=\!4t^2\!/U$ is the AF exchange coupling
constant, and $S_i$ is the spin operator. At half filling, as
charge excitations are gapped, we find at low-energy only spin
excitations governed by the AF Heisenberg Hamiltonian
$H\!=\!J\sum{\bf S}_i\cdot{\bf S}_j$ (the constant term
$-n_{i}n_{j}/4$ is usually neglected). Away from half filling, the
$t$-$J$ model describes the so called `doped AF', i.e., a system
of interacting spins and mobile holes. The latter acquire a
`magnetic dressing' because they are perturbing the correlations
of the spin background that they move through.

As we will see in more detail in Sec.\,\ref{sec:ns_ccoc}, the
ARPES work on the undoped insulator provides a starting point to
understand the doping evolution of the electronic structure of the
cuprate HTSCs and emphasizes a fundamental problem in the
theoretical description of the doped 2D AF: the Heisenberg model
is so strongly perturbed by the addition of mobile holes that,
above a certain doping level, some form of spin liquid may be a
better {\it ansatz} than the long range ordered N\'{e}el state.
This point is centrally important to high-$T_c$ superconductivity
and the HTSCs, which are poor conductors in the normal state with
a behavior fundamentally different from the FL paradigm and are
often regarded as doped AFs \cite{OrensteinJ:Advtph}. For this
reason the BCS theory \cite{Bardeen:1}, which was developed for
FL-like metals and has been so successful in describing
conventional superconductors, is generally considered not to have
the appropriate foundation for the description of the HTSCs. A new
approach may therefore be needed, and a necessary requirement for
a theory aiming to capture the essential physics of high-$T_c$
superconductivity might be the inclusion of the essential physics
of the doped AF: the competition between AF and Coulomb
interactions (which induce localization), and zero point kinetic
energy (which favors delocalization). Along this direction, the
most radical models seem to be those based on: (i) the resonating
valence bond (RVB) state and the related spin-charge separation
picture,\footnote{\textcite{ANDERSONPW:RESVBS,AFFLECKI:LARLTH,MaekawaS:Eleces,SuzumuraY:MeaftR,KotliarG:Supitl,LeePA:Gauttn,Xiao-GangWen:Theuc.,IoffeLB:EffvHs,LaughlinRB:Eviqdp,BalentsL:Nodltt,BalentsL:Duaopn,BalentsL:Duavts,AndersonPW:Souqph,LeeD-H:SupdMi}.}
(ii)
stripes,\footnote{\textcite{ZaanenJ:Chamdl,EmeryVJ:Frueps,EmeryV:1,SalkolaMI:Impcos,BianconiA:Dettll,KivelsonSA:Elel-c,IchiokaM:Elesst,TohyamaT:Effses,Zaanen:1,WhiteSR:Phassf,MarkiewiczRS:Disosp,FleckM:One-di,ChemyshevAL:Metstd,ZacherMG:Strdas,han:1}.}
and (iii) quantum
criticality.\footnote{\textcite{ChakravartyS:Two-di,VarmaCM:Phetns,VarmaCM:Non-Fe,LittlewoodPB:Phetns,SachdevS:Uniq-c,SokolA:Towump,EmeryVJ:SoloKa,CastellaniC:Sinqst,ChakravartyS:Hidotc}.}
Although very different, these theoretical approaches have one
common denominator: superconductivity is not simply caused by the
pairing of two quasiparticles, as in the BCS case, rather it is
the process in which the quasiparticle itself forms. Furthermore,
contrary to the standard theories of solids where any phase
transition into a long-range ordered state is driven by the gain
in potential energy, in cases (i) and (ii) the driving mechanism
for the superconducting phase transition is identified with the
gain in kinetic energy, a scenario which has recently received
direct support from the experimental investigation of the optical
conductivity of Bi$_2$Sr$_2$CaCu$_2$O$_{8+\delta}$
\cite{molegraaf:1}.  In the stripe or RVB models the hopping of
pairs of holes perturbs the AF spin background less than
individual holes.
\begin{figure}[t!]
\centerline{\epsfig{figure=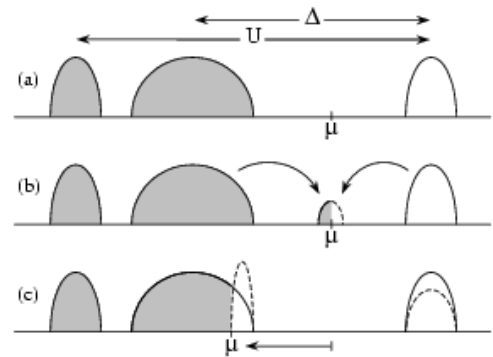,width=6.5cm,clip=}}
\vspace{.1cm}\caption{Doping of a charge-transfer insulator (a):
$\mu$ is pinned inside the charge-transfer gap and states move
towards the chemical potential (b); alternatively (c), $\mu$
shifts to the top of the valence band and spectral weight is
transferred as a consequence of correlations
\cite{vanVeenendaalMA:ElesB2}.} \label{mott}
\end{figure}
However, it is only when charge fluctuations become phase coherent
that the frustration of the kinetic energy is released, and
superconductivity sets in.

Upon further doping the system, AF correlations are reduced and a
metallic state appears. Eventually (i.e., in the optimum and
overdoped regime), the AF state is destroyed and a large LDA-like
FS emerges, with a volume which scales as ($1\!-\!x$) counting
electrons ($x$ is the concentration of doped holes for p-type
HTSCs), as expected within the FL approach. In this context, the
first important question to answer concerns the way the low energy
states emerge in the very underdoped regime (Fig.\,\ref{mott}).
For $x\!\ll\!1$, two alternative scenarios have been proposed
[see, e.g.,
\textcite{ALLENJW:RESPNN,DagottoE:DensdH,vanVeenendaalMA:ElesB2}]:
first, the chemical potential $\mu$ is pinned inside the
charge-transfer gap $\Delta$ as `in-gap states' are created
(Fig.\,\ref{mott}b); second, the chemical potential moves
downwards into the top of the valence band and states are
transferred from the upper to the lower Hubbard band because of
correlations (Fig.\,\ref{mott}c).

Another relevant question is how do the low-lying states evolve
upon going from the underdoped to the overdoped regime, where
FL-like behavior seems to recover? To better organize the
discussion, let us present an overview of some relevant
theoretical models. They can be classified as: (i) those that
preserve the underlying crystalline symmetry, and (ii) those that
break this symmetry (note that also the scenarios based on a
dynamical breaking of symmetry should be taken into account
because ARPES is sensitive to the latter, due to the relatively
high excitation energy).
\begin{figure}[t!]
\centerline{\epsfig{figure=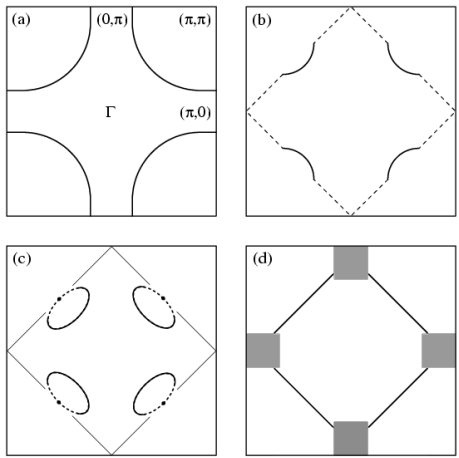,width=6.5cm,clip=}}
\vspace{.1cm}\caption{Calculated CuO$_2$ plane FS from: (a) LDA
(with next-nearest neighbor hopping), (b) truncation of a 2D FS
due to {\it umklapp} scattering \cite{FurukawaN:InsL-F}, (c)
RVB/flux-phase \cite{Xiao-GangWen:Theuc.}, and (d) vertical and
horizontal domains of disordered stripes
\cite{SalkolaMI:Impcos}.}\label{models}\end{figure}
The first models to be mentioned among (i) are the FL and band
structure perspectives \cite{pines:1,PickettWE:Elesth}, which
sever the connection to the undoped AF insulator by assuming that
the screening in the doped metal is strong enough for the FL
formalism to recover; in this case a well defined FS is expected
(Fig.\,\ref{models}a), with a volume proportional to ($1\!-\!x$)
in agreement with Luttinger's theorem \cite{luttinger:1}. An
alternative scenario considers the breakdown of FL theory due to
{\it umklapp} scattering
\cite{FurukawaN:Trut-d,FurukawaN:InsL-F,HonerkampC:BretL-}. As a
result, in the underdoped region of the phase diagram the FS is
truncated near the saddle points at ($\pi$,0) and (0,$\pi$),
because of the opening of spin and charge gaps. This results in
four disconnected arcs of FS centered at ($\pm\pi$/2,$\pm\pi$/2),
as shown in Fig.\,\ref{models}b. In agreement with a generalized
form of Luttinger's theorem, the area defined by the four arcs and
by the {\it umklapp} gapped FS (dashed lines in
Fig.\,\ref{models}b) encloses the full electron density.

Among the broken-symmetry models, we find the RVB/flux-phase
approach.\footnote{\textcite{AFFLECKI:LARLTH,Xiao-GangWen:Theuc.,ChakravartyS:Hidotc,MaekawaS:Eleces,SuzumuraY:MeaftR,KotliarG:Supitl}.}
This predicts a FS given by four hole-pockets close to
($\pm\pi$/2,$\pm\pi$/2) with a volume proportional to $x$, as in
Fig.\,\ref{models}c, which continuously evolve into a large FS
upon increasing the hole concentration. Note that this is very
similar in spirit to the spin-density wave picture, which also
assumes a dynamical breaking of symmetry
\cite{KampfAP:Spefps,KampfA:Pses-b}. Another model belonging to
(ii) is the stripe picture, which yields a momentum-space
distribution of low-lying excitations
\cite{SalkolaMI:Impcos,FleckM:One-di,MarkiewiczRS:Disosp}.
\begin{figure}[b!]
\centerline{\epsfig{figure=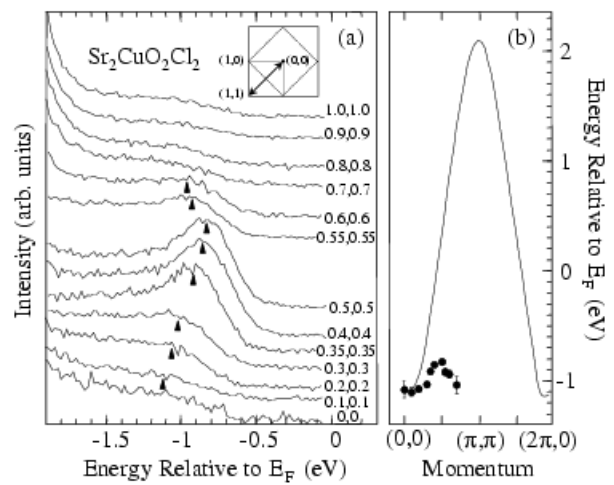,width=8cm,clip=}}
\caption{(a) ARPES data from SCOC along (0,0)-($\pi$,$\pi$) taken
at 350\,K [note that the N\'{e}el temperature is 256\,K on this
compound; however, for the correlation length indicated by neutron
studies \cite{GrevenM:Spict2}, it was argued that photoemission
should still be sensitive to the AF ordering]. (b) Comparison
between the experimental and the tight-binding quasiparticle
dispersions. After
\textcite{WellsBO:Evkrmb}.}\label{wells_EDC}\end{figure}
These are represented by the black patches in Fig.\,\ref{models}d,
where the results obtained for an equal number of vertical and
horizontal domains of disordered stripes are qualitatively
sketched (in this case the physics, together with the
superposition of domains, conspires to give the appearance of a
large LDA-like FS).

There is another meaningful way to differentiate the four models
discussed above: those depicted in
Fig.\,\ref{models}a,\,\ref{models}b, and\,\ref{models}c assume
that the system is spatially homogeneous [as far as
Fig.\,\ref{models}b is concerned, one could talk about phase
separation between insulating spin-liquid and metallic regions,
but only in momentum space
\cite{FurukawaN:Trut-d,FurukawaN:InsL-F,HonerkampC:BretL-}]. In
contrast, the model in Fig.\,\ref{models}d assumes that the system
is spatially inhomogeneous: the formation of stripes is defined as
the segregation of charge carriers into one-dimensional (1D)
domain walls which separate AF spin domains in antiphase with each
other \cite{TranquadaJM:Eviscs}. In Fig.\,\ref{models}d, in
particular, disordered stripes are considered
\cite{SalkolaMI:Impcos}.

Each of the above pictures seem to capture some of the
experimental aspects. Throughout the paper, we will try to compare
ARPES data from various systems with the results of these models,
in the hope of identifying the scenario that has the best overlap
with the experimental observations. This will also help us to
answer the question of whether different materials would favor
different scenarios, and to address the relevance of degrees of
freedom other than the electronic ones (e.g., lattice degrees of
freedom in the case of the stripe instability).

\section{Normal-state electronic structure}

\subsection{Sr$_2$CuO$_2$Cl$_2$ and Ca$_{2}$CuO$_2$Cl$_2$}
\label{sec:ns_ccoc}

The  $t$-$J$ model, briefly discussed in the previous section, is
of particular relevance to the low-energy ARPES features detected
on the cuprates. In fact, in ARPES experiments performed on the
insulating parent compounds of the HTSCs, as a result of the
photoemission process one photo-hole is injected in the CuO$_2$
plane.
\begin{figure}[b!]
\centerline{\epsfig{figure=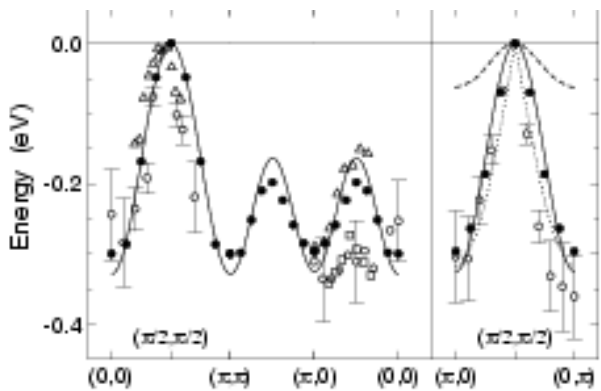,width=8cm,clip=}}
\caption{Electronic dispersion for SCOC ($E\!=\!0$ corresponds to
the top of the band). Open circles \cite{WellsBO:Evkrmb},
triangles \cite{LaRosaS:ElesC2}, and squares \cite{KimC:Systps}:
experimental data. Dashed line: $t$-$J$ model calculations
\cite{WellsBO:Evkrmb}. Solid circles: self-consistent Born
approximation (SCBA) for the $t$-$t'$-$t''$-$J$ model
($t$=0.35~eV, $t'$=$-$0.12~eV, $t''$=0.08~eV and $J$=0.14~eV).
Solid lines: fits of the SCBA data \cite{TohyamaT:Ang-re}. Dotted
line along ($\pi$,0)-(0,$\pi$): spinon dispersion
\cite{LaughlinRB:Eviqdp}.}\label{tohyama}\end{figure}
Therefore, this experiment is the practical realization of a
`single hole' in the AF insulator, and the comparison of ARPES
data and calculations based on the $t$-$J$ model is particularly
meaningful (the latter are typically performed on small clusters
and hence have difficulties treating charge ordering, which may
arise at finite doping; see Sec.\,\ref{sec:nslsco}).

\subsubsection{Single hole in the AF insulator}

ARPES spectra and the corresponding energy dispersion for
insulating Sr$_2$CuO$_2$Cl$_2$ (SCOC) along the nodal direction
(0,0)-($\pi$,$\pi$) are shown in Fig.\,\ref{wells_EDC}a,b
\cite{WellsBO:Evkrmb}. A more complete quasiparticle dispersion is
presented in Fig.\,\ref{tohyama} (note that throughout the paper
we will use terms like `quasiparticle' in a loose sense for
convenience, even though in most cases FL theory may not apply and
well defined quasiparticle peaks cannot be identified in the ARPES
spectra).  Along (0,0)-($\pi$,$\pi$) the dispersion is
characterized by a bandwidth $W\!\simeq\!0.3$\,eV. This is in very
good agreement with $t$-$J$ model calculations
\cite{DagottoE:Coreh-}, which show that, independent of the value
of $t\!\simeq\!350$ meV, the dressing of the hole moving in the AF
background reduces the quasiparticle bandwidth from the
square-lattice tight-binding value of $8t\!\simeq\!2.8$ eV
[\textcite{Kittel:1}; Fig.\,\ref{wells_EDC}b] to
2.2$J\!\simeq\!270$ meV [$J\!\simeq\!125$ meV for SCOC as
independently deduced from neutron scattering studies
\cite{GrevenM:Spict2}].

\begin{figure}[t!]
\centerline{\epsfig{figure=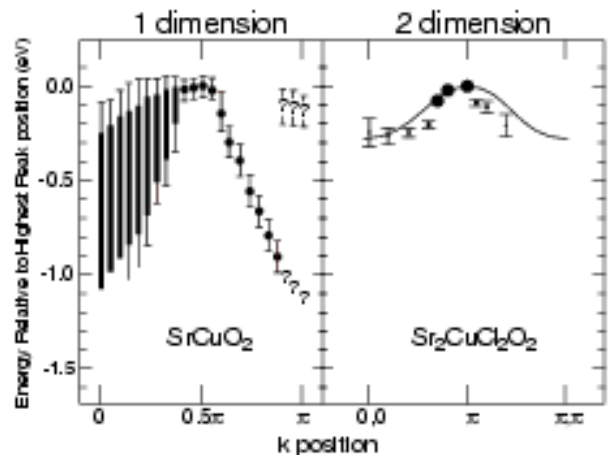,width=8cm,clip=}}
\caption{Experimental dispersion for quasi-1D \cite{KimC:Obss-c}
and quasi-2D insulating systems \cite{WellsBO:Evkrmb}.}
\label{kim_1d2d}
\end{figure}

The ability of $t$-$J$ model calculations to reproduce the
experimentally observed energy-scale renormalization from $t$ to
$J$ confirms the importance of many-body effects, and in
particular of the coupling between quasiparticles and magnetic
correlations, in the (undoped) cuprates. An additional proof of
this statement is provided by the comparison between the
experimentally determined dispersion for 1D and 2D systems with
almost identical structures and Cu-O-Cu bond lengths (e.g.,
SrCuO$_2$ and SCOC, respectively). The surprising aspect in the
data presented in Fig.\,\ref{kim_1d2d} is that the dispersion seen
in 1D systems is about three times as large as the one of the 2D
systems \cite{KimC:Obss-c,KimC:Sepsce,KimC:Spi-ch}. This violates
the non-interacting particle picture on a qualitative level, as
band theory would predict the dispersion in 2D ($8t$) to be twice
that of 1D ($4t$).
This catastrophic failure of the independent-particle picture can
be understood as a consequence of two factors: first, the
quasiparticle dispersion in 2D is strongly renormalized by the
magnetic interaction; second, in 1D spin-charge separation occurs,
and this frees the holon motion from the magnetic interaction
\cite{Lieb:1}.

On the other hand, the $t$-$J$ model predicts a relatively flat
dispersion \cite{ZhipingLiu:DynphH,DagottoE:Flaqdt} along the
($\pi$,0)-(0,$\pi$) direction (dashed line in
Fig.\,\ref{tohyama}), in contrast to the bandwidth
$W\!\simeq\!0.3$ eV observed by ARPES around ($\pi$/2,$\pi$/2)
independent of the direction. Also the poorly defined lineshape
and the spectral weight suppression observed at ($\pi$,0), which
indicate the lack of integrity of the quasiparticle at those
momenta, cannot be reproduced within the simple $t$-$J$ model
calculations \cite{KimC:Systps,KimC:Anotdt}. Better agreement
between the experimental dispersion and the calculations (solid
circles and solid line in Fig.\,\ref{tohyama}) is obtained by
adding second and third nearest-neighbor hopping ($t'$ and $t''$,
respectively) to the $t$-$J$
Hamiltonian.\footnote{\textcite{NazarenkoA:PhosS2,KyungB:Quadti,XiangT:Quaedd,BelinicherVI:Sin-ho,EderR:Dop-de,LeeTK:Dissht,LemaF:Quapid,LeungPW:Com3-s,SushkovOP:Holpic,KimC:Systps,TohyamaT:Spilsa,TohyamaT:Ang-re}.}
In fact, as $t'$ and $t''$ describe hopping within the same
magnetic sublattice, they do not alter the AF properties of the
system at half filling; at the same time, they are not strongly
renormalized by the AF correlations but contribute directly to the
coherent motion of the hole and, therefore, have a substantial
impact on the quasiparticle dispersion. Less can be said about the
lineshape because the broadening is not predicted by the theory,
which is a major limitation of this kind of approach.  Most
importantly, the inclusion of $t'$ and $t''$ accounts for the
suppression of the quasiparticle peak at ($\pi$,0), which may
reflect a reduction of AF spin correlations: the additional
hopping possibilities represented by $t'$ and $t''$ induce a
spin-liquid state around the photo-hole with ($\pi$,0) momentum
\cite{TohyamaT:Spilsa,TohyamaT:Ang-re}. As a consequence, one may
expect to find signatures of spin-charge separation in the ARPES
data \cite{ANDERSONPW:RESVBS}.  In this regard, it is interesting
to note that the full quasiparticle dispersion observed for SCOC
can be very well reproduced also by the spinon dispersion as
proposed by \textcite{LaughlinRB:Eviqdp} [the dotted line in
Fig.\,\ref{tohyama} shows the result along ($\pi$,0)-(0,$\pi$)],
who argued in favor of the decay of the photo-hole injected in the
2D\,AF CuO$_2$ plane into a spinon-holon pair. This is also
reminiscent of the flux phase physics,\footnote{See for example:
\textcite{MaekawaS:Eleces,SuzumuraY:MeaftR,KotliarG:Supitl,AFFLECKI:LARLTH,Xiao-GangWen:Theuc.,ChakravartyS:Hidotc}.}
an extension of the early RVB conjecture \cite{ANDERSONPW:RESVBS}.
We stress here that spin-charge separation in 2D, if realized,
would have a different impact on the dispersion, as suggested by
the comparison with the 1D case in Fig.\,\ref{kim_1d2d}: in 2D the
holon motion is much less coherent than in 1D.

\subsubsection{Remnant Fermi surface}
\label{sec:_ccocRFS}

We mentioned above that both the relatively isotropic dispersion
at ($\pi$/2,$\pi$/2) and the suppression of quasiparticle weight
at ($\pi$,0) observed by ARPES on SCOC and Ca$_2$CuO$_2$Cl$_2$
(CCOC), which is similar in many respects to SCOC
\cite{RonningF:PhoerF,ronning:1}, cannot be explained within the
nearest-neighbor hopping $t$-$J$ model. Better agreement with the
experiment is obtained by including in the model longer range
hopping terms. In this way, it is possible to reproduce also the
doping dependence of the quasiparticle band structure and, in
particular, of the ($\pi$,0) ARPES spectra \cite{EderR:Dop-de}.
However, the $t$-$J$ model, even in its more extended form, cannot
completely account for the strong momentum dependence of the ARPES
spectra from the undoped insulator \cite{EskesH:Hubmvt}. In
particular, although it predicts a decrease of intensity for the
lowest energy peak upon crossing the AF zone boundary
\cite{EskesH:Hubmvt,BulutN:Elepti}, this is not as sharp as
experimentally observed in SCOC (see Fig.\,\ref{wells_EDC}a) or
CCOC. This limitation of the $t$-$J$ model comes from having
projected out the doubly occupied states originally contained in
the Hubbard model: whereas the momentum occupation number $n(k)$
is a strongly varying function of $k$ in the intermediate-$U$
Hubbard model at half filling, it is trivially equal to 1/2 in the
$t$-$J$ model which, therefore, cannot describe the anomalous
distribution of spectral weight in the single particle spectral
function. This effect is accounted for by the complete large
U-limit of the Hubbard model, as shown on the basis of finite size
cluster calculations by \textcite{EskesH:Hubmvt}, and was referred
to by the same authors as a {\it pseudo-FS}.
\begin{figure}[t!]
\centerline{\epsfig{figure=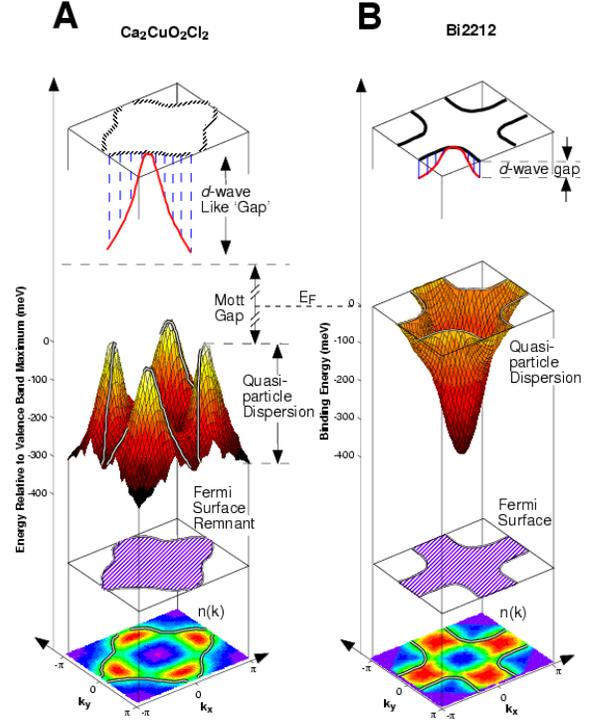,width=8cm,clip=}}
\caption{(bottom) remnant-FS and FS defined by the analysis of
$n(k)$ for insulating CCOC (a) and overdoped Bi2212 (b). While the
FS in Bi2212 is the isoenergetic contour located at $E_F$ (middle
right), the remnant-FS is away from $E_F$ (because of the presence
of the Mott gap), and a large $d$-wave-like dispersion (300 meV)
is found along its contour (middle left). The latter defines a
$d$-wave gap for the insulator (top left), similar to the $d$-wave
pseudo gap observed in the underdoped regime (top right). After
\textcite{RonningF:PhoerF}
[Color].}\label{Filip_cartoon}\end{figure}

A detailed experimental characterization of the $k$-dependence of
the ARPES spectral weight for the undoped insulator has been
presented by \textcite{RonningF:PhoerF}, on the basis of the
$n(k)$ mapping obtained by integrating the ARPES spectra from CCOC
over an energy window larger than the bandwidth. From the location
of the steepest drops in $n(k)$ (see Sec.\,\ref{sec:spectral}) a
{\it remnant-FS} was defined for CCOC, which closely follows the
AF Brillouin zone boundary (Fig.\,\ref{Filip_cartoon}). Note that
matrix element effects also influence the $k$-dependence of the
intensity and alter the profile of the remnant-FS
\cite{HaffnerS:Evisit,HaffnerS:Now-dr}. However, the $n(k)$ drop
observed at the AF Brillouin zone boundary both in the experiment
and in the numerical results appears to be a robust feature,
despite some uncertainties which indeed seem to be caused by
matrix element effects
\cite{ronning:1,DurrC:Ang-re,GoldenMS:elesch}. One has to realize
that the remnant-FS is not a real FS (the system has a Mott gap),
but identifies the same locus of rapid intensity drop referred to
as a pseudo-FS by \textcite{EskesH:Hubmvt}. It underscores the
fact that the system was driven into the insulating state by
strong electronic correlations. In addition, it does not even
correspond to an isoenergetic contour in the quasiparticle
dispersion (Fig.\,\ref{Filip_cartoon}), similarly to the FS
determined in the underdoped regime (Sec.\,\ref{sec:pseudogap}).
The relevance of this approach is that, once a remnant-FS has been
determined, it is also possible to identify a `gap' along its
contour (in addition to the Mott gap), and try to compare it to
the high energy pseudogap of the underdoped systems (see
Sec.\,\ref{sec:pseudogap}).

\subsubsection{Superconducting Ca$_{2-x}$Na$_x$CuO$_2$Cl$_2$}

Although these materials are very difficult to dope up to the
onset of superconductivity, recently doping dependent data from
the (Sr/Ca)$_2$CuO$_2$Cl$_2$ family have become available.
Following earlier success on polycrystalline samples
\cite{HiroiZ:Proh-d,HiroiZ:SynssC}, \textcite{kohsaka:1}, by a
flux method under a pressure of 4 GPa, have succeeded in growing
superconducting single crystals of Ca$_{2-x}$Na$_x$CuO$_2$Cl$_2$
(with maximum $T_c\!=\!28$ K at $x\!=\!0.15$). This achievement
has opened very exciting new opportunities in the systematic
investigation of the cuprates: the oxychlorides together with the
LSCO family represent the only materials in which the doping range
corresponding to the metal-insulator transition can be accessed.
Furthermore, since the best ARPES data from the insulator have
been obtained on this family of compounds, this makes the direct
comparison between data from the metal and the insulator more
informative.
\begin{figure}[t!]
\centerline{\epsfig{figure=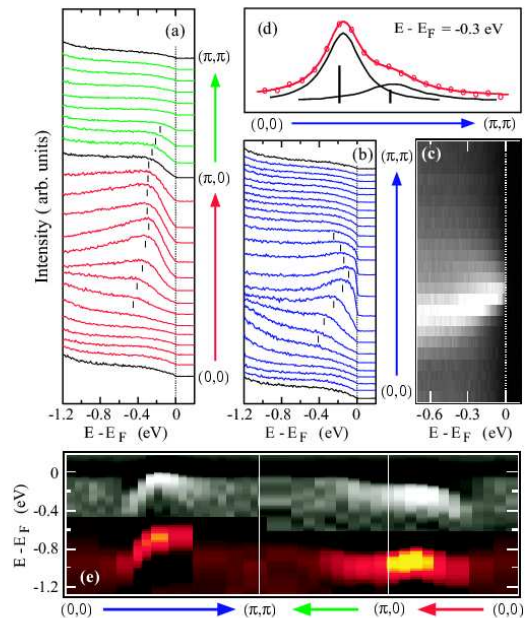,width=7.5cm,clip=}}
\caption{(a,b) ARPES spectra from 10\% Na-doped CCOC, taken at
$T\!=\!10$\,K. (c) Intensity plot and (d) constant energy spectra
from (b). (e) Second derivative with respect to the binding energy
of the ARPES spectra from Na-free (color) and doped (grey) CCOC.
After \textcite{kohsaka:1} [Color].}\label{NArawdata}\end{figure}

Fig.\,\ref{NArawdata} shows the energy distribution curves
measured at 10 K with 25.5 eV photons on 10\% Na-doped CCOC
($T_c\!=\!13$ K), along the high symmetry directions. In going
from (0,0) to ($\pi$,$\pi$), a broad feature disperses up to the
Fermi level as indicated by the clear Fermi edge seen in the
spectra near $E_F$ (Fig.\,\ref{NArawdata}b). After the band has
crossed $E_F$, one can still follow a feature dispersing backwards
(see tick-marks) in a fashion very similar to what is expected for
the magnetic {\it shadow bands} due to spin fluctuations diverging
at the AF zone
boundary.\footnote{\textcite{KampfAP:Spefps,DagottoE:Coreh-,HaasS:Antipb,PreussR:Quadt2,ChubukovAV:Quasna,Xiao-GangWen:Theuc.}}
In the same figure, the constant-energy or momentum distribution
curves are also plotted (Fig.\,\ref{NArawdata}d). It appears that
these cannot be fitted to a single Lorentzian but two Lorentzian
peaks are needed. This is probably the most convincing case for
the presence of shadow bands as in this system complications due
to structural distortions are not present and, therefore, the
origin of the effect is likely magnetic. Fig.\,\ref{NArawdata}e
compares the dispersion of insulating CCOC and of the Na-doped
superconductor: although in the insulator the top of the valence
band is located $\sim\!0.7$ eV below the chemical potential, there
is a striking similarity in the dispersion of the two compounds.
This suggests that upon doping the system the chemical potential
shifts from inside the gap of the insulator to the top of the
valence band near ($\pi$/2,$\pi$/2), which is also supported by
the similarities between the ($\pi$,0) ARPES spectra from Na-free
and doped CCOC \cite{ronning:2}. Note also that, while according
to band theory the dispersive feature in Fig.\,\ref{NArawdata}a
should reach $E_F$ in going from ($\pi$,0) to ($\pi$,$\pi$), an
unambiguous Fermi crossing is not seen and near ($\pi$,0) the band
is $\sim\!200$ meV below $E_F$, a point that will be discussed in
greater details in Sec.\,\ref{sec:pseudoNa} (pseudogap).

\begin{figure}[t!]
\centerline{\epsfig{figure=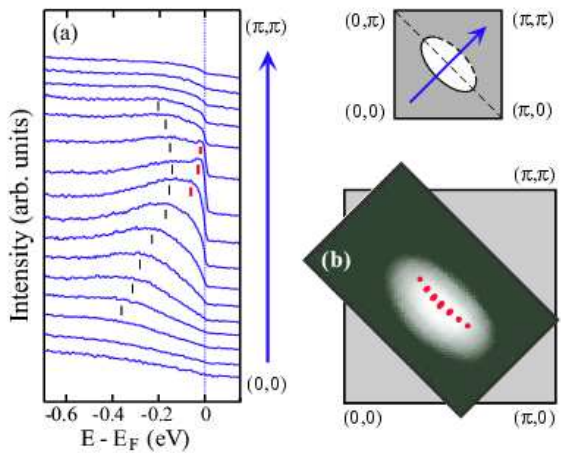,width=8cm,clip=}}
\caption{(a) ARPES spectra from 10\% Na-doped CCOC near
($\pi$/2,$\pi$/2). (b) Integrated intensity (over 100 meV); the
red dots indicate those momenta at which a sharp peak as been
observed close to $E_F$. After \textcite{kohsaka:1}
[Color].}\label{NAcont}\end{figure}

The results from Na-doped CCOC are consistent with a shift of the
chemical potential (as in Fig.\,\ref{mott}c) and are very
suggestive of the two scenarios depicted in Fig.\,\ref{models}b or
Fig.\,\ref{models}c, as emphasized in particular by the contour
map of the low-lying excitations presented in Fig.\,\ref{NAcont}b.
The task of conclusively distinguishing between the scenarios of
Fig.\,\ref{models}b and Fig.\,\ref{models}c is rather difficult.
The key point is whether or not a shadow FS is detected (marked by
the dash lines in Fig.\,\ref{models}c), whose presence is however
not clear in the contour map of Fig.\,\ref{NAcont}b. This can
either mean that the shadow FS is not present or simply too weak
to be detected [the data in Fig.\,\ref{NArawdata}b-d show the
presence of a shadow band that disperses away from the Fermi level
when passing ($\pi$/2,$\pi$/2), but this is not clear at $E_F$].
Even more difficult would be, at this stage, to distinguish
between the scenario of Fig.\,\ref{models}c in which the FS
pockets are not exactly centered at ($\pi$/2,$\pi$/2), and the
simplest antiferromagnetic/spin-density wave picture [see, e.g.,
\textcite{fuldeP:1}], in which the excitation spectrum is
symmetric with respect to the AF zone boundary (as in Brillouin
zone sketch shown in Fig.\,\ref{NAcont}).

A more detailed analysis of the ARPES spectra along the nodal
direction reveals additional information \cite{kohsaka:1}. As
shown in Fig.\,\ref{NAcont} the energy distribution curves near
the Fermi energy actually consist of two components: a broader
component at higher binding energy that appears to disperse
backwards past ($\pi$/2,$\pi$/2) defining the shadow band, and a
sharp (although weak) component that crosses the Fermi level (its
$k$-space location is indicated by the red dots in
Fig.\,\ref{NAcont}b). Note that the sharper low-energy peak is not
directly related to superconductivity as it is seen in the 20 K
data of Fig.\,\ref{NAcont} ($T_c\!=\!13$ K) and, as a matter of
fact, was still present at temperature as high as 75 K
\cite{ronning:2}. On the one hand, these two components may be
thought of as the coherent and incoherent parts of the spectral
function, as discussed in Sec.\,\ref{sec:spectral}. On the other
hand, these results may also indicate that the doping evolution
cannot be fully accounted for in terms of a rigid shift of the
chemical potential.

\subsubsection{Summary}

It must be stressed that so far only 10\% Na-doped samples were
studied \cite{kohsaka:1}. Whether the relatively simple behavior
(i.e., `rigid-band' like) seen for Na-doped CCOC remains true at
lower dopings (near the metal-insulator transition boundary) is
still an open question, and is of particular relevance to those
theories which lead to $d$-wave superconductivity starting from a
rigid band picture [see, e.g., \textcite{DagottoE:AntvHs}]. As we
will see in the next section, at first glance the physical picture
emerging from LSCO is quite different and is more consistent with
the scenario in which at low doping levels new states are created
inside the gap (Fig.\,\ref{mott}b). This could be naturally
understood if the system becomes inhomogeneous at intermediate
dopings: in fact two electronic components are observed in LSCO in
the underdoped region (5-10\%), and only one in the overdoped
regime. However, without data from the very underdoped regime one
may also interpret the LSCO doping evolution in terms of a
quasi-rigid band shift with some modifications near the ($\pi$,0)
region. Therefore, until more data on Na-CCOC are available, one
cannot conclude whether or not the difference with LSCO is of
qualitative or quantitative nature.

\subsection{La$_{2-x}$Sr$_x$CuO$_4$}
\label{sec:nslsco}

In order to study the doping evolution of the low-energy
electronic properties over the full doping range the most suitable
system is La$_{2-x}$Sr$_x$CuO$_4$ (LSCO). The hole concentration
in the CuO$_2$ plane can be controlled and determined by the Sr
content $x$, from the undoped insulator ($x\!=\!0$) to the heavily
overdoped metal ($x\!\sim\!0.35$). In addition, LSCO has a simple
crystal structure with a single CuO$_2$ layer
(Fig.\,\ref{bz_lsco}), and none of the complications due to
superstructure which are found in Bi2212
(Sec.\,\ref{sec:ns_Bi2212}). Another interesting aspect is the
suppression of $T_c$ at $x\!=\!1/8$ which, together with the
incommensurate AF long-range order observed by inelastic neutron
scattering \cite{SuzukiT:Obsmml}, has been discussed as evidence
for fluctuating stripes in LSCO [similar AF order accompanied by
charge ordering was interpreted as a realization of `static
stripes' in La$_{1.48}$Nd$_{0.4}$Sr$_{0.12}$CuO$_4$
\cite{TranquadaJM:Eviscs}].

\subsubsection{Metal-insulator transition}

\begin{figure}[b!]
\centerline{\epsfig{figure=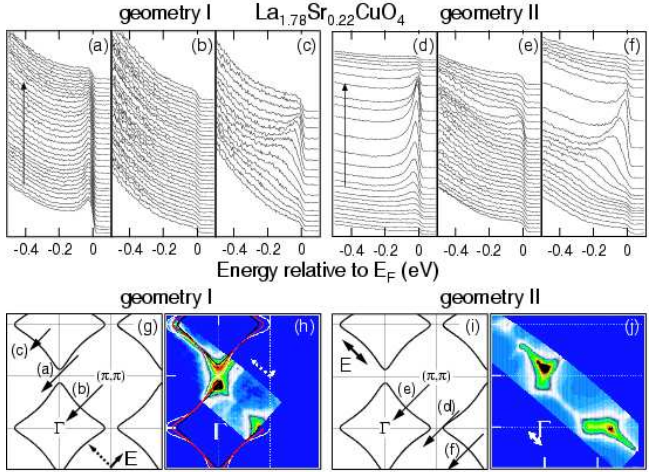,width=0.99\linewidth,clip=}}
\vspace{.1cm}\caption{(a-f) La$_{1.78}$Sr$_{0.22}$CuO$_4$ ARPES
spectra, measured at 20 K with 55.5\,eV photons in two different
experimental geometries. The cut orientation in $k$-space and
field polarization are indicated in (g,i) together with the
electronlike FS. (h,j) $E_F$ intensity map (30 meV integration
window). White and black curves are the $k_z\!=\!0$ and $\pi/c$
FSs from band calculations \cite{XuJ-H:Domrt2}, and red curves the
FS from a tight-binding fit of the data. After
\textcite{YoshidaT:EleFsr} [Color].}\label{teppei_edc}\end{figure}

We start our discussion with data from $x\!=\!0.22$ overdoped LSCO
\cite{YoshidaT:EleFsr}.  Because many-body effects such as charge
ordering are expected to be weak in the overdoped regime, the
results from this sample can be used to test the overall
reliability of the ARPES data from LSCO. Fig.\,\ref{teppei_edc}
presents ARPES spectra from two Brillouin zones taken with 55.5 eV
photons under different polarization conditions (while in geometry
II the electric field was fixed and almost parallel to the sample
surface, in geometry I it had only a small in-plane component
which increases at larger momenta). It appears that the spectra
from this overdoped sample are very sharp, contrary to the typical
data from underdoped LSCO that we will discuss later.
Indeed the sharpness of the spectra from the second zone is
comparable to that observed in Bi2212, at similar doping.
Fig.\,\ref{teppei_edc}h and\,\ref{teppei_edc}j present the
spectral intensity integrated over a 30 meV energy window at
$E_F$. In general, the $E_F$ intensity map derived from high
resolution data gives a good evaluation of the FS, provided that
matrix element effects are carefully considered (see
Sec.\,\ref{sec:spectral} and\,\ref{sec:stateart}).
\textcite{YoshidaT:EleFsr} have attempted to account for the
photoionization cross section effects for LSCO by varying the
experimental geometry and by performing simulations of the
spectral weight distribution, which qualitatively reproduce some
of the variation observed in different geometries. From the
comparison (Fig.\,\ref{teppei_edc}h) between the experimental
$E_F$ intensity maps and the FSs obtained from a tight-binding fit
to the data and from LDA calculations \cite{XuJ-H:Domrt2},
\textcite{YoshidaT:EleFsr} concluded that the FS of overdoped LSCO
consists of an electron pocket centered at (0,0), in agreement
with band-structure calculations. These results, namely the
observation of sharp spectral features and an LDA-like FS, as well
as the fact that low energy electron diffraction investigations do
not detect any evidence of surface reconstruction, give strong
credence to the ARPES data from the LSCO system. Since the quality
of the LSCO single crystals improves upon decreasing doping $x$,
these data suggest that the broad spectral features observed in
underdoped LSCO are of intrinsic nature.

\begin{figure}[t!]
\centerline{\epsfig{figure=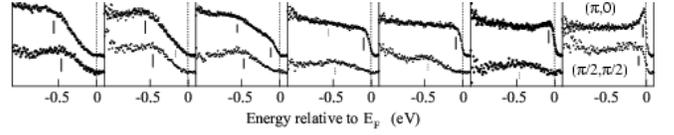,width=1\linewidth,clip=}}
\vspace{.1cm}\caption{ARPES data from LSCO for different dopings
at ($\pi$,0) and ($\pi/2$,$\pi/2$). The spectra were normalized to
the integrated intensity of the valence bands, and at
($\pi/2$,$\pi/2$) were multiplied by a factor of two
\cite{InoA:ElesL2}.} \label{Akihiro_Dop}
\end{figure}

Let us then move to the low doping region, near the
metal-insulator transition boundary. Fig.\,\ref{Akihiro_Dop}
presents ARPES spectra at ($\pi$,0) and ($\pi$/2,$\pi$/2) as a
function of doping \cite{InoA:ElesL2}. The data were recorded
under identical experimental geometry so that the photoionization
matrix elements are the same. For the insulating samples
($x\!=\!0$), the data are characterized by a high binding energy
feature [$\sim\!0.5$ eV at ($\pi/2$,$\pi/2$), and $\sim\!0.7$ eV
at ($\pi$,0)], and are consistent with what we have discussed in
Sec.\,\ref{sec:ns_ccoc} for insulating SCOC
\cite{WellsBO:Evkrmb,KimC:Systps}, albeit the features are now
broader. The remarkable result is that for $x\!=\!0.05$ two
features can be identified in the ARPES spectra at the ($\pi$,0)
point (some additional weight at low energy is already observable
for $x\!=\!0.3$): in addition to the high-energy one, reminiscent
of the Zhang-Rice singlet band of the AF insulator, a second
shoulder is observable close to $E_F$. Upon further doping the
system with holes, a systematic transfer of spectral weight from
the high to the low-energy feature takes place, and a well-defined
quasiparticle peak develops near optimal doping.
On the other hand, the results obtained at ($\pi/2$,$\pi/2$) are
very different: first of all, the data show an overall suppression
of weight as compared to ($\pi$,0) [the spectra plotted in
Fig.\,\ref{Akihiro_Dop} for ($\pi/2$,$\pi/2$) have been multiplied
by a factor of 2]; second, in the nodal region [i.e., along
(0,0)-($\pi$,$\pi$)], a quasiparticle peak is observable only for
$x\!\geq\!0.15$. As we will discuss later, with different
experimental geometries more spectral weight is detected near
$E_F$ in the nodal region, but the overall trend of the doping
dependence of the electronic structure is robust.

The dispersion of the LSCO spectral features for different doping
levels is summarized in Fig.\,\ref{Akihiro_Der}, which presents
the second derivative with respect to the binding energy of the
ARPES spectra  \cite{InoA:ElesL2}. Upon increasing doping, we can
observe the building of near-$E_F$ weight first at ($\pi$,0), and
then at ($\pi$/2,$\pi$/2). One can clearly observe the presence of
the flat-band saddle point extensively discussed in the
literature. Furthermore, the second derivative emphasizes the
presence of the high-energy feature in the heavily underdoped
samples (coexisting with the low-energy one, at least for
$x\!=\!0.05$). This has a 200 meV lower energy at
($\pi$/2,$\pi$/2) than at ($\pi$,0), in qualitative agreement with
what is observed in the undoped insulator SCOC
\cite{WellsBO:Evkrmb}.

The ARPES results from LSCO, and in particular the presence of two
electronic components, suggest that the effects of doping on the
electronic structure of the correlated insulator cannot be
accounted for by a simple shift of the Fermi level in a rigid band
model \cite{InoA:ElesL2}. This is in agreement with another
observation by \textcite{InoA:Chepso}: from the analysis of direct
and inverse angle-integrated photoemission spectra, it was
concluded that the chemical potential $\mu$ is pinned inside the
charge-transfer gap for $x\!\leq\!0.10$, and starts shifting
downwards significantly only for $x\!\geq\!0.15$. The above
results seem to indicate that, in the case of LSCO, in-gap states
are created upon doping the insulator (Fig.\,\ref{mott}b).

\subsubsection{Nano-scale phase separation}

\textcite{InoA:ElesL2} suggested that the ARPES results from LSCO
may be understood within the stripe picture, which is a particular
form of nano-scale phase separation. This would explain the
pinning of the chemical potential for $x\!\leq\!0.125$ (i.e., the
so called 1/8 doping) as a consequence of the segregation of doped
holes into metallic domain walls, which gives rise to the
appearance of in-gap states. Furthermore, the suppression of nodal
intensity at $E_F$ would be consistent with the vertical and
horizontal orientation of the metallic stripes (orthogonal domains
are expected for this nano-scale phase separation). Here the
conjecture is that charge fluctuations would be suppressed along
directions crossing the stripes, and is supported by finite-size
cluster calculations \cite{TohyamaT:Effses}. The increase of $E_F$
weight for $x\!\geq\!1/8$ in the nodal region may indicate that
above this doping level the holes overflow from the saturated
stripes into the CuO$_2$ planes.

Concerning the relevancy of the stripe scenario to the ARPES data
from the HTSCs, more insights could come from the investigation of
Nd-LSCO, a model compound for which the evidence of spin and
charge stripe-ordering is the strongest \cite{TranquadaJM:Eviscs}.
High resolution ARPES data from
La$_{1.28}$Nd$_{0.6}$Sr$_{0.12}$CuO$_4$ were reported by
\textcite{ZhouXJ:One-di}. Remarkably, the low-energy spectral
weight is mostly concentrated in narrow and straight regions along
the (0,0)-($\pi$,0) and (0,0)-(0,$\pi$) directions and is confined
between lines crossing the axes at $\pm\,\pi/4$.
\begin{figure}[b!]
\centerline{\epsfig{figure=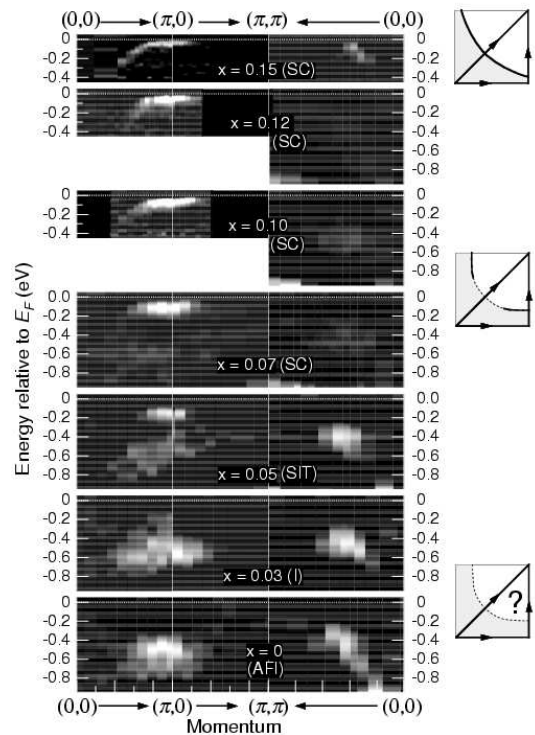,width=8cm,clip=}}
\vspace{.1cm}\caption{Second derivative, with respect to the
binding energy, of the ARPES spectra from LSCO along the
high-symmetry directions for many doping levels
\cite{InoA:ElesL2}.}\label{Akihiro_Der}\end{figure}
These straight patches of high intensity are suggestive of almost
perfectly nested 1D FS segments. \textcite{ZhouXJ:One-di}
interpreted these results as a signature of a 1D electronic
structure related to the presence of static 1/4-filled charge
stripes which, as indicated by neutron and x-ray experiments
\cite{TranquadaJM:Eviscs}, at 1/8 doping are separated by AF
regions resulting in a pattern with periodicity 4$a$ (and
characterized by macroscopic orthogonal domains). This
interpretation would also explain the origin of the two components
seen in the ARPES spectra from LSCO near the metal-insulator
transition boundary (Fig.\,\ref{Akihiro_Dop}): the signal from the
AF insulating regions would be pushed to high binding energies
because of the Mott gap, while the charge stripes would be
responsible for the component near $E_F$. In this sense, the
stripe interpretation is rather appealing
\cite{MarkiewiczRS:Disosp}.
\begin{figure}[t!]
\centerline{\epsfig{figure=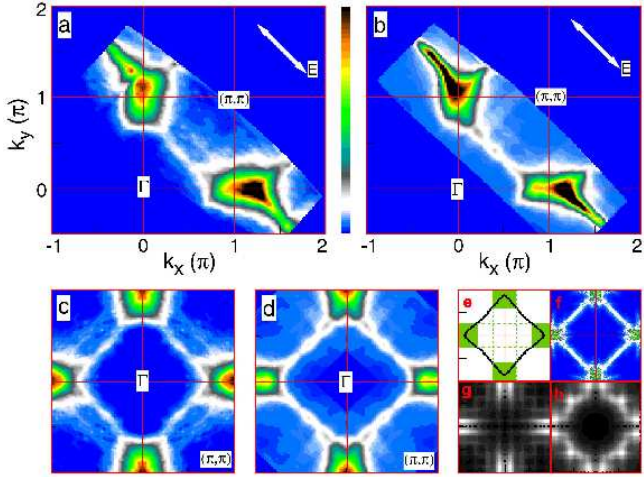,width=0.99\linewidth,clip=}}
\vspace{0cm}\caption{(a,b) Intensity maps obtained by integrating
over 30 meV at $E_F$ the ARPES spectra from $x\!=\!0.15$ Nd-LSCO
and LSCO, respectively, and (c,d) corresponding four-fold
symmetrization of the first-zone data. White arrows indicate the
field polarization. Other maps: (e) cartoon of the experimental
results; calculations for (f) orthogonal domains of disordered or
fluctuating stripes [as in Fig.\,\ref{models}d;
\textcite{SalkolaMI:Impcos}], and (g) site-centered and (h)
bond-centered stripes \cite{ZacherMG:Strdas}. After
\textcite{ZhouXJ:Duante} [Color].}\label{dual2}\end{figure}
On the other hand, there are some aspects which cannot be
satisfactorily explained within the idealized stripe picture. For
example, in both LSCO (Fig.\,\ref{Akihiro_Der}) and Nd-LSCO
\cite{ZhouXJ:One-di}, the quasiparticle bands along the
(0,0)-($\pi$,0) and (0,0)-(0,$\pi$) directions are characterized
by a considerably fast dispersion. This is counter to what one
would expect for an ideal 1D system even in presence of orthogonal
domains, as the electronic dispersion should be observed only
within and perpendicularly to each of the two superimposed 1D FSs
[for more on this point see \textcite{ZhouXJ:One-di}].
Furthermore, also possible artifacts due to matrix element effects
have to be cautiously considered when interpreting the ARPES data
and especially modulations of the integrated spectral weight
(Sec.\,\ref{sec:spectral} and\,\ref{sec:matele}).

In order to gain more insight into these issues, in particular in
relation to the straight segments of FS observed in Nd-LSCO and to
the suppression of the nodal state, \textcite{ZhouXJ:Duante}
extended the measurements to the second zone, and used different
polarizations and orientations of the incoming electric field.
Fig.\,\ref{dual2} presents data from 15\% doped LSCO and Nd-LSCO
recorded in the same geometry as the data of
Fig.\,\ref{teppei_edc}j from 22\% doped LSCO. These results
confirm the presence of the flat bands in extended regions around
($\pi$,0) and (0,$\pi$). On the other hand, in the second zone
appreciable spectral weight at the Fermi level is detected in the
nodal region, which appears to become more intense upon increasing
the Sr concentration in both Nd-LSCO and LSCO, and is stronger in
Nd-free LSCO for a given Sr content (Fig.\,\ref{dual2}). This
nodal weight was not found in the earlier study of 12\% doped
Nd-LSCO \cite{ZhouXJ:One-di}, and is not expected in calculations
considering only rigid stripes
\cite{IchiokaM:Elesst,MarkiewiczRS:Disosp}.

Possible descriptions of these results within the stripe context
are summarized in Fig.\,\ref{dual2}c-h where the low-lying
spectral weight from LSCO (c) and Nd-LSCO (d) is compared to the
results of model calculations. First, the experimental FS composed
of straight patches at ($\pi$,0) and (0,$\pi$) connected by `nodal
segments' (as emphasized in the sketch of Fig.\,\ref{dual2}e)
closely resembles the one arising from disorder or fluctuation of
the stripes \cite{SalkolaMI:Impcos}, which was sketched in
Fig.\,\ref{models}d and is here exactly reproduced in
Fig.\,\ref{dual2}f. Alternatively, the experimental FS may result
from the coexistence of site-centered (Fig.\,\ref{dual2}g) and
bond-centered (Fig.\,\ref{dual2}h) stripes \cite{ZacherMG:Strdas}.
Both scenarios, as well as more recent calculations performed for
various inhomogeneous models for the cuprates \cite{Eroles:1},
suggest that the stripe picture catches the essence of the
low-lying physics for the LSCO system. This is also indicated by
dynamical mean-field theory results, which are characterized by a
near-$E_F$ 1D dispersive band and a high-energy incoherent feature
in the very underdoped regime \cite{FleckM:One-di}, as well as by
numerical studies of the spin-fermion model, which indicate that
both stripe-induced midgap states and valence band states
contribute to the FS \cite{MoraghebiM:Ferssf}, consistent with the
two electronic components detected by ARPES on LSCO.

\subsubsection{From hole to electronlike FS}

It was noted that the presence of a sharp structure in momentum
space such as a FS-like feature does not necessarily imply well
defined one-electron states and, in particular, is not
incompatible with the stripe scenario \cite{OrgadD:Eviefp}.
\begin{figure}[b!]
\centerline{\epsfig{figure=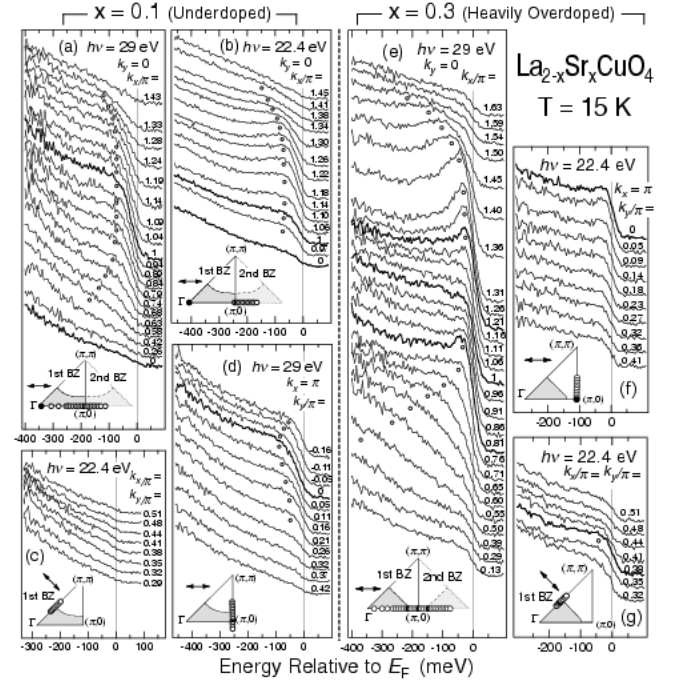,width=1\linewidth,clip=}}
\vspace{0cm}\caption{(a-d) ARPES spectra for underdoped
($x\!=\!0.1$) and (e-g) heavily overdoped ($x\!=\!0.3$) LSCO.
Insets: measured $k$-space points and incident light polarization
\cite{InoA:Fersbd}.}\label{Akihiro_EDC}\end{figure}
For instance in a non FL system like the 1D electron gas, in which
the energy distribution curves are typically completely
incoherent, the near-$E_F$ momentum distribution curves (as
defined in Fig.\,\ref{valla1}) might be very sharp because of more
stringent kinematic constraints. Anyway, given that the striped
system does exhibit a FS-like feature (Fig.\,\ref{dual2}), it is
still meaningful to investigate its evolution as a function of
doping. In doing so, one must bear in mind that the FS concept is
here referred to in a loose manner: it identifies the
momentum-space location where a very broad feature (and not a well
defined peak in energy) crosses $E_F$, after having defined some
kind of band dispersion.

ARPES spectra for underdoped ($x\!=\!0.1$) and overdoped
($x\!=\!0.3$) LSCO are shown in Fig.\,\ref{Akihiro_EDC}
\cite{InoA:Fersbd}. Although the spectral features from the
underdoped samples tend to be broad, which as discussed may be
related to charge inhomogeneity, some clear Fermi crossings are
observable in part of the Brillouin zone especially in overdoped
samples. For $x\!=\!0.1$, along the direction
(0,0)-($\pi$,0)-(2$\pi$,0) at 29 eV photon energy
(Fig.\,\ref{Akihiro_EDC}a), a broad quasiparticle peak emerges
from the background, disperses towards $E_F$ without crossing it,
and then pulls back in the second Brillouin zone.  Similar results
are obtained at 22.4 eV (Fig.\,\ref{Akihiro_EDC}b), the only
difference being a decrease of intensity in the first Brillouin
zone due to matrix element effects specific to this photon energy
\cite{InoA:Fersbd}. Along ($\pi$,0)-($\pi$,$\pi$) the
quasiparticle peak, with maximum binding energy at ($\pi$,0),
disperses almost up to $E_F$, loses intensity, and disappears
(Fig.\,\ref{Akihiro_EDC}d). The leading-edge midpoint never
reaches $E_F$ because of the superconducting gap ($\sim\!8$ meV)
opened along the FS at this temperature (as clearly shown in
Fig.\,\ref{pseudoLSCO2}). Following \textcite{CampuzanoJC:Dirop-},
the underlying FS crossing is thus identified by the locus of the
minimum gap and is located at ($\pi$,0.2$\pi$). Along the nodal
direction no clear peak can be identified
(Fig.\,\ref{Akihiro_EDC}c), as discussed above. However, having
detected a band below $E_F$ at ($\pi$,0), \textcite{InoA:Fersbd}
concluded that for $x\!=\!0.1$ the FS of LSCO is holelike in
character and centered at ($\pi$,$\pi$). By comparing the spectra
from heavily overdoped and underdoped LSCO
(Fig.\,\ref{Akihiro_EDC}e and\,\ref{Akihiro_EDC}a), we see a
striking difference: for $x\!=\!0.3$ the quasiparticle peak along
this cut has almost disappeared at ($\pi$,0). The decrease of
intensity, together with a leading-edge midpoint located above
$E_F$, provides evidence for the quasiparticle peak crossing $E_F$
just before ($\pi$,0). The FS thus determined for heavily
overdoped LSCO is electronlike and centered at (0,0), in agreement
with what discussed in relation to Fig.\,\ref{teppei_edc}. Careful
investigations by \textcite{InoA:Dop-de} indicated that the FS
changes from holelike to electronlike for $x\simeq0.15$-0.2 as the
saddle point moves from below to above $E_F$, consistent with LDA
calculations.

\subsubsection{Summary}
\label{sec:lscodisc}

In summary, what has emerged from the study of the LSCO system is
a very complex and intriguing picture, characterized by some
contrasting aspects: neither a simple stripe model, nor any of the
other models proposed in Fig.\,\ref{models} can provide a
satisfactory explanation for the complete body of available data.
As we have discussed, the stripe picture, when
disorder/fluctuations and realistic charge disproportionation are
considered, has the advantage of qualitatively explaining the data
over the entire doping range, including the presence of two
electronic components, the straight FS segments, and the lack of a
chemical potential shift in the very underdoped regime. On the
other hand, on a more quantitative level, there are still many
open questions. In particular, in order to gain more reliable
insights, the role of matrix element effects on the ARPES data
should be investigated in more detail. In fact, {\it ab initio}
calculations of matrix elements are still unavailable for LSCO,
and tight binding fits do not reproduce the results to a
satisfactory degree.

As a last remark, it should be mentioned that the very underdoped
regime was very recently studied by \textcite{yoshida:2}. The
two-component electronic structure discussed above was observed
also for these samples, hinting again at a tendency towards
nano-scale phase separation. However, the low-energy component
(i.e., from the hole-rich region) defines a FS arc reminiscent of
an LDA FS being gapped in the ($\pi$,0) region. Despite this
seemingly more conventional behavior, one should not abandon a
many-body description for LSCO because a simple band picture
cannot explain the basic observation of a two-component low-energy
electronic structure.

\subsection{Bi$_2$Sr$_2$CaCu$_2$O$_{8+\delta}$}
\label{sec:ns_Bi2212}

In proceeding with the comparative study of the cuprate
superconductors, let us now turn our attention to
Bi$_2$Sr$_2$CaCu$_2$O$_{8+\delta}$ (Bi2212) which is the HTSC most
intensively investigated by ARPES thanks to the availability of
large high-quality single crystals and to the presence of a
natural cleavage plane between the BiO layers, and especially to
the richness of information that can be obtained by ARPES both in
the normal and superconducting state. Due to sample quality issues
most of the Bi2212 experiments were carried out near optimal
doping, and there is very limited or almost no information on the
electronic structure near the metal-insulator transition boundary.
Here we concentrate on cases with a doping of 10\% or higher.
Therefore, we cannot answer the questions of whether the
two-component electronic structure observed in LSCO for 5-7\%
doping is also present in the Bi2212 case, or how the metallic
state emerges in the latter system. Core level spectroscopy
results, however, suggest a behavior intermediate between the one
seen on LSCO and the simple linear shift of the chemical potential
\cite{harima:2}.

\subsubsection{Fermi surface topology}

\begin{figure}[t!]
\centerline{\epsfig{figure=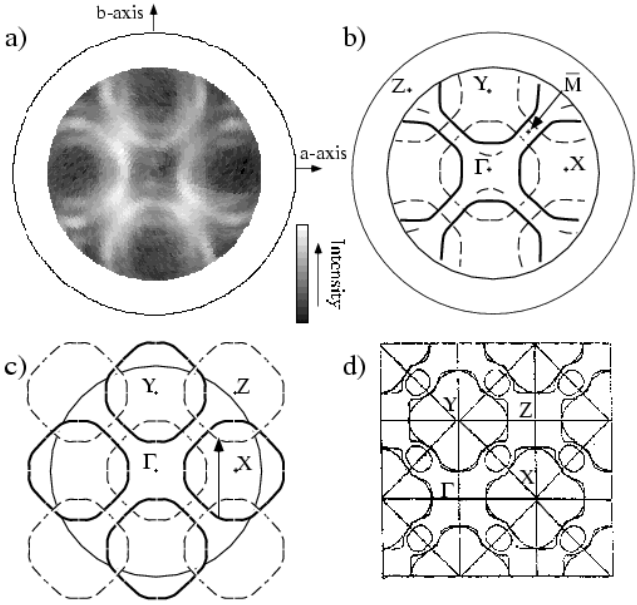,width=0.99\linewidth,clip=}}
\caption{(a) Integrated intensity map (10 meV window centered at
$E_F$) obtained for Bi2212 at 300 K with 21.2 eV photons (HeI
line). (b,c) Sketch emphasizing the superposition of the main FS
(thick lines) and of its ($\pi$,$\pi$) translation (thin lines)
due to the backfolded shadow bands. (d) FS calculated by
\textcite{MassiddaS:ElespB}. After \textcite{AEBIP:COMFMB1}.
}\label{aebi}\end{figure}

\begin{figure}[b!]
\centerline{\epsfig{figure=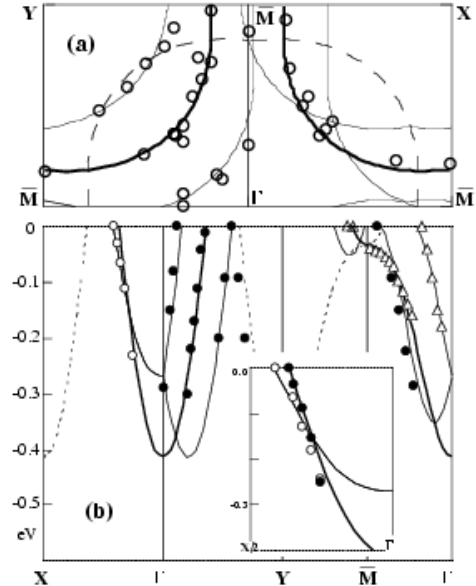,width=6.3cm,clip=}}
\caption{Bi2212 ($T_c\!=\!87$\,K) FS (a) and dispersion (b)
measured in the normal state at 95 K (the various symbols denote
different polarizations conditions). Thick lines are the result of
a tight binding fit of the main band dispersion. Thin and dashed
lines represent {\it umklapp} and shadow bands, respectively. The
inset shows a blowup of $\Gamma$-$X$
\cite{DingH:EleeB2}.}\label{ding2}\end{figure}

\begin{figure}[b!]
\centerline{\epsfig{figure=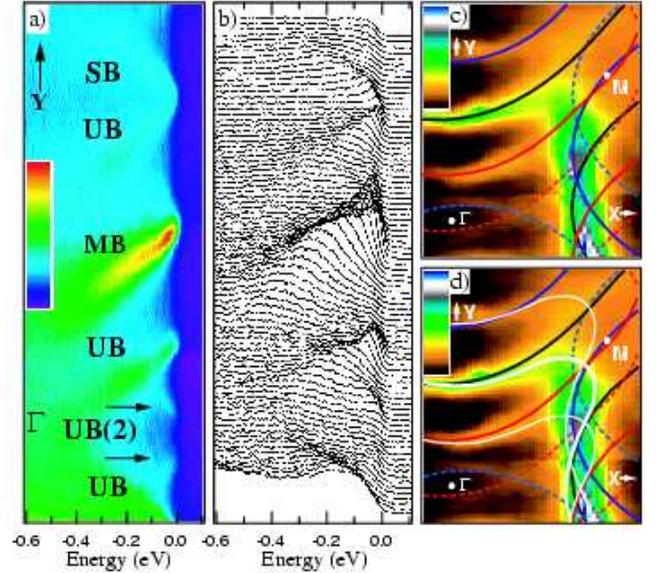,width=8.5cm,clip=}}
\caption{(a) Intensity $I({\bf k},\omega)$ and (b) corresponding
ARPES spectra from optimally doped Bi2212 ($T_c\!=\!90$ K) taken
along $\Gamma$-$Y$ with 33\,eV photons at 40\,K. MB, UB, UB(2),
and SB refer to main, {\it umklapp}, second-order {\it umklapp},
and shadow bands, respectively. (c) Integrated intensity map
($\pm\!100$ meV at $E_F$) together with holelike MB (black), UB
(blue/red), UB(2) (dashed blue/red) FSs obtained from a
tight-binding fit of the data \cite{FretwellHM:FersB2}. (d) Same
as (c) but with in addition electronlike MB and UB FSs indicated
by white thin and thick lines \cite{gromko:1}
[Color].}\label{fretwell1}\end{figure}

In the following we will illustrate the normal state electronic
properties of Bi2212 focusing in particular on the FS topology, a
topic which has been plagued by an intense controversy since the
very beginning of its investigation. The main reason for this can
be identified in the complexity of the electronic structure in the
($\pi$,0) region of momentum space, which is where the
quasiparticle dispersion as determined by ARPES is weakly momentum
dependent giving rise to the so called {\it flat bands}
\cite{ABRIKOSOVAA:EXPOES,GofronK:OccvHs,DessauDS:Keyftm}. The
complications arise from the detection of several additional
features besides those related to the primary electronic
structure. Firstly, {\it shadow bands}, which give rise to a
replica of the main FS shifted by the wave vector ($\pi$,$\pi$) in
the 2D Brillouin zone [\textcite{AEBIP:COMFMB1}; see
Fig.\,\ref{aebi},\,\ref{ding2},\,and \ref{fretwell1}]. These
correspond to the backfolding of the main electronic structure
with respect to the mirror planes perpendicular to $\Gamma$-$X$ or
$\Gamma$-$Y$ (in the following we will make use of the momentum
space notations defined for Bi2212 in Fig.\,\ref{bz_tot}). After
the early explanation by \textcite{AEBIP:COMFMB1} in terms of
short-range and dynamic AF correlations resulting in a
$(\!\sqrt{2}\times\!\sqrt{2})$R45$^\circ$ superstructure
\cite{KampfAP:Spefps}, several alternatives have been
proposed.\footnote{\textcite{ChakravartyS:ComCFs,SinghDJ:Strmbc,DingH:EleeB2,SalkolaMI:Impcos,HaasS:Antipb,kordyuk:2};
see also \textcite{lynch:1}, pg.\,248.}
These include a structural origin reflecting the presence of two
inequivalent Cu sites per CuO$_2$ plane in the face-centered
orthorhombic unit cell \cite{SinghDJ:Strmbc,DingH:EleeB2}, but a
possible although not yet well-defined link to superconductivity
has also been proposed: \textcite{kordyuk:2} observed that the
intensity of the shadow bands shows a correlation with $T_c$, as a
function of doping. However, no conclusive picture has been
reached so far.

\begin{figure}[t!]
\centerline{\epsfig{figure=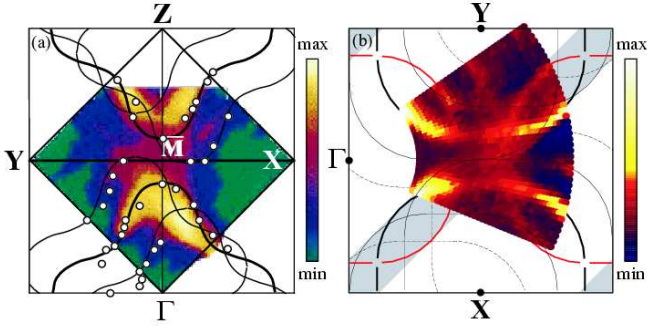,width=1\linewidth,clip=}}
\caption{(a) Comparison between integrated intensity map [50\,meV
at $E_F$; \textcite{SainiNL:Toptps}] and FS crossings (white
circles) determined by several methods from 33\,eV data on
slightly overdoped Bi2212. Thick and thin black lines indicate the
main electronlike FS and its {\it umklapp} replicas, respectively
\cite{ChuangY-D:Reetes}. (b) Normal state $E_F$ intensity map for
Bi2212 obtained with 21.2\,eV photons by normalizing the ARPES
spectra with respect to the total integrated intensity. Main and
shadow FSs are indicated by thick black and red lines; first and
second-order {\it umklapp} FSs by thin and dashed black lines
\cite{BorisenkoSV:JoypFs} [Color].}\label{chuang1}\end{figure}

The second additional feature are the {\it umklapp bands}, which
are commonly referred to as originating from the diffraction of
the photoelectrons off the superstructure present in the BiO
layers, but should rather be considered as a modification of the
intrinsic in-plane electronic structure due to the incommensurate
distortion of the approximate tetragonal symmetry
\cite{OsterwalderJ:Ang-re,AEBIP:COMFMB2,SchwallerP:StrFsm,DingH:EleeB2}.
This distortion is characterized by a periodicity of 27 \AA\ along
the $b$ axis and results in a unit cell approximately five times
larger along this direction
\cite{WithersRL:eledgt,YamamotoA:Rieatm}. This gives rise to two
{\it umklapp} replicas of the FS shifted by $\pm(0.21\pi,0.21\pi)$
as shown in Fig.\,\ref{ding2}a,\,\ref{chuang1},
and\,\ref{fretwell1}, and even to higher order ones shifted by
$\pm(0.42\pi,0.42\pi)$ and labelled UB2 in Fig.\,\ref{fretwell1}a
\cite{OsterwalderJ:Ang-re,FretwellHM:FersB2}.

The last complications come from the so-called {\it bilayer band
splitting}, namely the splitting of the CuO$_2$ plane derived
electronic structure in bonding (BB) and antibonding (AB) bands
due to the presence of CuO$_2$ bilayer blocks in the unit cell of
Bi2212. On the basis of symmetry arguments the splitting is
expected to be maximum at ($\pi$,0), while it vanishes along the
(0,0)-($\pi$,$\pi$) direction
\cite{ANDERSENOK:LDAELH,ChakravartyS:Inttga}. As a result, around
($\pi$,0) two main bands, two shadow bands, and four {\it umklapp}
bands cross the Fermi level.

Fig.\,\ref{aebi} reproduces the FS data and analysis by
\textcite{AEBIP:COMFMB1} for near optimally doped Bi2212. Panel
(a) shows a plot of the spectral weight integrated within an
energy window of 10 meV near the Fermi level (data were taken with
an energy resolution of 30-40 meV). This is the first time that a
detailed momentum-space $E_F$ map has been carried out on cuprate
superconductors. The authors concluded that the FS is a hole like
piece, as indicated by the thick lines in Fig.\,\ref{aebi}b. The
complications in the intensity map stem from the superposition of
shadow FSs that are offset from the main FS by ($\pi$,$\pi$), as
shown in Fig.\,\ref{aebi}c and discussed above.
\textcite{AEBIP:COMFMB1} also concluded that there are no BiO FS
pockets, which had been predicted on the basis of band structure
calculations as shown in Fig.\,\ref{aebi}d
\cite{MassiddaS:ElespB}. As a matter of fact, one additional
problem with Bi2212 is that there are no reliable band
calculations: all theoretical results predict a BiO FS that has
not been convincingly observed \cite{ShenZX:Elesps}. The work by
\textcite{AEBIP:COMFMB1} also represents the beginning of a very
different approach in the field: instead of analyzing the data in
terms of energy scans at fixed momenta or {\it energy distribution
curves} (EDCs) as usually done, \textcite{AEBIP:COMFMB1} for the
first time made use of momentum scans at fixed energy or {\it
angular distribution curves} [nowadays more commonly referred to
as {\it momentum distribution curves} (MDCs) after
\textcite{VallaT:Eviqcb} who used this approach to extract
self-energy corrections from the ARPES data; see
Fig.\,\ref{valla1} and Sec.\,\ref{sec:selfmodes}]. In a series of
papers, Aebi, Osterwalder, and coworkers used this approach to
identify the superstructure periodicity, and the intensity and
dispersion of main and shadow bands
\cite{AEBIP:COMFMB1,OsterwalderJ:Ang-re,SchwallerP:StrFsm}.

The results by \textcite{AEBIP:COMFMB1} are different from the
early work by \textcite{DessauDS:Keyftm}, which indicated another
main FS crossing in the ($\pi$,0) region: the detection of two
pieces of FS, one electronlike and the other holelike, was
interpreted as a signature of bonding-antibonding splitting in a
bilayer system due to the presence of two CuO$_2$ planes per unit
cell \cite{DessauDS:Keyftm}. Subsequently, in a set of two papers
\textcite{DingH:EleeB2,DingH:EvotFs} argued against the bilayer
splitting [a result which was taken as key supporting evidence for
the inter-layer tunnelling mechanism for high temperature
superconductivity
\cite{AndersonPW:Inttmh,andersonPW:1,AndersonPW:c-axie}], and in
favor of a FS characterized by a simple holelike barrel.
They attributed the additional FS crossing identified by
\textcite{DessauDS:Keyftm} along the (0,0)-($\pi$,0) line to
complications arising from the presence of a superstructure in the
BiO layers (i.e., {\it umklapp} bands), as shown in
Fig.\,\ref{ding2}. In particular, main and {\it umklapp} bands
were identified in the complicated ($\pi$,0) region on the basis
of their different polarization dependence. In contrast with the
case of LSCO where a crossover from a hole to electronlike FS is
observed near optimal doping \cite{InoA:Fersbd,InoA:Dop-de},
\textcite{DingH:EvotFs} suggested that a single holelike sheet of
FS is characteristic of all the studied doping levels. In
addition, no shadow band was detected for the underdoped sample
($T_c\!=\!15$\,K), which was taken as evidence to rule out the FS
pocket idea presented in Fig.\,\ref{models}c, and also the
explanations in terms of AF correlations for the shadow bands
observed near optimal doping by \textcite{AEBIP:COMFMB1}. However,
the ARPES spectra from the $T_c\!=\!15$\,K underdoped Bi2212
sample are characterized by a very broad lineshape indicating
strong many-body effects on the spectral function which, as a
consequence, is completely incoherent. The FS extracted from those
data should be cautiously considered, in particular as far as the
lack of bilayer splitting and of the FS pockets (or shadow band)
is concerned.

After the initial debate, an agreement has been reached concerning
the absence of the BiO pockets predicted for Bi2212 by band
structure calculations. Furthermore, the general consensus was in
favor of a holelike FS centered at ($\pi$,$\pi$) from under to
overdoped samples ($T_c$ between 15 and 67 K), with a volume
consistent with the electron density in accordance with the
Luttinger's theorem \cite{luttinger:1}. Recently, however, other
reports questioned this simple picture arguing that this may not
be a complete characterization of the low-lying excitations in
Bi2212
\cite{FengDL:1,ChuangY-D:Reetes,gromko:1,BogdanovPV:PhosPd}. These
studies revealed a pronounced photon energy dependence of the
low-energy spectral weight in the ($\pi$,0) region and suggested
an electronlike FS centered at the $\Gamma$-point as shown in
Fig.\,\ref{chuang1}a \cite{ChuangY-D:Reetes,gromko:1}; or
alternatively two co-existing electronic components resulting in
electron and holelike FSs \cite{FengDL:1}, similar to what
originally proposed by \textcite{DessauDS:Keyftm}. These
suggestions were opposed by other groups which claimed that only a
universal holelike FS is supported by the ARPES data once the
effects of the photon energy dependence of the matrix elements in
the ($\pi$,0) region are taken into account
\cite{MesotJ:DettFs,FretwellHM:FersB2,BorisenkoSV:JoypFs,BorisenkoSV:Estm-e}.
As a matter of fact, first-principles simulations of the ARPES
spectra performed within the one-step model, for an idealized
tetragonal Bi2212 structure (i.e, no superstructure modulation),
indicate a strong non-trivial dependence of the low-energy ARPES
intensity on photoelectron momentum, and energy/polarization of
the incident photons \cite{BansilA:Impmet}. Therefore,
photoemission matrix elements may largely dominate the momentum
dependence of the ARPES intensity and should be taken into
account, especially when attempting to estimate the momentum
distribution function $n(k)$ from the integrated photoemission
intensity (see also Sec.\,\ref{sec:spectral}).

Owing to the superior momentum resolution, the complexity of the
ARPES results from Bi2212 is beautifully shown by the new
generation of photoemission data (Fig.\,\ref{chuang1}b
and\,\ref{fretwell1}). In particular along $\Gamma$-$Y$ all the
features characterizing the ARPES spectra from Bi2212 can be
unambiguously identified in the energy distribution curves, as it
is along this direction that the different bands are more clearly
separated (see Fig.\,\ref{fretwell1}a,b, where the dispersive
bands are labelled as indicated in the caption).
\begin{figure}[t!]
\centerline{\epsfig{figure=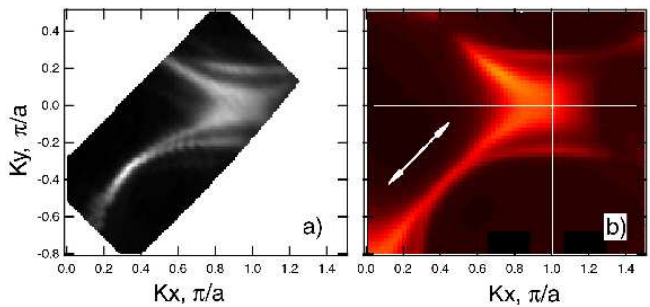,width=0.99\linewidth,clip=}}
\vspace{.0cm} \caption{(a) Intensity map at 12\,meV binding energy
from overdoped Bi2212, obtained with 22\,eV photons at 20\,K with
the superconducting gap already opened \cite{BogdanovPV:PhosPd}.
(b) First-principles simulation for the same experimental geometry
\cite{BansilA:Impmet} [Color].}\label{pasha_bilayer}\end{figure}
However, because of the superposition of several contributions and
especially because of matrix element effects, not even the most
detailed intensity maps are very clear around ($\pi$,0), as shown
in Fig.\,\ref{chuang1}, and in particular in
Fig.\,\ref{fretwell1}c,d where holelike and electronlike scenarios
are compared with respect to the same set of data. In turn, any
interpretation necessarily rests on the extrapolation of results
obtained away from this region.

Very recently it became evident that one of the key ingredients in
the debate upon the FS topology of Bi2212 is actually the
unresolved bilayer splitting of the in-plane electronic structure.
In this regard, as it will be discussed in greater detail in the
next section, an essential piece of information comes from ARPES
data obtained with improved angular resolution in the very
overdoped regime: two FS sheets are resolved, which originate from
the bonding-antibonding splitting of the in-plane electronic
structure caused by the interaction between the layers
\cite{FengDL:Bilste,ChuangY-D:Doutbo}. In this context, the
cleanest experimental results have been obtained on very overdoped
Pb-Bi2212 ($T_c\!=\!70$\,K), in which Pb substitutes into the BiO
planes suppressing the superstructure and in turn the intensity of
the {\it umklapp} bands \cite{BogdanovPV:PhosPd}. As shown in
Fig.\,\ref{pasha_bilayer}, where the ARPES data are compared to
the theoretical simulations by \textcite{BansilA:Impmet} for the
same experimental geometry, two different sheets of Fermi surface
are clearly resolved. With respect to the $\Gamma\!=\!(0,0)$
point, the outer sheet is the bonding FS while the inner piece is
the antibonding one. The bonding pocket is holelike, similar to
what was reported many times before for Bi2212. The antibonding
FS, on the other hand, has its van Hove singularity near ($\pi$,0)
and is thus very close to the holelike/electronlike boundary.
Therefore, the experiments indicating a single holelike or
electronlike FS in Bi2212 might simply be more sensitive to the
bonding or antibonding sheet, respectively, as a consequence of
sample doping and/or experimental conditions (such as the
polarization and the energy of the incident photons).

\subsubsection{Bilayer band splitting}
\label{sec:bilayer}

Independently of the controversy on the FS of Bi2212, whether the
electronic structure of the cuprates can be influenced by the
interaction between the CuO$_2$ planes has been an issue of
interest for a long time, as one of the proposed mechanisms for
high-temperature superconductivity is via interlayer pair
tunnelling \cite{ChakravartyS:Inttga,andersonPW:1}.
\begin{figure}[b!]
\centerline{\epsfig{figure=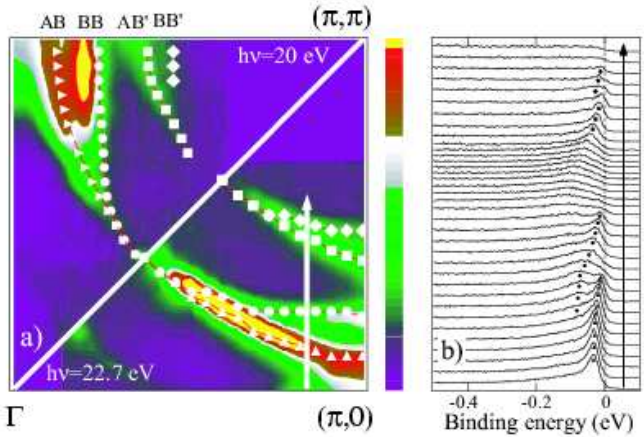,clip=,angle=0,width=1\linewidth}}
\caption{(a) Integrated $E_F$ intensity map (integration window
[-20\,meV, 10\,meV]) from overdoped Bi2212 ($T_c\!=\!65$\,K). The
data were measured above $T_c$ at 22.7\,eV (lower-right,
$T\!=\!75$\,K) and 20\,eV (upper-left, $T\!=\!80$\,K). Symbols are
the FS crossings estimated directly from the dispersion of the
peaks. BB and AB refer to primary bonding and antibonding bands,
and BB$^\prime$ and AB$^\prime$ to the corresponding
superstructure replicas. (b) ARPES spectra along the cut indicated
by the white arrow in (a). After \textcite{FengDL:Bilste}
[Color].} \label{bilayerfs}
\end{figure}
Within this model, interlayer hopping is suppressed because of
correlations and only electron pairs can tunnel between the
planes: coherent pair-tunnelling will drive the system into the
superconducting state via the lowering of kinetic energy
perpendicular to the planes. As discussed above, an early ARPES
study on Bi2212 claimed the absence of bilayer splitting
\cite{DingH:EleeB2}.  This report attracted much interest as it
provided a crucial support for the interlayer tunnelling model,
although with hindsight the momentum resolution was not sufficient
to conclusively address this issue.  Recently, new ARPES data
suggested that the bilayer splitting can be detected for overdoped
Bi2212 with the improved angular resolution of the Scienta
spectrometer \cite{FengDL:Bilste,ChuangY-D:Doutbo}. Below, we will
discuss this issue in more detail. As for the interlayer
tunnelling model, it has to be mentioned that strong evidence
against it was reported on the basis of direct imaging of magnetic
flux vortices by scanning SQUID microscopy \cite{MolerKA:ImaiJv},
and by optical measurements on Tl$_2$Ba$_2$CuO$_{6+\delta}$
\cite{TsvetkovAA:Glolmt}. On the other hand, a subsequent optical
study provided direct evidence for the lowering of the kinetic
energy parallel to the planes across the superconducting
transition, supporting the view that superconductivity in the
cuprates is indeed unconventional \cite{molegraaf:1}.

\begin{figure}[t!]
\centerline{\epsfig{figure=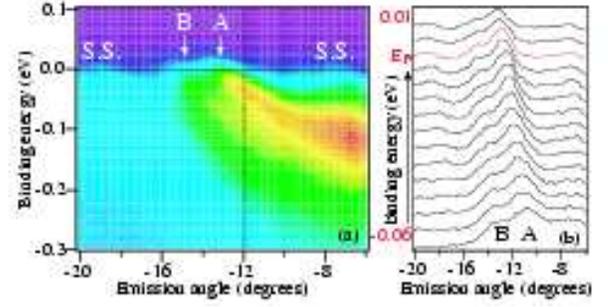,clip=,angle=0,width=8.5cm}}
\caption{(a) $E$-$k$ image plot of the ARPES intensity and (b)
corresponding MDCs taken on overdoped Bi2212 ($T_c\!=\!85$\,K)
near the ($\pi$,0) region \cite{ChuangY-D:Doutbo} [Color].}
\label{bilayerchuang}
\end{figure}

\begin{figure}[b!]
\centerline{\epsfig{figure=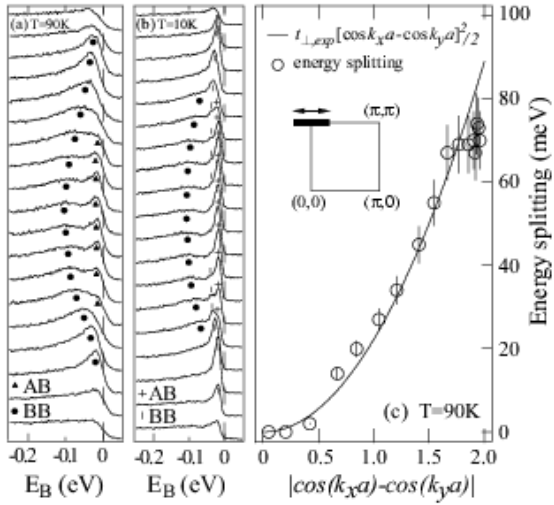,clip=,angle=0,width=8cm}}
\caption{(a) Normal and (b) superconducting state ARPES spectra
from overdoped Bi2212 ($T_c\!=\!65$\,K) taken with 21.2\,eV
photons (HeI) along $(-0.24\pi,\pi)$-$(0.24\pi,\pi)$. (c)
Magnitude of the bilayer splitting along the antibonding FS
(normal state data). The curve is a fit to
$t_{\bot,exp}[\cos(k_xa)-\cos(k_ya)]^2/2$ obtained with
$t_{\bot,exp}\!=\!44\pm5$\,meV. After \textcite{FengDL:Bilste}.}
\label{bilayerpi0edc}
\end{figure}

Fig.\,\ref{bilayerfs}a presents the normal-state data from a
$T_c\!=\!65$\,K overdoped Bi2212 sample \cite{FengDL:Bilste}. Two
FSs are observed for the main electronic structure (BB and AB) as
well as its superstructure replica (BB' and AB'). These FSs
correspond to four well resolved quasiparticle peaks
(Fig.\,\ref{bilayerfs}b), whose dispersion can be clearly followed
in the ARPES spectra taken along the $k$-space cut indicated by
the arrow in Fig.\,\ref{bilayerfs}a. As expected on the basis of
symmetry arguments \cite{ANDERSENOK:LDAELH,ChakravartyS:Inttga} no
splitting has been detected along the nodal direction (even with
the best achievable angular resolution of about $0.12^\circ$); the
splitting increases upon approaching ($\pi$,0) where it exhibits
the largest value of about 88\,meV. The different photon energy
dependence of BB and AB bands, evidenced by the relative intensity
change in the two FS mappings presented in Fig.\,\ref{bilayerfs}a
(top-left and bottom-right), is consistent with the different
symmetry of the two bands along the $c$-axis (odd and even for AB
and BB, respectively). Analogous conclusions were independently
obtained by \textcite{ChuangY-D:Doutbo} from the analysis in terms
of momentum distribution curves of similar ARPES data [see
Fig.\,\ref{bilayerchuang}, where BB and AB bands can be observed
both in the image plot of the ARPES intensity near ($\pi$,0) and
in the corresponding MDCs]. All together, the different studies
indicate a normal state bilayer splitting of the order of 70-100
meV in overdoped Bi2212
\cite{FengDL:Bilste,ChuangY-D:Doutbo,BogdanovPV:PhosPd}.
As shown in Fig.\,\ref{bilayerpi0edc}c where the magnitude of the
normal state bilayer splitting along the AB FS is plotted, the
observed energy splitting agrees well with the functional form
$2t_\bot({\bf k})\!=\!t_\bot[\cos(k_xa)-\cos(k_ya)]^2/2$ predicted
on the basis of bilayer LDA calculations \cite{ANDERSENOK:LDAELH}.
However, the experimental value of 88 meV ($2t_{\bot,exp}$) is
much smaller than the 300 meV ($2t_{\bot,LDA}$) predicted by the
bilayer LDA calculations. It is actually in better agreement with
bilayer Hubbard model calculations \cite{LiechtensteinAI:Quabsb},
which predicted an energy splitting of similar momentum dependence
and with a maximum value of about 40 meV. This smaller value
reflects the reduced hopping and is a direct consequence of the
on-site Coulomb repulsion $U$ explicitly included in the model.
The fact that the splitting is underestimated as compared to the
experiment suggests that a smaller value of  $U$ should be used
for this system.

\textcite{FengDL:Bilste} also investigated the effect of the
superconducting transition on the ARPES spectra at ($\pi$,0),
where the bilayer splitting is stronger. The data measured above
and below $T_c$ are presented in Fig.\,\ref{bilayerpi0edc}a
and\,\ref{bilayerpi0edc}b, respectively. Note that the
peak-dip-hump structure in Fig.\,\ref{bilayerpi0edc}a is totally
unrelated to superconductivity (Sec.\,\ref{sec_suppeak}), as the
data were taken well above $T_c$; instead it corresponds to the BB
and AB split bands (indicated by circles and triangles,
respectively). In the superconducting state
(Fig.\,\ref{bilayerpi0edc}b), the two bands reach the same minimum
binding energy (16\,meV), indicating the opening of a
superconducting gap with similar amplitude for both FSs
(Sec.\,\ref{sec_supgap}). Furthermore, while no change was seen
for the AB band at high-binding energy (see black circles in
Fig.\,\ref{bilayerpi0edc}a,b), a double peak structure was
detected close to $E_F$ at certain momenta (see crosses and bars
in Fig.\,\ref{bilayerpi0edc}b). The appearance of a second sharp
peak near $E_F$ below $T_c$ presumably corresponds to the
sharpening of a structure that could not be resolved in the normal
state. A possible origin of this effect may be the coupling
between the BB state and a low-energy collective mode (either
phononic or electronic in nature), which would result in the
transfer of spectral weight to the characteristic energy of the
mode (see Sec.\,\ref{sec:selfmodes}).

At this point the obvious question is why has a clear bilayer
splitting never been observed before by ARPES, despite the intense
effort? Obviously this achievement would have not been possible
without the recent improvements in energy and momentum resolution.
However, in this case the major breakthrough can probably be
identified in the recent progress in high pressure annealing
techniques. This allowed the synthesis of heavily overdoped single
crystals of Bi2212, which appear to exhibit much better defined
quasiparticle peaks, suggesting a more FL-like behavior.
\begin{figure}[t!]
\centerline{\epsfig{figure=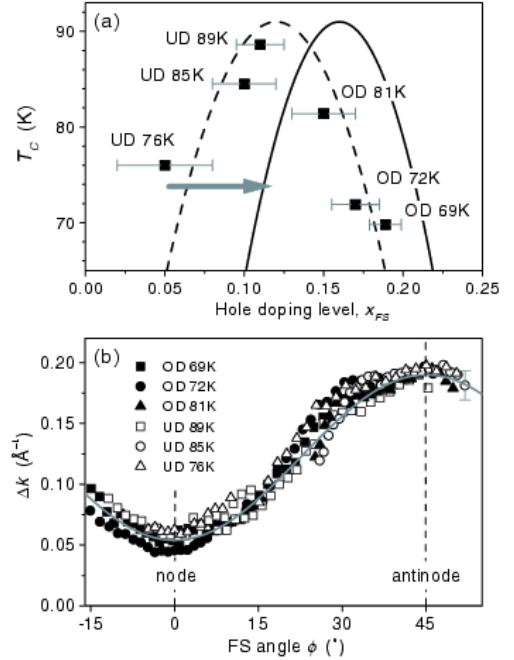,clip=,angle=0,width=7.2cm}}
\caption{(a) $T_c$ plotted versus the hole concentration $x_{FS}$
as inferred from the volume of the FS determined by ARPES
(symbols). The solid line is the universal empirical $T_c$-$x$
relation \cite{TallonJL:Genspb}. (b) MDC width at $E_F$ from along
the 300\,K FS ($\phi$ is the angle defined with respect to the
nodal direction). The solid grey line is a fit to a functional
form proportional to $\sin^2(2\phi)$. After \textcite{kordyuk:2}.}
\label{bilayer_kord}
\end{figure}
One could then speculate that the absence of two well-defined
features in the spectra of less overdoped samples does not
necessarily indicate the complete absence of bilayer splitting
effects, which may be hidden in the breadth of the lineshapes.
Indeed, this speculation is supported by many recent ARPES results
\cite{FengDL:3,ChuangY-D:1,kordyuk:1,kordyuk:2}. For example,
using the HeI line as the excitation energy, \textcite{kordyuk:2}
could only observe one large holelike FS centered at $(\pi,\pi)$.
However, as shown in Fig.\,\ref{bilayer_kord}a the experimentally
determined FS volume would correspond to a doping level which is
5-7\% off the expected value (see gray arrow in
Fig.\,\ref{bilayer_kord}a). This apparent violation of Luttinger's
theorem \cite{luttinger:1} can be understood by considering that
the second sheet of FS resulting from bilayer splitting might have
escaped detection do to matrix elements.
\begin{figure}[b!]
\centerline{\epsfig{figure=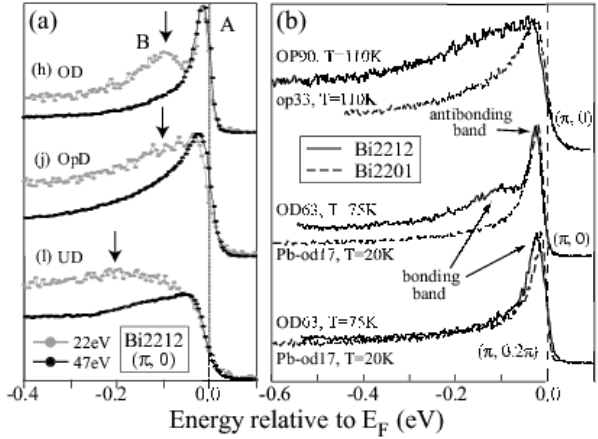,clip=,angle=0,width=1\linewidth}}
\caption{(a) ARPES spectra from Bi2212 at ($\pi$,0), for different
doping levels and photon energies \cite{ChuangY-D:1}. (b) ARPES
spectra from Bi2212 and Bi2201 measured with 22.7\,eV photons near
($\pi$,0), for various dopings \cite{FengDL:3}. All data were
taken in the normal state.} \label{bilayer_mix}
\end{figure}
This is further corroborated by the results reported for the
evolution of the MDC width along the FS. As shown in
Fig.\,\ref{bilayer_kord}b, the angular dependence of the width
does not change with doping and is thus unlikely to be due to
many-body effects; in addition, its functional form is
proportional to $\sin^2(2\phi)$, which is in agreement with a
bilayer splitting of the form $[\cos(k_xa)\!-\!\cos(k_ya)]^2$, as
determined in the overdoped regime
\cite{FengDL:Bilste,ChuangY-D:Doutbo}.

Evidence for bilayer splitting at under and optimal doping is also
provided by systematic ARPES investigations performed as a
function of doping and photon energy
\cite{FengDL:3,ChuangY-D:1,kordyuk:1,kordyuk:2,gromko:2}. As shown
in Fig.\,\ref{bilayer_mix}a, the BB and AB bands clearly resolved
in overdoped Bi2212 with 22\,eV photons become indistinguishable
upon underdoping the system. However, in spectra taken with
different photon energy at a given doping the centroid is found at
different positions; it moves from close to $E_F$ to higher
binding energies upon changing the photon energy from 47 to 22\,eV
(Fig.\,\ref{bilayer_mix}a).  This behavior indicates the presence
of two different components in the spectra, with opposite photon
energy dependence as expected for BB and AB bands, which have
opposite symmetry along the $c$-axis with respect to the midpoint
between the two CuO$_2$ planes \cite{ChuangY-D:1,FengDL:3}.
Additional signatures of bilayer splitting can be observed by
comparing spectra from Bi2212 and Bi2201 at similar doping levels
(Fig.\,\ref{bilayer_mix}b). While in the nodal region the spectra
have very similar lineshapes (not shown), the situation is much
more complicated close to ($\pi$,0). Note that in
Fig.\,\ref{bilayer_mix}b, while the ($\pi$,0) spectrum from
overdoped Bi2212 (OD63) shows both BB and AB peaks, the one at
($\pi$,0.2$\pi$) reduces to only the BB peak because the AB band
is already above $E_F$. As a consequence, the spectra from Bi2212
and Bi2201 match almost perfectly at ($\pi$,0.2$\pi$) but not at
($\pi$,0). In this context, the mismatch between optimally doped
Bi2212 and Bi2201 spectra at ($\pi$,0) and, in particular, the
additional spectral weight at high energy observed in Bi2212 can
be naturally attributed to the presence of the BB band. On the
basis of these results, it was concluded that the broad lineshape
observed in the normal state for under and optimally doped Bi2212
at ($\pi$,0) has significant contributions from bilayer splitting
effects \cite{ChuangY-D:1,FengDL:3}. It was also proposed that BB
and AB bands may become directly detectable in the superconducting
state due to the sharpening of the peaks
\cite{kordyuk:1,gromko:2}, which provides a possible explanation
for the so-called peak-dip-hump structure typically detected only
below $T_c$ at those doping levels (see Sec.\,\ref{sec_suppeak}
and\,\ref{sec:selfdiscu}).

\subsubsection{Summary}

Despite the controversy over the FS, many of the ARPES results
from Bi2212 are still reliable and the same holds for the
qualitative descriptions that were developed on their basis. This
is particularly true with regard to doping dependent studies
performed under identical experimental conditions. For example,
the magnitude and symmetry of the superconducting gap
(Sec.\,\ref{sec_supgap}) and of the normal state excitation gap or
pseudogap (Sec.\,\ref{sec:pseudogap}), are relatively insensitive
to the details of the FS topology.

Due to the limited doping range investigated, the ARPES results
from Bi2212 are insufficient to conclude in favor of any of the
four scenarios summarized in Fig.\,\ref{models}. While the FS of
optimally doped Bi2212 resembles the one reported in
Fig.\,\ref{models}a and\,\ref{models}d (let us here neglect the
issue of bilayer splitting), in the underdoped region, due to the
opening of the pseudogap along the underlying FS
(Sec.\,\ref{sec:pseudogap}), the data are reminiscent of the
models depicted in Fig.\,\ref{models}b and\,\ref{models}c. The
distinction between Fig.\,\ref{models}a and\,\ref{models}d has to
be determined on the basis of the behavior near ($\pi$,0), while
between the models described in Fig.\,\ref{models}b
and\,\ref{models}c it lies in the detection of the shadow FS given
by the dashed line in Fig.\,\ref{models}c. Although it has been
argued that the case of Fig.\,\ref{models}c does not apply to
Bi2212 \cite{DingH:EvotFs}, in contrast to an earlier report
\cite{AEBIP:COMFMB1}, no consensus has yet been reached.

\subsection{Bi$_2$Sr$_2$CuO$_{6+\delta}$}

The investigation of the low-energy electronic structure should be
an easier task for the single-layer system
Bi$_2$Sr$_2$CuO$_{6+\delta}$ (Bi2201) than for Bi2212. First, the
ARPES spectra are free from possible complications arising from
bilayer splitting. Second, Bi2201 has a lower $T_c$ which allows
measurements at lower temperature (i.e., less thermal broadening),
without complications due to the opening of the superconducting
gap. Furthermore, as in the case of Bi2212, also in Bi2201 the
partial substitution of Bi with Pb introduces additional disorder
in the BiO planes which, in turn, suppresses the superstructural
modulation and the intensity of the {\it umklapp} bands.
\begin{figure}[b!]
\centerline{\epsfig{figure=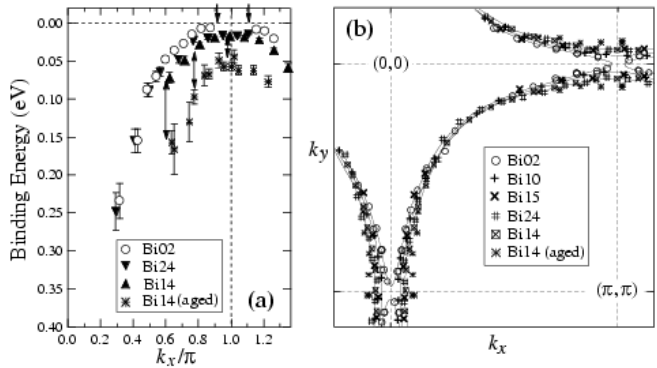,width=1\linewidth,clip=}}
\caption{(a) Dispersion along (0,0)-($\pi$,0) and (b) FS crossings
for Pb-Bi2201 at various dopings (numbers in the legends indicate
the values of $T_c$). An electronlike FS is observed for the
overdoped sample with $T_c\!<\!2$\,K. After
\textcite{TakeuchiT:ToptFs}.}\label{takeuchi2}\end{figure}
All these characteristics would make Bi2201 an ideal candidate to
study the evolution of the FS topology with doping, providing a
valuable comparison for the more complex case of Bi2212. On the
other hand, similarly to the Bi2212 case, no data are available in
the extremely underdoped regime and, therefore, the ARPES
investigation of Bi2201 does not provide any more detailed
information about the metal-insulator transition.

Recently, two independent investigations of the low energy
electronic structure of Pb-doped Bi2201 for various dopings were
reported by \textcite{TakeuchiT:ToptFs} and by
\textcite{SatoT:Ferssg,SatoT:Evih-l}. The ARPES spectra are
characterized by a well defined quasiparticle band whose
dispersion can be easily followed and is reported for different
doping levels in Fig.\,\ref{takeuchi2}a, while panel b presents
the FS determined by various methods: by symmetrizing the spectra
with respect to $E_F$ \cite{NormanMR:Phetl-}, and by dividing out
the Fermi distribution broadened by the experimental resolution
\cite{GreberT:PhoaFl}. In underdoped and optimally doped samples
the FS is characterized by a single hole barrel centered at
($\pi$,$\pi$), in good agreement with the expectation
(Fig.\,\ref{takeuchi2}b).
\begin{figure}[t!]
\centerline{\epsfig{figure=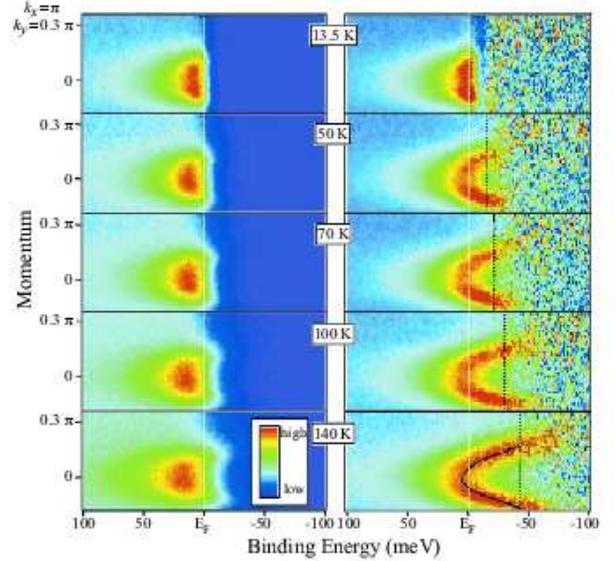,width=8.5cm,clip=}}
\caption{Raw ARPES intensity (left panels) and divided by the
Fermi-Dirac distribution function (right panels) for
Bi$_{1.80}$Pb$_{0.38}$Sr$_{2.01}$CuO$_{6-\delta}$, along the
($\pi$,0)-($\pi$,$\pi$) direction, at several temperatures. The
black dotted lines indicate the energy at which the Fermi-Dirac
function has a value of 0.03, and the black solid line for
$T\!=\!140$\,K represents the peak positions obtained by fitting
the data \cite{SatoT:Evih-l} [Color].}\label{sato2201}\end{figure}
Upon increasing doping, the FS volume also increases (counting
holes). In the overdoped regime and, in particular, for the overdoped
Pb-Bi2201 sample characterized by $T_c\!<\!2$ K, the quasiparticle
band crosses $E_F$ before ($\pi$,0) as shown in
Fig.\,\ref{takeuchi2}a, defining an electronlike FS centered at (0,0),
in agreement with the rigid band picture (Fig.\,\ref{takeuchi2}b). The
results reported by \textcite{TakeuchiT:ToptFs} represents the first
direct observation of a continuous doping evolution of the FS topology
of Bi2201 from holelike to electronlike, similar to that seen for
LSCO.

This topological crossover was not observed by
\textcite{SatoT:Ferssg,SatoT:Evih-l}, possibly due to a slightly
lower doping level of the samples (instead in the underdoped and
optimally doped regime, the results from the two different groups
are essentially the same). To determine whether the saddle point
at ($\pi$,0) is above or below $E_F$, \textcite{SatoT:Evih-l}
performed a detailed temperature dependence study. The left panels
of Fig.\,\ref{sato2201} present data which indicate the presence
of spectral weight below the Fermi level. From the bare data it is
hard, however, to judge whether the quasiparticle band is indeed
below $E_F$ or the intensity corresponds to the broadened tail of
a band located just above $E_F$. This issue can be addressed in a
more conclusive way by dividing the spectra by the Fermi-Dirac
distribution function at each temperature. The results are
presented in the right panels of Fig.\,\ref{sato2201} and show the
actual dispersion in a much clearer fashion. It is now obvious
that the saddle point is located approximately 5\,meV below $E_F$
(which indicates that the difference between the results reported
by the two groups is very subtle). As a more general remark, it
should be emphasized that this kind of analysis appears to be a
better method than the simple tracking of the ARPES intensity in
determining the position of a quasiparticle band very close to the
Fermi energy, especially in borderline cases such as in proximity
to a topological transition.

\subsection{Bi$_2$Sr$_2$Ca$_2$Cu$_3$O$_{10+\delta}$}
\label{sec:nsBi2223}

\begin{figure}[t!]
\centerline{\epsfig{figure=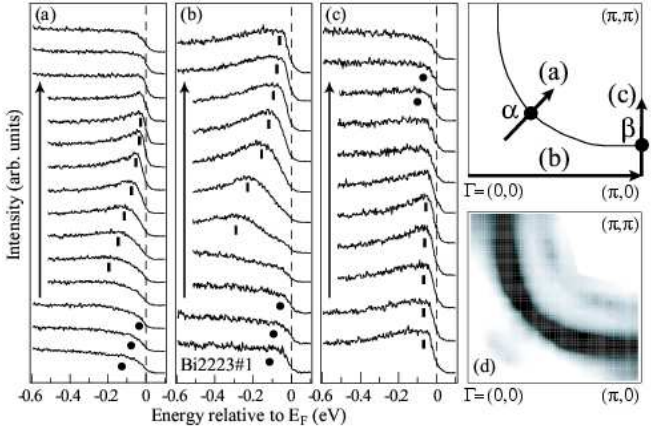,width=1\linewidth,clip=}}
\caption{(a-c) Normal state Bi2223 ARPES spectra taken along the
high-symmetry lines as shown in the Brillouin zone sketch
($T\!=\!125$\,K, $h\nu\!=\!21.2$\,eV). Main ({\it umklapp}) bands
are marked with bars (circles). (d) Momentum space map of the
spectral weight at $E_F$ ($\pm10$\,meV). After
\textcite{FengDL:2}.}\label{dl2223}\end{figure}

Very recently, the first ARPES studies of the trilayer cuprate
superconductors Bi$_2$Sr$_2$Ca$_2$Cu$_3$O$_{10+\delta}$ (Bi2223)
have been carried out by several groups
\cite{FengDL:2,sato:1,muller:1}. As shown in Fig.\,\ref{dl2223}
where normal-state data from nearly-optimally doped Bi2223 are
presented, dispersive quasiparticle peaks were detected by ARPES
(including the {\it umklapp} features due to the superstructure of
the BiO layers, already discussed for Bi2212 and Bi22201). The FS
obtained by integrating the ARPES spectra in a $\pm10$\,meV energy
window at $E_F$ is a large holelike pocket centered at the
($\pi$,$\pi$) point (see Fig.\,\ref{dl2223}d). At present, no
clear evidence of additional sheets of FS due to interlayer
coupling was reported, contrary to what is theoretically
predicted. In particular, one would expect the FS to be comprised
of bonding (BB), nonbonding (NB), and antibonding (AB) sheets
[see, e.g., \textcite{mori:1}]. Whether or not the three FS sheets
would have the same topology depends on the details of the system
(such as, e.g., the doping level). In particular one could expect
the BB and NB FSs to be holelike in character, while the AB could
be electronlike.
However, because of the short escape depth of the outgoing
photoelectrons, the photoemission intensity is expected to be
dominated by the signal from the outer CuO$_2$ plane and, in turn,
from the NB sheet of FS (the NB wavefunction has the highest
probability density on the outer CuO$_2$ plane because, contrary
to the BB and AB ones, it is the antisymmetric combination of the
wavefunctions of {\it only} the two outer planes within a trilayer
block). This could be the reason why a single holelike FS was
observed by ARPES for Bi2223 (Fig.\,\ref{dl2223}d). The above
speculation, however, needs to be tested by more extensive
experiments.

\subsection{YBa$_2$Cu$_3$O$_{7-\delta}$}
\label{sec:nsYBCO}

\begin{figure}[b!]
\centerline{\epsfig{figure=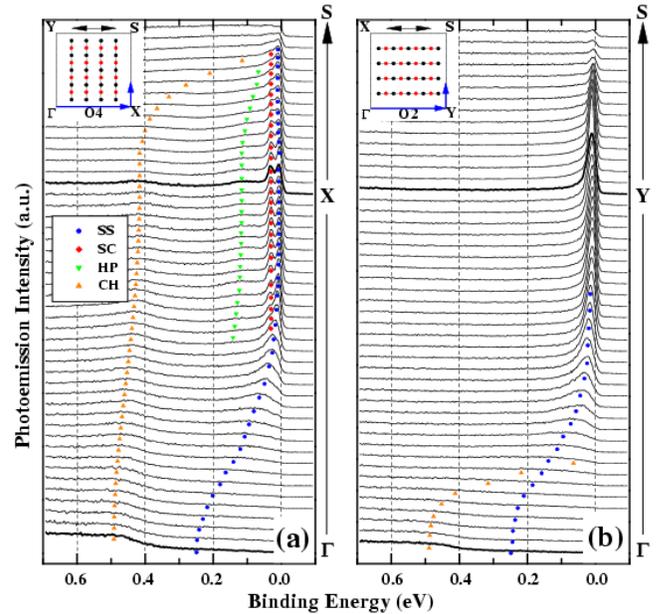,width=8.5cm,clip=}}
\caption{ARPES spectra from untwinned Y123 taken at 10 K with 28
eV photons. Polarization (black arrows) and CuO chain directions
are indicated in the insets. The labels SS, SC, HP, and CH, refer
to surface state, superconducting peak, hump, and chain state,
respectively \cite{LuDH:Supgsi} [Color].} \label{lu_edc}
\end{figure}

The Y-based copper oxides are probably the most important family
of high-$T_c$ superconductors. In fact YBa$_2$Cu$_3$O$_{7-\delta}$
(Y123) was the first compound for which superconductivity with
$T_c\!>\!77$\,K (i.e., the liquid nitrogen boiling point) was
discovered \cite{WuMK:Sup9Kn}. Furthermore, it is on the Y-based
cuprates that some of the most important experiments have been
preformed. One specific feature of Y123 is that, contrary to the
other cuprate HTSCs, the reservoir layers in these materials
contain CuO chains aligned along the $\Gamma$-$Y$ direction (see
Fig.\,\ref{bz_tot} or\,\ref{lu_edc} for the Brillouin zone
notations). Because of the presence of the CuO chain layer Y123
possesses an orthorhombic crystal structure
\cite{JorgensenJD:Oxyoo-}, which according to band calculations
should result in a significant anisotropy of the in-plane
electronic structure \cite{ANDERSENOK:LDAELH}. Compared to other
compounds, in recent years relatively little photoemission work
has been done on Y123, in particular as far as the doping
evolution of the electronic structure is concerned
[\textcite{RongLiu:Elesaf}; see also
\textcite{ShenZX:Elesps,lynch:1}]. This is partly due to the
natural twinning of the Y123 crystals (i.e., domains orthogonal to
each other with respect to the CuO chain orientation are present),
and to generic sample quality issues which are responsible for the
poor quality of the cleaved surface
\cite{EdwardsHL:Enegss,SchabelMC:AngpuY}. However, the most
serious problem in performing photoemission experiments on Y123 is
represented by the presence of a very narrow and intense surface
state peak located 10 meV below $E_F$ at both $X$ and $Y$
\cite{SchabelMC:AngpuY}. This feature, which was initially
interpreted as an extended van Hove singularity analogous to the
one observed in the Bi-based superconductors
\cite{GofronK:ObsEVH}, dominates over the low-energy bulk signal
hindering the investigation of the intrinsic electronic structure.

More detailed studies of untwinned samples were carried out only
near optimal doping. The most important results are the
identification of the surface state, $k_z$ dispersion, and the
superconducting gap
\cite{Schabel:1,SchabelMC:Ang-re,SchabelMC:AngpuY}.
The establishment of the surface state nature of the intense peak
just below $E_F$ [originally proposed on the basis of the
dependence of its intensity on photon energy/polarization, as well
as on oxygen, Co, and Pr doping \cite{SchabelMC:AngpuY}] is
crucial in order to understand the data, and in particular to
extract information on the superconducting state
(Sec.\,\ref{sec:ybcosuper}).

Recently, \textcite{LuDH:Supgsi} reported significant progress in the
investigation of Y123 by ARPES, made possible by the improvement in
sample quality \cite{LiangRuixing:GrohqY} and instrumental resolution.
\begin{figure}[t!]
\centerline{\epsfig{figure=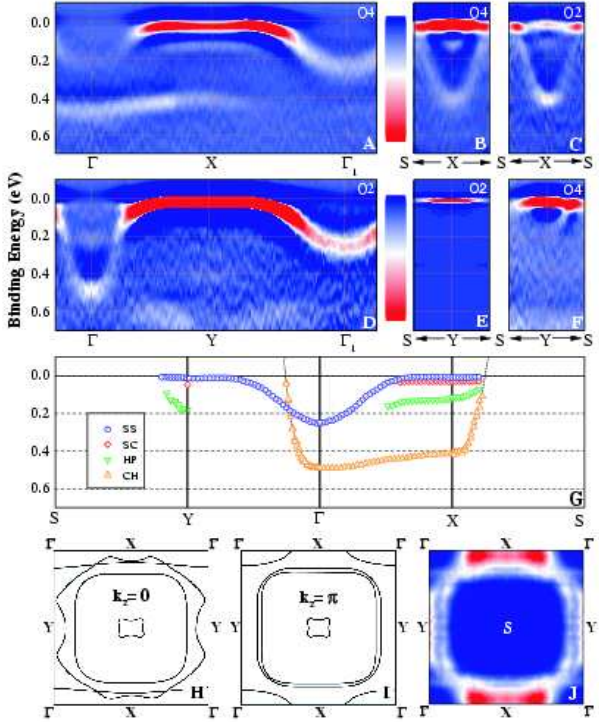,width=8.5cm,clip=}}
\caption{(a)-(f) Second derivative with respect to the binding
energy of the raw ARPES spectra from Y123 (sample orientations as
in Fig.\,\ref{lu_edc}). (g) Summary of all the dispersive bands.
(h,i) FS of Y123 from band structure calculations
\cite{ANDERSENOK:ELEPTI,ANDERSENOK:PLADSB}, and (j) integrated
photoemission intensity map (50 meV at $E_F$). After
\textcite{LuDH:Supgsi} [Color].} \label{lu_disp}
\end{figure}
Fig.\ \ref{lu_edc} presents the energy distribution curves from
Y123 along the high symmetry directions in the Brillouin zone, for
polarization perpendicular (Fig.\ \ref{lu_edc}a) and parallel
(Fig.\,\ref{lu_edc}b) to the CuO chains direction.
Fig.\,\ref{lu_disp}a-f present the intensity plots of the second
derivative with respect to the binding energy of the ARPES
spectra, which allow one to follow the dispersion of the broader
and weaker electronic bands. \textcite{LuDH:Supgsi} identified the
four dispersive features summarized in Fig.\,\ref{lu_disp}g:
surface state (SS), superconducting peak (SC), hump (HP), and
chain state (CH). As one can see in Fig.\,\ref{lu_edc}, the
surface state at $X$ and $Y$ is the most pronounced feature in the
spectra. It has to be emphasized that while the surface state
nature of this peak has received strong support from its high
sensitivity to surface degradation at elevated temperatures
\cite{LuDH:Supgsi}, the precise origin of this state is not well
understood at present: on the one hand, the scenario of a
chain-related surface state proposed by
\textcite{SchabelMC:AngpuY} appears to be consistent with recent
scanning tunnelling microscopy data \cite{DerroDJ:Nano-d}; on the
other hand, this assignment is difficult to reconcile with the 2D
dispersion of this feature along $\Gamma$-X and $\Gamma$-Y (blue
symbols in Fig.\,\ref{lu_disp}g). The broad feature at 400 meV
below $E_F$ at the X point (yellow symbols in
Fig.\,\ref{lu_disp}g) was assigned by \textcite{LuDH:Supgsi} to a
bulk chain-derived state, rather than to a bonding $\sigma$ state
as previously suggested \cite{SchabelMC:AngpuY}. This
interpretation is based on the quasi 1D character of this feature
(i.e, much larger dispersion along the chain direction than
perpendicular to it), and the good agreement with band structure
calculations \cite{ANDERSENOK:LDAELH}. Two more features are
clearly detected at the $X$ point and are marked with red and
green symbols in Fig.\,\ref{lu_edc}a and\,\ref{lu_disp}g. These
are related to the electronic structure of the CuO$_2$ planes and
will be further discussed in Sec.\,\ref{sec:ybcosuper}.

In summary, the ARPES results from Y123 reveal bulk bands
originating from the CuO chains and the CuO$_2$ planes, and a
surface state band. As shown in Fig.\,\ref{lu_disp}h-j, the FS
estimated from the near-$E_F$ integrated intensity is consistent
with the band-structure results
\cite{ANDERSENOK:ELEPTI,ANDERSENOK:PLADSB}: a large holelike FS
centered at $S$ is observed, along with a 1D FS related to the
bulk CuO chains (obviously this can only be an approximate
comparison as the experimental data were taken in the
superconducting state and the intensity at $X$ and $Y$ is
significantly affected by the presence of the surface state). This
results in two FS crossings along $\Gamma$-$S$, one being due to
the chain and the other to the plane electronic bands. Lastly, it
has to be stressed that because of the reassignment of the 400 meV
dispersive peak to a bulk chain state rather than a CuO$_2$ plane
state, at present there is no consensus for bilayer splitting in
Y123 contrary to what had been previously suggested by
\textcite{SchabelMC:AngpuY}. Also, no evidence of shadow bands is
seen in Y123, which possibly supports a structural rather than a
magnetic origin for these additional features in Bi2212.

\subsection{Nd$_{2-x}$Ce$_x$CuO$_4$}
\label{sec:normalNCCO}

After having discussed the ARPES results from many hole-doped (or
p-type) systems, in this final section of our overview of the
doping evolution of the normal-state electronic structure of the
cuprates we will focus on the electron-doped (or n-type)
superconductor Nd$_{2-x}$Ce$_x$CuO$_4$ (NCCO). As shown in
Fig.\,\ref{rmp_pd}a, the undoped material is an AF insulator. Upon
the substitution of Ce for Nd, the AF ordering temperature drops
precipitously approaching $x\!=\!0.13$, and superconductivity
occurs between  $0.14\!<\!x\!<\!0.18$ for oxygen reduced samples.
Note that there is only an approximate symmetry in the
temperature/doping phase diagram about the zero doping line
between p-type and n-type systems, as the AF phase is much more
robust in the electron doped material and persists to much higher
doping levels. At the same time, superconductivity occurs in a
much narrower doping range. The relevance of a comparative study
of n and p-type cuprate HTSCs is that of a verification of the
symmetry, or lack thereof, between doping the AF insulator with
electrons or holes. This issue has important theoretical
implications as most models which are thought to capture the
essence of high-$T_c$ superconductivity implicitly assume
electron-hole symmetry.

After the early investigations by \textcite{KingDM:Ferses} and
\textcite{AndersonRO:LutFsm}, high energy and momentum resolution
ARPES data on superconducting NCCO ($x\!=\!0.15$, $T_c\!=\!22$ K)
were reported only very recently
\cite{ArmitageNP:ElesN1,ArmitageNP:SupgaN,ArmitageNP:Anoesp,armitage:1,SatoT:Obsdx}.
\begin{figure}[t!]
\centerline{\epsfig{figure=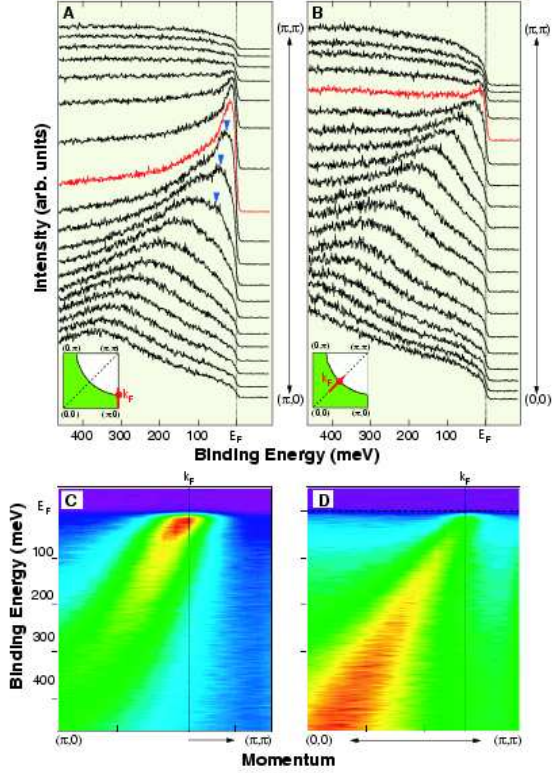,width=8cm,clip=}}
\vspace{0cm} \caption{(a-b) Nd$_{1.85}$Ce$_{0.15}$CuO$_{4}$ ARPES
spectra measured at 10 K with 21.2 eV photons (HeI), and (c-d)
corresponding image plots. After \textcite{SatoT:Obsdx}
[Color].}\label{Sato_EDC}\end{figure}
Fig.\,\ref{Sato_EDC} presents the ARPES spectra (a,b) and the
corresponding image plots (b,c) reported by
\textcite{SatoT:Obsdx}. The results are analogous to those
obtained by \textcite{ArmitageNP:Anoesp}. Along the
(0,0)-($\pi$,$\pi$) direction, NCCO shows a dispersion which is
ubiquitous among the cuprates, with a quasiparticle peak
dispersing quickly towards $E_F$ and crossing it at about
($\pi$/2,$\pi$/2). Similarly, a peak disperses and crosses $E_F$
also along ($\pi$,0)-($\pi$,$\pi$), in agreement with the expected
holelike FS (see insets of Fig.\,\ref{Sato_EDC}a-b). However,
while in the p-type cuprates at ($\pi$,0) the quasiparticle peak
forms a flat band just below $E_F$, in NCCO the flat band is
located at 300 meV below $E_F$.

\begin{figure}[b!]
\centerline{\epsfig{figure=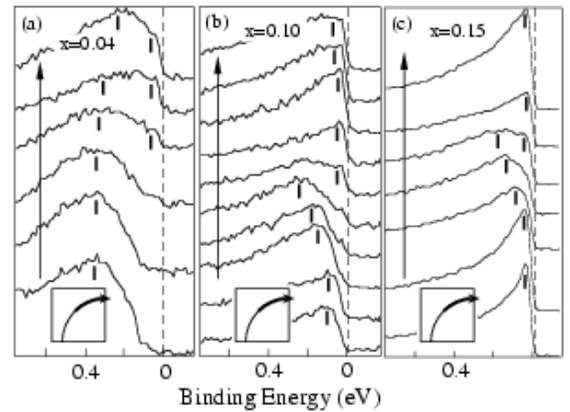,width=8cm,clip=}}
\caption{(a-c) ARPES spectra measured with 16.5\,eV photons on
4\%, 10\%, and 15\% doped NCCO, respectively, from along the
putative band-structure FS
\cite{armitage:1}}\label{Peter_EDC}\end{figure}

\begin{figure}[t!]
\centerline{\epsfig{figure=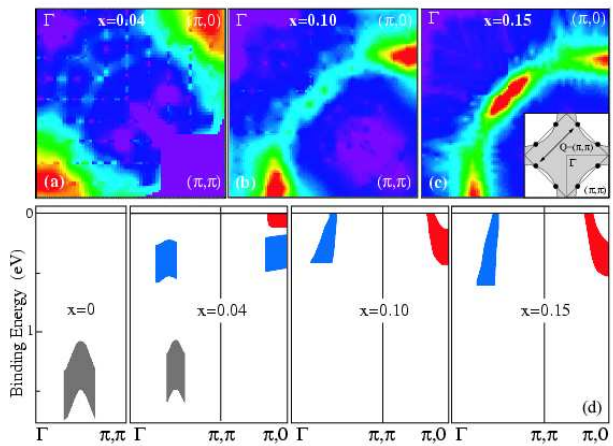,width=1\linewidth,clip=}}
\caption{(a-c) Momentum space maps of the low-lying spectral
weight integrated between $-40$ and 20\,meV, for $x\!=\!0.04$,
0.10, and 0.15. Inset of panel (c): regions of the FS that can be
coupled by ($\pi$,$\pi$) scattering. (d) Energy dispersion of the
detected features. After \textcite{armitage:1}
[Color].}\label{Peter_FST}\end{figure}

Studies of the doping evolution of the electronic structure were
recently carried out by \textcite{armitage:1}. As a starting
point, these authors have identified a charge-transfer band
feature in the undoped insulator Nd$_2$CeO$_4$, which has the same
dispersion as that found in the parent compounds of the p-type
HTSCs (such as Ca$_{2}$CuO$_2$Cl$_2$; Sec.\,\ref{sec:ns_ccoc}).
This indicates that the electronic structure of the CuO$_2$ plane
in these two compounds is the same, although one can be doped with
electrons and the other with holes. More interestingly, the
charge-transfer band in Nd$_2$CeO$_4$ is at about 1.2\,eV below
$E_F$, which is comparable to the charge-transfer gap energy if
the chemical potential is pinned close to the conduction band
minimum. Data for different dopings are presented in
Fig.\,\ref{Peter_EDC}, which shows the ARPES spectra from along
the FS contour expected from band structure calculations. At low
doping (4\%), the spectra are gapped near ($\pi$/2,$\pi$/2) over
an energy scale of about 200\,meV and exhibit a broad maximum
around 300\,meV (Fig.\,\ref{Peter_EDC}a). Moving towards
($\pi$,0), the 300\,meV feature remains more or less at the same
energy, and additional spectral weight develops near $E_F$. Upon
increasing the doping level, the ARPES features become sharper and
the one at ($\pi$/2,$\pi$/2) moves closer to $E_F$. At optimal
doping, (15\%) the spectra are characterized by a sharp
quasiparticle peak at $E_F$ both at ($\pi$,0) and
($\pi$/2,$\pi$/2), but not in the region in between where the
lineshape is still very broad (Fig.\,\ref{Peter_EDC}c). The
momentum dependence of the $E_F$ spectral weight (integrated from
$-\!40$ to $+\!20$\,meV) is displayed for the different doping
levels in Fig.\,\ref{Peter_FST}.
At 4\% the low-energy weight is concentrated in a small pocket
around ($\pi$,0), with a volume approximately consistent with the
nominal doping level $x$. Upon increasing doping, one can observe
both the modification of the pockets and the emergence of
low-lying spectral weight around ($\pi$/2,$\pi$/2); eventually
(15\%), the $E_F$ intensity evolves into a large holelike FS
centered at ($\pi$,$\pi$) with a volume given by $(1\!+\!x)$. In
the lower panels of Fig.\,\ref{Peter_FST} one can follow the
evolution of the overall electronic structure across the
(charge-transfer) insulator to metal transition. In the insulator
the valence band maximum is found at ($\pi$/2,$\pi$/2). At 4\%
doping, spectral weight develops inside the gap (blue) in addition
to the low-energy feature (red) which gives rise to the ($\pi$,0)
pockets discussed above; one can still observed some remanent of
the charge-transfer band of the insulator. At higher doping values
the latter disappears and the incoherent spectral weight evolves
into dispersive quasiparticle bands which define the large
holelike FS centered at ($\pi$,$\pi$).

It should be emphasized that although 15\% doped NCCO appears to
have a large FS with a volume given by $(1\!+\!x)$ counting
electrons, in agreement with Luttinger's theorem
\cite{luttinger:1}, the data still present some unexpected
features. What catches one's eye in Fig.\,\ref{Peter_FST}c are the
distinct regions of reduced intensity along the FS contour near
(0.65$\pi$,0.3$\pi$) and (0.3$\pi$,0.65$\pi$), in which the ARPES
spectra are extremely broad even at $E_F$
(Fig.\,\ref{Peter_EDC}c). This behavior appears to be an intrinsic
property of the spectral function of NCCO, as evidenced by the
comparison of data obtained for different photon energies and
experimental geometries \cite{ArmitageNP:Anoesp}.
Although some effects due to the energy and polarization
dependence of the photoemission matrix elements are present, the
overall systematics does not change. As noted by
\textcite{ArmitageNP:Anoesp}, the regions of suppressed weight
correspond to the intersections between the Luttinger FS and the
AF Brillouin zone boundaries (see inset of
Fig.\,\ref{Peter_FST}c). These regions, connected by a
$Q\!=\!(\pi,\pi)$ scattering vector, are referred to in the
literature as {\it hot spots}
\cite{PinesD:NeaaFl,SchmalianJ:Weapbt,KampfA:Pses-b,KampfAP:Spefps}.
The $(\pi,\pi)$ scattering experienced by the charge carriers has
been proposed to be the origin of the pseudogap observed in the
underdoped p-type HTSCs near ($\pi$,0). What is seen in NCCO is
reminiscent of that behavior (Sec.\,\ref{sec:pseudogap}), although
in the present case the effects are detected at
(0.65$\pi$,0.3$\pi$). This could merely be a consequence of the
different FS volume and, in turn, of the intersections between FS
and AF zone being pushed away from ($\pi$,0) for NCCO. In this
view, magnetic fluctuations would be a natural source of
$(\pi,\pi)$ scattering. More generally, however, any {\it umklapp}
scattering would result in a breakdown of FL theory and in a
segmented FS, as in Fig.\,\ref{models}b
\cite{FurukawaN:Trut-d,FurukawaN:InsL-F,HonerkampC:BretL-}.

It should be stressed that the doping dependence of the ARPES data
from NCCO suggests the absence of particle-hole symmetry in the
quasiparticle dispersion of the cuprates. In fact, the lowest
energy states in p-type materials are located at
($\pi$/2,$\pi$/2), while they appear at ($\pi$,0) in the n-type
systems. This constitutes the first direct evidence for the Mott
gap not being a direct gap. This is consistent with calculations
performed within the $t$-$t'$-$t''$-$J$ model (successfully used
to describe the undoped insulator data; Sec.\,\ref{sec:ns_ccoc}),
in which the breaking of particle-hole symmetry results from the
inclusion of the $t'$ term. In particular, the asymmetry seen by
ARPES between undoped and 4\% doped NCCO is well reproduced by a
mean field solution of the $t$-$t'$-$t''$-$U$ model
\cite{kusko:1}. Of course, the dynamic aspects of the data (e.g.,
the development of spectral weight inside the gap) cannot be
obtained within a mean-field approach; rather, the transfer of
spectral weight upon doping is generally found in cluster model
calculations.

\subsection{Discussion}
\label{sec_ns}

The most complete ARPES investigations of the doping evolution of
the normal state electronic structure have been performed on LSCO,
which can be studied over a wide doping range. The results from
this system have been extensively interpreted in the stripe
scenario, which provides a possible explanation for many of the
experimentally observed features: (i) the two-component electronic
structure seen in the very underdoped regime that is suggestive of
the creation of new electronic states inside the Mott gap; (ii)
the lack of chemical potential shift in the underdoped regime;
(iii) the straight FS segments observed under certain experimental
geometries that are indicative of 1D electronic behavior. However,
there are aspects of the data that are difficult to reconcile with
a naive stripe picture. One such aspect is the Fermi-arc  feature
observed in ARPES experiments at very low doping levels, defined
by the lowest-energy feature of the two-component electronic
structure \cite{yoshida:2}. Furthermore, as doping is increased,
LSCO becomes more of a band-like system with an LDA-like FS, and
the signatures of stripes weaken; near optimal doping, the
distinction between the pictures reported in Fig.\,\ref{models}a
and\,\ref{models}d would become blurred.

For the Bi-based cuprates, on the other hand, there are no
extensive reliable data in the underdoped regime and, in
particular, near the metal-insulator transition boundary. To date,
those features which have been discussed as possible signatures of
a charge ordered state in extremely underdoped LSCO have not been
observed in other systems. In the case of Bi2212, long straight
segments of FS have been observed for the bonding band along the
direction (0,0)-($\pi$,0). This was already noted in the early low
resolution data \cite{FengDL:1}, but it is more clearly displayed
by the recent data taken with improved momentum resolution
\cite{kordyuk:2}. The presence of nested FS segments could give
rise to a charge-density wave instability.  This scenario is
however different from the stripe interpretation of somewhat
similar features observed in LSCO. In the latter case the straight
FS segments would be the result and not the cause of the formation
of quasi-1D stripes. Furthermore, it has to be emphasized that the
persistence of the partial FS nesting well into the overdoped
regime in Bi2212 makes a connection with a simple stripe
instability less realistic \cite{kordyuk:2}. On the other hand,
why the FS nesting does not lead to a charge-density wave
instability is an interesting problem. In this regard, the
presence of the antibonding sheet of FS might have an important
role, which may be worthwhile studying.

At this stage of the research on the HTSCs and their undoped
parent compounds, it does not seem possible to firmly conclude in
favor of one particular comprehensive theoretical model, in spite
of the considerable progress made in recent years. This situation
is emphasized by the longstanding puzzle concerning a fundamental
question, i.e., how does the doping of a Mott insulator take place
(Fig.\,\ref{mott}). On the one hand, contrary to the early report
by \textcite{ALLENJW:RESPNN}, very recent angle-integrated
photoemission results from NCCO \cite{steeneken:1,harima:1} favor
a scenario based on the shift of the chemical potential, which is
also consistent with the ARPES data from NCCO. Actually, this
would be in agreement with $t$-$t'$-$t''$-$J$ model calculations,
which reproduce the substantial deformation of the quasiparticle
band structure upon doping, and suggest a unifying point of view
for both the undoped insulator and the HTSC
\cite{EderR:Dop-de,kusko:1}. The chemical-potential shift scenario
is supported also by the data available on Na-CCOC, which show a
quasiparticle dispersion strikingly similar to the one of undoped
CCOC. On the other hand, the lack of chemical potential shift
observed in LSCO in the underdoped regime and the detection of
multiple electronic components support the formation of in-gap
states upon doping the systems and, consequently, the need for a
completely new approach.  Further scrutiny is required to
establish if the evolution from Mott insulator to HTSC is truly
accounted for by one of the existing models or whether a different
approach, maybe beyond a purely electronic description, is
required (e.g., in which other factors, such as the underlying
structural distortions, are explicitly included).

\section{Superconducting gap}
\label{sec_supgap}
\begin{figure}[t!]
\centerline{\epsfig{figure=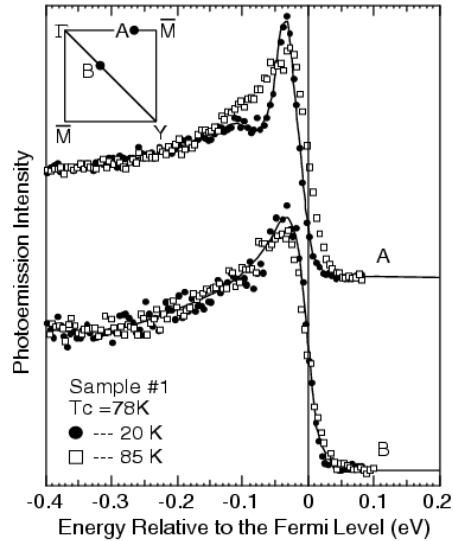,width=6.5cm,clip=}}
\vspace{0cm} \caption{Temperature dependent ARPES spectra from
Bi2212 ($T_c\!=\!88$\,K) measured in the nodal region (B) and
close to the ($\pi$,0) point (A), as sketched in the inset
\cite{ShenZ-X:Anolga}.}\label{shen1}\end{figure}

The ability of ARPES to detect spectral changes across the
superconducting phase transition is a remarkable testimony to the
improvement in resolution over recent years, and is the key to the
success of this technique in the study of the cuprate HTSCs. The
most important results obtained in the superconducting state are
the detection of: (i) the anisotropic  $d$-wave gap along the
normal-state FS, which contributed to the debate on the pairing
mechanism [for a recent review on the pairing symmetry in cuprate
HTSCs see \textcite{TsueiCC:Paiscs}]; (ii) the dramatic changes in
the spectral lineshape near ($\pi$,0). In the current section, we
will review point (i) for several systems while we will come back
to (ii) later, within the discussion of the superconducting peak
(Sec.\,\ref{sec:speak_Bi2212}) and of the self-energy corrections
(Sec.\,\ref{sec:selfmodes}).

\subsection{Bi$_2$Sr$_2$CaCu$_2$O$_{8+\delta}$}
\label{sec:supgapBi2212}

Fig.\,\ref{shen1} shows the early ARPES data from an overdoped
Bi2212 sample at two different momenta in the Brillouin zone
\cite{ShenZ-X:Anolga}. In the nodal region (B), the spectra taken
above and below $T_c$ are very similar indicating a small or
vanishing gap. Near the ($\pi$,0) point (A), on the other hand,
normal and superconducting-state spectra are clearly very
different: in addition to the obvious lineshape evolution, note
also the shift of the leading edge, which reflects the opening of
a sizable energy gap. These results strongly suggest that the
superconducting gap is anisotropic and, in particular, consistent
with a $d$-wave order parameter \cite{ScalapinoDJ:casdx}.
\begin{figure}[t!]
\centerline{\epsfig{figure=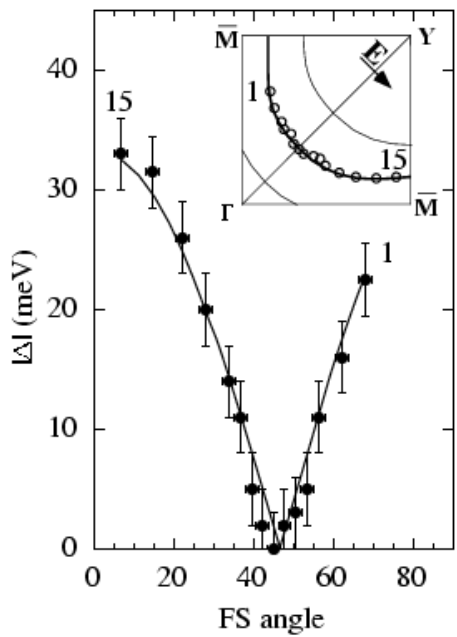,width=6.1cm,clip=}}
\vspace{0cm} \caption{Superconducting gap measured at 13\,K on
Bi2212 ($T_c\!=\!87$\,K) plotted versus the angle along the
normal-state FS (see sketch), together with a $d$-wave fit
\cite{DingH:Ang-re}.}\label{sgap2}\end{figure}
Together with the microwave penetration depth results
\cite{HardyWN:Premtt}, this direct evidence for the gap anisotropy
played a major role in the early debate on the pairing symmetry
\cite{levi:1}.

Initially the magnitude of the gap was quantified simply on the
basis of the position of the leading-edge midpoint of the ARPES
spectra, which has since become a standard procedure
\cite{tinkham:1}. In particular, one could either follow the
leading-edge shift of the spectra measured on the superconducting
material above and below $T_c$, or compare at the same temperature
below $T_c$ the position of leading edges for the superconductor
and a polycrystalline noble metal like Pt or Au (a caveat here is
that the comparison between the nontrivial lineshape measured on a
HTSC single crystal and the structureless Fermi edge from a
polycrystalline metal may be quantitatively inaccurate).
Subsequently, some effort has been invested in trying to perform a
more detailed analysis of the superconducting gap. In principle,
the gap $\Delta_k$ could be estimated by measuring the
quasiparticle dispersion in the vicinity of the Fermi level which,
within the BCS framework \cite{schrieffer:1}, is predicted to
change from $\epsilon_k$ to
$E_k\!=\!\sqrt{\epsilon_k^2\!+\!\Delta_k^2}$ upon entering the
superconducting state. However, the determination of $\Delta_k$ on
the basis of the change in the dispersion is a very difficult task
because the cuprates are not conventional FL metals in the normal
state and, as a consequence, $\epsilon_k$ cannot be determined
with sufficient accuracy. Furthermore, this analysis is also
complicated by the dramatic change in quasiparticle lineshape that
takes place across the phase transition (Fig.\,\ref{shen1}).
Nonetheless, by fitting energy distribution curves at different
momenta with a phenomenologically broadened BCS spectral function,
\textcite{DingH:Momdts,DingH:1,DingH:Ang-re} managed to obtain the
momentum dependence of the gap along the normal state FS
(Fig.\,\ref{sgap2}). The results agree with the $d_{x^2-y^2}$
functional form $\Delta({\bf
k})\!=\!\Delta_0[\cos(k_xa)\!-\!\cos(k_ya)]$ extremely well.

It has to be emphasized that such a good fit to the $d$-wave gap
functional form as the one of Fig.\,\ref{sgap2} is the exception
rather than the norm. In most cases, and especially in the
underdoped regime, instead of the ``V''-like shape cusp seen in
the data of Fig.\,\ref{sgap2}, one finds an extended area around
the nodal region characterized by gapless
excitations.\footnote{\textcite{DessauDS:Keyftm,DingH:Ang-re,MesotJ:Supgaq,HarrisJM:Anossg,NormanMR:DestFs}.}
In this regard, it is important to mention that also recent high
momentum resolution results from underdoped Pb-Bi2212
($T_c\!=\!77$\,K), for which the bilayer splitting could be
resolved making the analysis more reliable, indicated a ``U''
rather than a ``V''-like shape for the superconducting gap along
the bonding FS \cite{BorisenkoSV:1}.\footnote{Note that although
an identical gap was detected for both bilayer split FSs in the
antinodal region, the momentum dependence along the antibonding
one could not be investigated in detail.} In order to account for
this behavior \textcite{MesotJ:Supgaq} suggested that higher
harmonics consistent with $d$-wave symmetry, such as
$[\cos(2k_xa)\!-\!\cos(2k_ya)]$, should be included in the
expansion of the gap function $\Delta({\bf k})$, in addition to
the simple ``V''-like $[\cos(k_xa)\!-\!\cos(k_ya)]$ first term
\cite{Wenger:1}. Presently, however, this is still an open issue
that deserves further scrutiny for its significant theoretical
implications.

\subsection{Bi$_2$Sr$_2$CuO$_{6+\delta}$}

The angular dependence of the superconducting gap in Bi2201 was
first investigated by \textcite{HarrisJM:Meaaeg}. Similar to the
case of Bi2212, the results from Bi2201 appear to be consistent
with a $d$-wave pairing symmetry, albeit a larger region of
gapless excitations was observed close to the nodal direction. The
overall magnitude of the superconducting gap ($\sim\!10$\,meV at
optimal doping) is approximately a factor of three smaller than
for Bi2212 at similar doping levels, consistent with the reduction
in transition temperature. More recently, the superconducting gap
was investigated with improved resolution by
\textcite{SatoT:Ferssg}. By comparing the data obtained in the
nodal and antinodal regions, these authors also concluded in favor
of a $d$-wave symmetry order parameter with a maximum size of
10-15\,meV at ($\pi$,0).

\subsection{Bi$_2$Sr$_2$Ca$_2$Cu$_3$O$_{10+\delta}$}
\label{sec:Bi2223_supgap}

\begin{figure}[t!]
\centerline{\epsfig{figure=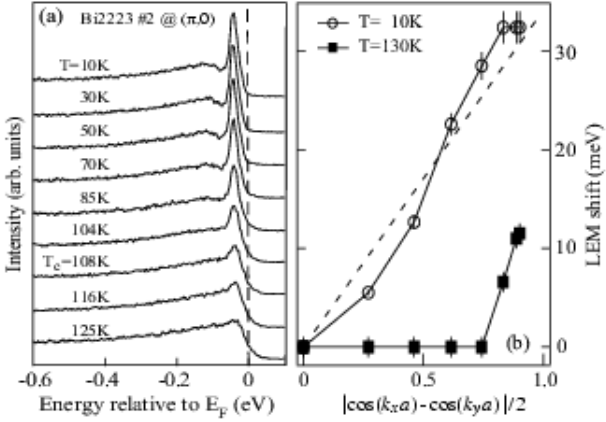,width=1\linewidth,clip=}}
\vspace{0cm} \caption{(a) Temperature dependent ($\pi$,0) ARPES
spectra measured on nearly optimally doped Bi2223
($T_c\!=\!108$\,K) with 21.2\,eV photons. (b) Position of the
leading-edge midpoint above and below $T_c$ along the normal state
FS. The dashed line represents the d-wave functional form
\cite{FengDL:2}.}\label{Bi2223_GAP}\end{figure}

The superconducting gap in Bi2223 was recently measured by several
groups \cite{FengDL:2,sato:1,muller:1}. In
Fig.\,\ref{Bi2223_GAP}a, where the temperature dependence of the
($\pi$,0) ARPES spectra is presented, one can clearly observe the
remarkable change of quasiparticle lineshape that takes place for
$T\!<\!T_c$, as well as the opening of the superconducting gap.
Fig.\,\ref{Bi2223_GAP}b shows the angular dependence of the
superconducting gap as determined from the shift of the
leading-edge midpoint along the normal-state FS.
The results are consistent with a $d$-wave symmetry order
parameter, with a magnitude which is the largest so far reported
among the Bi-based cuprate superconductors.

\subsection{La$_{2-x}$Sr$_x$CuO$_4$}

The superconducting gap in LSCO is much smaller than in Bi2212
and, so far, no detailed investigations have been reported. As we
already mentioned in Sec.\,\ref{sec:nslsco} when discussing the
data of Fig.\,\ref{Akihiro_EDC}, the presence of a finite
superconducting gap was evidenced by the shift of the leading-edge
midpoint of the spectra along the normal-state FS
\cite{InoA:Fersbd}. On the basis of this behavior
\textcite{InoA:Fersbd} estimated a superconducting gap for
optimally doped LSCO of about 10-15\,meV, in agreement with the
results from other techniques
\cite{ChenXK:Invtsg,YamadaK:Diromg,NakanoT:Corbdd}. Furthermore,
it appears that the measured gap is qualitatively consistent with
a $d$-wave scenario. Similar conclusions were reached by
\textcite{SatoT:PseodL} on the basis of a detailed fit of
angle-integrated photoemission data.

\subsection{YBa$_2$Cu$_3$O$_{7-\delta}$}
\label{sec:ybcosuper}

Information on the superconducting state of Y123 by ARPES can be
obtained only after the identification of the surface state
(Sec.\,\ref{sec:nsYBCO}), which dominates the photoemission
spectra at both $X$ and $Y$ points in the Brillouin zone (note
that it is at these momenta that a gap, if $d$-wave like, would
have its maximum amplitude). By investigating the $X$ region,
where the surface state was weaker in that specific experimental
geometry, \textcite{Schabel:1} provided supporting evidence for a
superconducting gap whose momentum dependence is consistent with
the $d$-wave form.
\begin{figure}[b!]
\centerline{\epsfig{figure=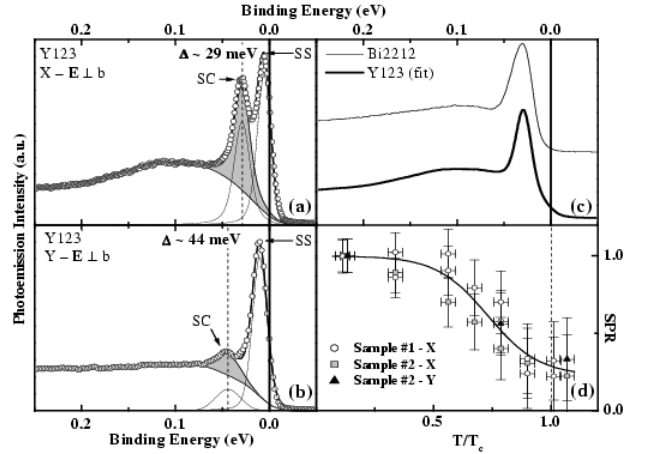,width=1\linewidth,clip=}}
\vspace{.3cm} \caption{(a,b) ARPES spectra from overdoped Y123
($T_c\!=\!89$\,K) measured at $X$ and $Y$ ($T\!=\!10$\,K), with
corresponding fitting curves. (c) Comparison of the Y123 fit at
$X$ (surface state peak subtracted) with the ($\pi,0$) spectrum
from overdoped Bi2212 ($T_c\!=\!84$\,K). (d) Temperature
dependence of the normalized superconducting-peak intensity
\cite{LuDH:Supgsi}.} \label{ybco_gap}
\end{figure}
In that study, however, due to sample quality issues the surface
was not stable enough to allow cycling the temperature during the
experiments.

Recently, the substantial improvement in sample quality and
instrumental resolution allowed the convincing detection of
superconductivity related features on Y123 \cite{LuDH:Supgsi}. As
shown in Fig.\,\ref{ybco_gap}a,b, after subtracting the surface
state contribution by means of a phenomenological fitting
procedure, the remainder of the fitted function looks strikingly
similar to the lineshape typically observed in Bi2212 at ($\pi,0$)
in the superconducting state (see Fig.\,\ref{ybco_gap}c and
Sec.\,\ref{sec_suppeak}). Direct evidence for the second peak
being related to superconductivity comes from measurements
performed by cycling the temperature, which show that this feature
consistently disappears above $T_c$. This behavior is summarized
for several overdoped samples ($T_c\!=\!89$\,K) and momenta in
Fig.\,\ref{ybco_gap}d, where the normalized superconducting peak
intensity is plotted (note that in Fig.\,\ref{ybco_gap}d SPR
refers to {\it superconducting peak ratio}, which will be defined
in Sec.\,\ref{sec:speak_Bi2212}).

An interesting finding, which is specific to Y123, is the strong
$a$-$b$ anisotropy of the in-plane electronic structure. In fact,
as determined by \textcite{LuDH:Supgsi}, the energy positions of
peak and hump for overdoped Y123 are: 29 ($\Delta_x$) and 120 meV
($\omega_x$) at X, and 44 ($\Delta_y$) and 180 meV ($\omega_y$) at
Y. These results, in agreement with those from other techniques
\cite{PolturakE:Meatat,LimonovMF:Sup-in}, indicate a 50\%
difference in the gap magnitude $\Delta$ between X and Y, which
represents a remarkable deviation from the ideal $d_{x^2-y^2}$
pairing symmetry.

\subsection{Nd$_{2-x}$Ce$_x$CuO$_4$}

The improved experimental resolution also allowed the successful
detection of the superconducting gap in the electron doped
superconductor NCCO \cite{ArmitageNP:SupgaN,SatoT:Obsdx}, a task
that was not possible before
\cite{KingDM:Ferses,AndersonRO:LutFsm}.
\begin{figure}[b!]
\centerline{\epsfig{figure=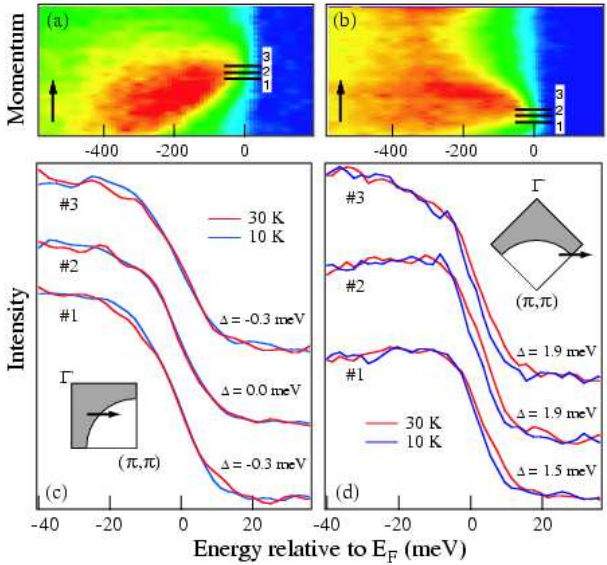,width=1\linewidth,clip=}}
\vspace{0cm} \caption{(a,b) Image plots of the ARPES spectra
measured on NCCO ($T_c\!=\!24$\,K) near ($\pi$/2,$\pi$/2) and
($\pi$/2,0.3$\pi$) along the cuts indicated by arrows in the
Brillouin-zone sketches of (c) and (d), respectively. (c,d) Normal
and superconducting-state spectra from the $E$-$k$ regions marked
by the black bars in (a) and (b), respectively. Gap values
obtained by fitting are also indicated. After
\textcite{ArmitageNP:SupgaN}
[Color].}\label{NCCOpet_GAP1}\end{figure}
This investigation is of particular relevance because, whereas for
the hole-doped HTSCs it is now generally accepted that the order
parameter has a majority component of $d$-wave symmetry [see,
e.g., \textcite{TsueiCC:Paiscs}], for the electron doped systems
this issue is less clear. On the one hand, an isotropic
superconducting gap was supported by early tunnelling
\cite{HuangQ:Tunepe}, microwave penetration depth
\cite{WangSH:EnegN1,WuDH:Temdpd}, and Raman experiments
\cite{StadloberG:IsN2-x}. On the other hand, this picture has been
seriously questioned by recent scanning SQUID microscopy
measurements on tricrystal films, which provided direct evidence
for an order parameter with a large $d$-wave component
\cite{TsueiCC:Pha-se}.

In order to investigate the superconducting gap by ARPES,
\textcite{ArmitageNP:Anoesp} focused on the FS crossings near
($\pi$/2,$\pi$/2) and ($\pi$,0.3$\pi$) in the Brillouin zone (see
Fig.\,\ref{NCCOpet_GAP1}a,b), where a $d_{x^2-y^2}$
superconducting gap is expected to be zero and maximum,
respectively. The blue and red curves in
Fig.\,\ref{NCCOpet_GAP1}c-d are data collected below and above
$T_c\!=\!24$\,K, respectively. Near the ($\pi$/2,$\pi$/2) region
(Fig.\,\ref{NCCOpet_GAP1}c), one does not see a shift of the
leading edge of the spectra with temperature, indicating minimal
or zero gap. Near ($\pi$,0.3$\pi$) on the other hand, all curves
show a clear shift below $T_c$ of about 1.5-2\,meV, in agreement
with the opening of a superconducting gap
(Fig.\,\ref{NCCOpet_GAP1}d). Systematic estimates for the
leading-edge midpoint shift were obtained by means of a
phenomenological fitting procedure and are indicated, for the
different spectra, in Fig.\,\ref{NCCOpet_GAP1}c-d.
\textcite{ArmitageNP:SupgaN} concluded that, although an
anisotropic $s$-wave order parameter with a small amplitude in the
nodal region cannot be completely excluded, these results provide
support for a $d$-wave pairing state in NCCO. Furthermore, as the
leading-edge analysis can underestimate the superconducting gap
value by as much as a factor of 2
\cite{DingH:Momdts,LoeserAG:Temddt}, a gap of about 4\,meV would
be consistent with the ARPES data, in agreement with other
techniques
\cite{HuangQ:Tunepe,WangSH:EnegN1,WuDH:Temdpd,StadloberG:IsN2-x,KokalesJD:Micee-,ProzorovR:Evinqe}.

Similar results were also reported by \textcite{SatoT:Obsdx}. In
this case, the authors did not perform temperature dependent
measurements on NCCO across the phase transition but relied on the
comparison, at one fixed temperature below $T_c$, between the
Fermi edges measured on polycrystalline Pt and on a single-crystal
of NCCO. \textcite{SatoT:Obsdx} detected for NCCO a leading edge
shift of about 2-3 meV at the Fermi crossing along
($\pi$,0)-($\pi$,$\pi$) and no shift at the Fermi crossing along
(0,0)-($\pi$,$\pi$). Furthermore, by fitting the data with a
phenomenologically broadened BCS spectral function
\cite{DingH:Momdts,DingH:1,DingH:Ang-re}, they concluded in favor
of a $d$-wave gap of about 5 meV, in good agreement with the
estimate by \textcite{ArmitageNP:SupgaN}.

\subsection{Discussion}
\label{sec:sgdiscussion}

\begin{figure}[t!]
\centerline{\epsfig{figure=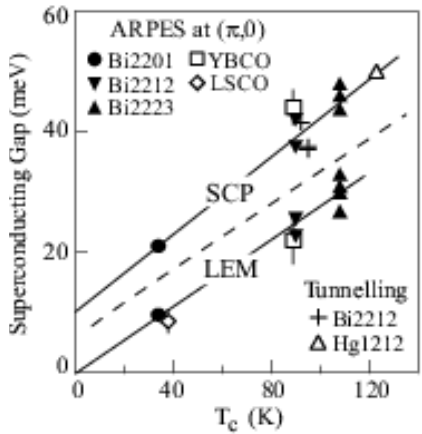,width=6cm,clip=}}
\vspace{0cm} \caption{Superconducting gap magnitude estimated from
the superconducting peak position and the leading-edge midpoint
shift (separated by the dashed line), plotted versus $T_c$ for
various optimally doped materials
\cite{FengDL:2}.}\label{Bi2223_GAPtot}\end{figure}

From the results presented in this section it is clear that the
cuprate HTSCs are characterized by an overall $d$-wave pairing
symmetry, although further scrutiny is required for specific
issues such as the existence of an area of gapless excitations in
the nodal region, which would hint to significant deviations from
a pure $d_{x^2-y^2}$ symmetry. Furthermore, as summarized in
Fig.\,\ref{Bi2223_GAPtot} where ARPES as well as tunnelling
spectroscopy results are presented for several families of
cuprates, at optimal doping the gap magnitude of the different
systems scales linearly with the corresponding $T_c$
\cite{FengDL:2,SatoT:Ferssg,sato:1}. In particular, the gap versus
$T_c$ data obtained from the position of the leading-edge midpoint
can be fitted by a line across the origin corresponding to the
ratio $2\Delta_0/k_BT_c\!\simeq\!5.5$ \cite{FengDL:2}.

At this stage, a system that seems to stand out is Y123, as
evidenced by the 50\% anisotropy between the gap amplitude at $X$
and $Y$. As discussed by \textcite{LuDH:Supgsi}, this anisotropy
could originate from the orthorhombicity of the CuO$_2$ planes in
Y123 which, according to band structure calculations, should
affect the normal state electronic structure
\cite{ANDERSENOK:LDAELH}.
Alternatively, it could be a consequence of the presence of the
CuO chains which, for strong chain-plane hybridization, can
directly affect the CuO$_2$-plane-derived electronic structure
\cite{AtkinsonWA:DiscsY}. Anyway it seems that for Y123, rather
than a pure $d_{x^2-y^2}$ symmetry, alternative models should be
considered \cite{AtkinsonWA:DiscsY,wu:1}.

Although not detected by ARPES, significant deviations from a
simple $d_{x^2-y^2}$ symmetry may be present in the case of NCCO.
In fact, a non monotonic $d$-wave superconducting order parameter
has been recently proposed on the basis of Raman experiments
\cite{BlumbergG:Nondso}. These results, whose interpretation
however was recently questioned \cite{Venturini:1}, were taken as
an indication that the absolute maximum in gap magnitude may be
located at the intersection between the holelike FS and the AF
zone boundary (the so called hot spots indicated in the inset of
Fig.\,\ref{Peter_FST}c), rather than at ($\pi$,0) as in the p-type
cuprate superconductors. In this regard, as for the possible
deviations from a ``V''-like cusp in the nodal region for the
p-type materials \cite{MesotJ:Supgaq}, it was proposed that high
order harmonics should be included in the expansion of the gap
function \cite{Guinea:1}. This would account for a doping
evolution of the gap anisotropy and, in particular, for the
maximum gap amplitude being pushed away from ($\pi$,0) in
optimally doped NCCO. However, in order to conclusively address
these issues, more detailed ARPES studies of the momentum
dependence of the superconducting gap in both n and p-type HTSCs
are required.

\section{Superconducting peak}
\label{sec_suppeak}

\begin{figure}[b!]
\centerline{\epsfig{figure=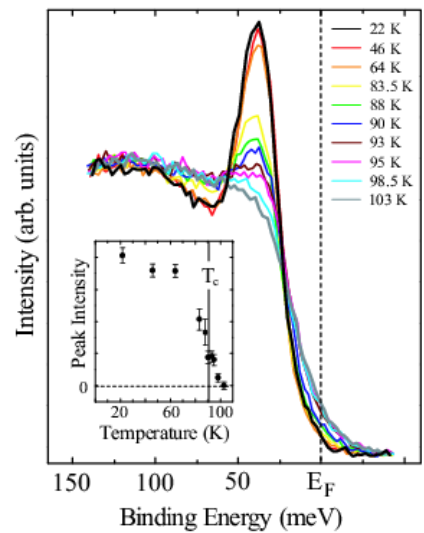,width=6cm,clip=}}
\caption{Temperature dependent photoemission spectra from
optimally doped Bi2212 ($T_c\!=\!91$\,K), angle integrated over a
narrow cut at ($\pi$,0). Inset: superconducting peak intensity
versus temperature. After \textcite{FedorovAV:Temdps}
[Color].}\label{fedorov1}\end{figure}

We now move on to the second well-known phenomenon observed in the
ARPES data below $T_c$, namely the dramatic change in lineshape of
the ($\pi$,0) spectra, first seen in Bi2212
\cite{DessauDS:Anoswt,HwuY:Elesthh}: a sharp quasiparticle peak
develops at the lowest binding energies followed by a dip and a
broader hump, giving rise to the so called {\it peak-dip-hump}
structure. This evolution can be clearly seen in
Fig.\,\ref{fedorov1}, or in Fig.\,\ref{shen1} where the normal and
superconducting state spectra from ($\pi$,0) and ($\pi$/2,$\pi$/2)
are compared. While the opening of the superconducting gap has
been observed in many different cuprate HTSCs, for a long time the
emergence of a sharp peak below $T_c$ appeared to be a phenomenon
unique to Bi2212.
Only very recently, a similar feature has been detected below
$T_c$ also on Y123 and Bi2223 (see Sec.\,\ref{sec:ybcosuper}
and\,\ref{sec:speakBi2223}, respectively). However, as it is on
Bi2212 that the most extensive body of work for doping,
temperature, and momentum dependence of this feature is available,
in this section we will mainly focus on that material.

Because the emergence of the peak-dip-hump structure in the Bi2212
($\pi$,0) spectra is the most remarkable effect seen across $T_c$,
a lot of work has been devoted to the experimental and theoretical
investigation of this behavior. However, as we will discuss in
more detail below and in Sec.\,\ref{sec:selfmodes}, the lineshape
analysis of the ARPES spectra from Bi2212 is a very controversial
subject and is currently matter of an intense debate. In
particular, the recent observation of bilayer band splitting in
Bi2212 (Sec.\,\ref{sec:bilayer}) suggests that the peak-dip-hump
structure near ($\pi$,0) needs to be completely re-explored. It is
indeed possible that its main features could be interpretable as
bilayer splitting effects. Nonetheless, due to the present
uncertainty and the inherent interest of this issue, in this
section we will review previous analyses, focusing mainly on the
experimental phenomenology.

\subsection{Bi$_2$Sr$_2$CaCu$_2$O$_{8+\delta}$}
\label{sec:speak_Bi2212}

The connection between the emergence of the peak-dip-hump
structure and the onset of superconductivity is evidenced by
detailed temperature dependent
investigations.\footnote{\textcite{LoeserAG:Temddt,FranzM:Phafsp,FedorovAV:Temdps,millis:1,FengDL:Sigsdt,DingH:Cohqwi}.}
Although it starts to manifest itself slightly above $T_c$, it is
below $T_c$ that the quasiparticle peak really stands out. This
can be clearly observed in the high-resolution (8\,meV) data from
optimally doped Bi2212 presented in Fig.\,\ref{fedorov1}, which
indicate an intrinsic width for this peak of about 14\,meV
\cite{FedorovAV:Temdps}.
\begin{figure}[t!]
\centerline{\epsfig{figure=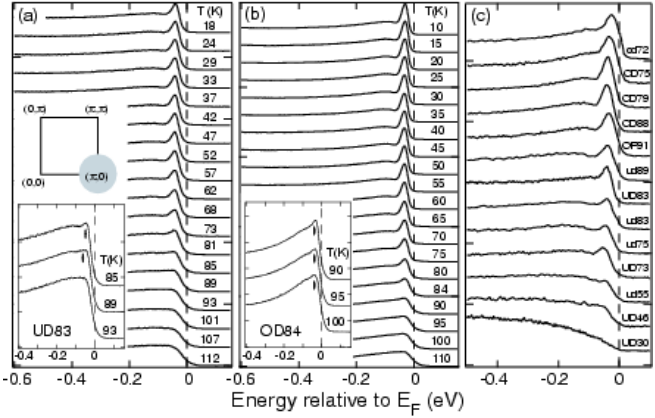,width=1\linewidth,clip=}}
\caption{(a,b) $T$-dependent photoemission spectra from under
($T_c\!=\!83$ K) and overdoped ($T_c\!=\!84$ K) Bi2212,
respectively, angle-integrated over the area around ($\pi$,0)
shaded in grey in the Brillouin zone sketch. Insets: enlarged view
of spectra taken just above $T_c$. (c) Doping dependence, from
underdoped to overdoped (bottom to top), of the
superconducting-state ($\pi$,0) spectra from Bi2212, for
$T\!\ll\!T_c$ \cite{FengDL:Sigsdt}.}\label{DL_EDC}\end{figure}
Temperature dependent data from underdoped and overdoped Bi2212
are presented in Fig.\,\ref{DL_EDC}a,b \cite{FengDL:Sigsdt}. As
shown in particular in the insets of Fig.\,\ref{DL_EDC}, the
superconducting peak turns on slightly above $T_c$ in striking
contrast to the pseudogap which behaves very differently in under
and overdoped samples (e.g., it opens up well above $T_c$ on the
underdoped side, as discussed in Sec.\,\ref{sec:pseudogap}). The
results obtained at low temperatures ($\sim\!10$ K) for different
doping levels are displayed in Fig.\,\ref{DL_EDC}c. The peak, not
resolved in the very underdoped samples (bottom curve in
Fig.\,\ref{DL_EDC}c), grows with doping.

At the experimental level, the momentum dependence of the
peak-dip-hump structure is a more complicated and controversial
issue than the doping or temperature dependence. As we will see
below, this is mostly a consequence of the superstructure and
bilayer splitting effects (Sec.\,\ref{sec:ns_Bi2212}), which are
particularly pronounced in the ($\pi$,0) region of momentum space.
\begin{figure}[b!]
\centerline{\epsfig{figure=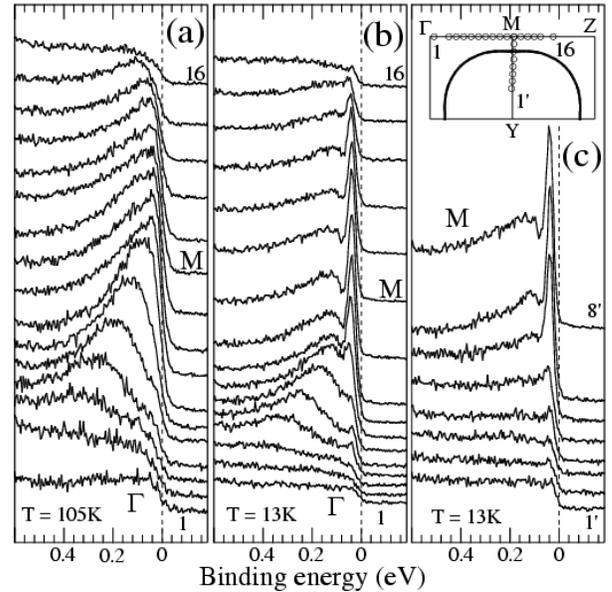,width=8cm,clip=}}
\caption{(a) Normal and (b,c) superconducting state ARPES spectra
from overdoped Bi2212 ($T_c\!=\!87$\,K), measured at the
$k$-points indicated in the inset of (c). Momentum-space notations
are defined as in Fig.\,\ref{bz_tot}. After
\textcite{NormanMR:Unudls}.}\label{norma2}\end{figure}
The other reason is that this investigation requires many spectra
to be recorded as a function of momentum and temperature, which is
technically more challenging especially if one is going after
small effects. For instance, \textcite{ShenZ-X:Tem-in} reported
2-5\% temperature induced spectral weight transfer with momentum
${\bf Q}\!\simeq\!(0.45\pi,0)$. This was observed in samples with
lower $T_c$, reflecting the presence of impurities, but not in
those characterized by the highest $T_c$; it was interpreted as a
consequence of the stripe formation. It was later found that this
effect may more likely be an experimental artifact due to the
aging of the sample surface \cite{Campuzano:1,white:1}, as
evidenced by ARPES experiments performed on samples containing
impurities whose surface was intentionally and systematically aged
\cite{white:1}.

\begin{figure}[t!]
\centerline{\epsfig{figure=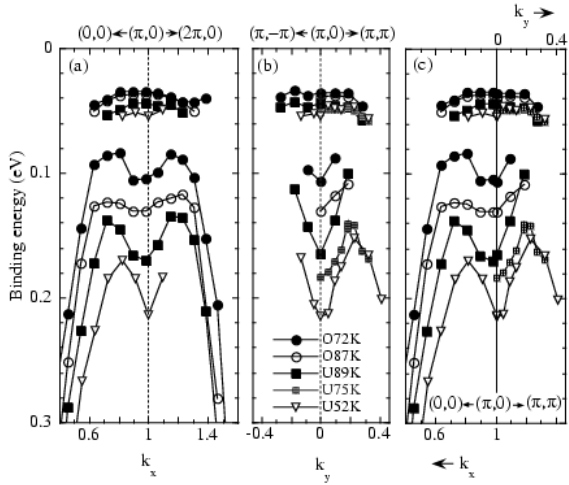,width=7.5cm,clip=}}
\caption{Peak and hump dispersion in the superconducting state of
Bi2212, for several dopings
\cite{CampuzanoJC:Elestr}.}\label{pg6}\end{figure}

Fig.\,\ref{norma2} presents ARPES spectra from Bi2212 along the
high symmetry lines \cite{NormanMR:Unudls}, which suggest that the
sharp superconducting peak persists in a very large momentum space
region around ($\pi$,0). However, as later recognized the
low-energy peak observed at those $k$-points, at which the hump
has dispersed to binding energies larger than 200\,meV, stems from
the superstructure contamination discussed in
Sec.\,\ref{sec:ns_Bi2212}. The actual experimental phenomenology
is that the superconducting peak occurs only in those momentum
space regions where the normal state band is within 100-150 meV of
$E_F$, and has a very weak dispersion. This can be seen in
Fig.\,\ref{pg6}, which presents the quasiparticle dispersion
reported for different doping levels by
\textcite{CampuzanoJC:Elestr}.

The detection of the peak-dip-hump structure has generated a
tremendous amount of interest as it is thought to carry critical
information about the superconducting transition.\footnote{See,
e.g.,
\textcite{HarrisJM:Anossg,LoeserAG:Temddt,NormanMR:Unudls,ShenZ-X:Momtdd,ShenZ-X:Tem-in,FranzM:Phafsp,ChubukovAV:Spefsc,FedorovAV:Temdps,CampuzanoJC:Elestr,FengDL:Sigsdt,DingH:Cohqwi,EschrigM:Neurmp,KaminskiA:Rensls}.}
In particular, it gave the main impetus for a phenomenological
description of the single-particle excitations in terms of an
interaction between quasiparticles and collective modes, which is
of fundamental relevance to the nature of superconductivity and of
the pairing mechanism in the HTSCs [see, e.g.,
\textcite{EschrigM:Review}]. However, as we will discuss in more
detail in Sec.\,\ref{sec:selfmodes}, different interpretative
scenarios have been proposed, which are still matter of intense
debate.
In this regard, it is important to point out that the main
constituents of the peak-dip-hump structure have been interpreted
also as possible fingerprints of the bilayer splitting of the
electronic structure of Bi2212, with peak and hump corresponding
to antibonding and bonding bands respectively. As shown in
Sec.\,\ref{sec:bilayer}, this assignment was particularly obvious
for both normal and superconducting state data from overdoped
Bi2212 (see Fig.\,\ref{bilayerpi0edc}). For the optimally and
underdoped cases it has been suggested that in the normal state,
due to the breadth of the lineshapes, the two bands can only be
recognized on the basis of the different photon energy dependence
(see Fig.\,\ref{bilayer_mix}); in the superconducting state on the
other hand (see also Sec.\,\ref{sec:selfdiscu}), bonding and
antibonding bands can be resolved directly due to the sharpening
and/or spectral weight increase of the peaks
\cite{kordyuk:1,gromko:2}. In this context, it should be noted
that also the complex lineshape observed in the ($\pi$,0) spectra
from YBCO (Fig.\,\ref{ybco_gap}) and Bi2223
(Fig.\,\ref{Bi2223_GAP}) may originate from multilayer splitting
effects.

Despite the degree of uncertainty on the lineshape analysis of the
peak-dip-hump structure as a whole, and independent of the bilayer
splitting effects, the detailed study of the temperature and
doping dependence of this feature may still provide very relevant
information.
\begin{figure}[b!]
\centerline{\epsfig{figure=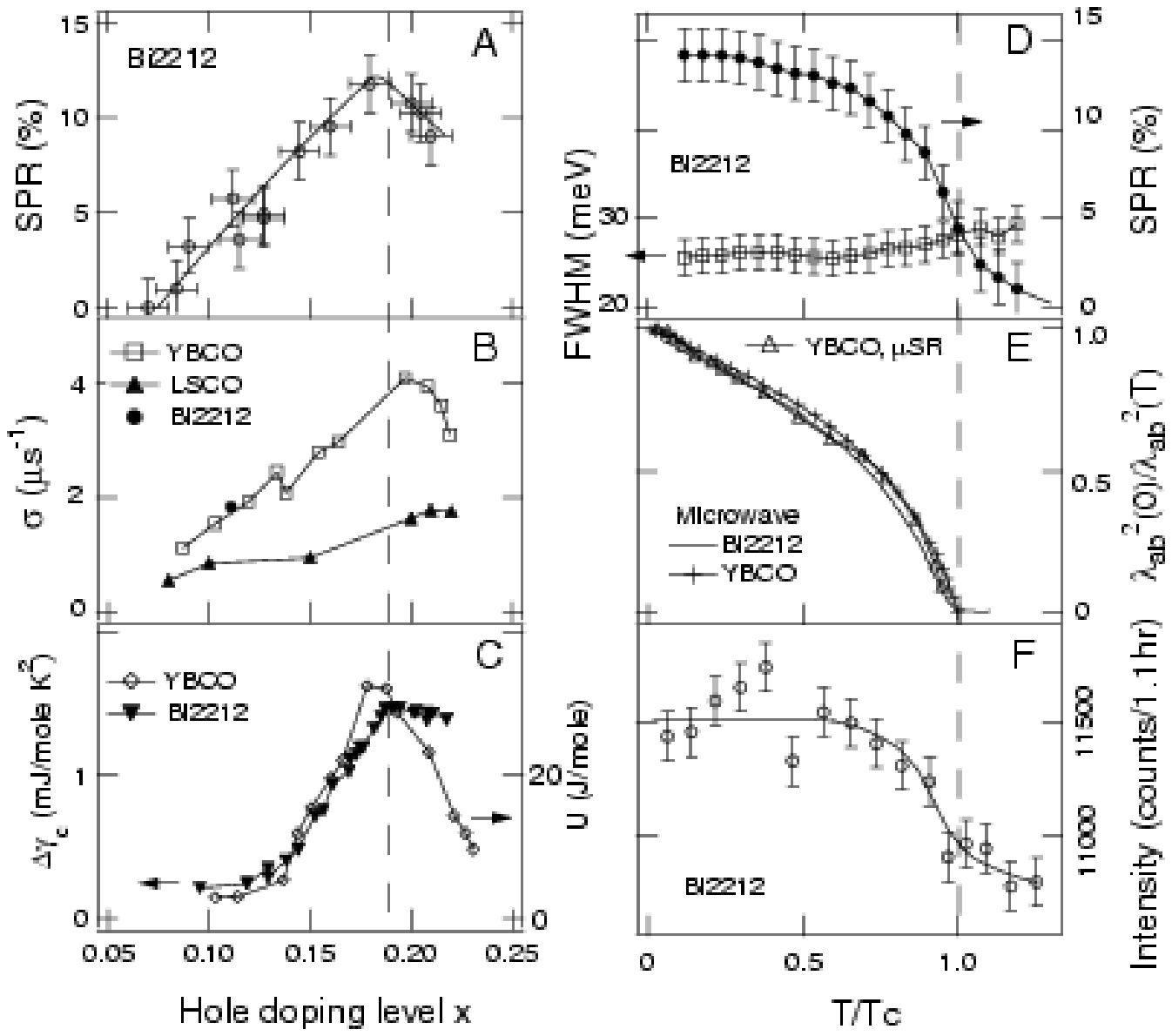,width=1\linewidth,clip=}}
 \caption{Doping dependence, for $T\!<\!T_c$: (a) SPR for the Bi2212
 spectra of Fig.\,\ref{DL_EDC}c; (b) $\mu$SR relaxation rate
 ($\sigma\!\propto\!n_s$) \cite{TallonJL:Cridoh,UemuraYJ:Bassac}; (c)
 Bi2212 specific-heat coefficient jump
 $\Delta\gamma_c\!=\!\gamma(T_c)\!-\!\gamma(120K)$
 \cite{TallonJL:Facaod}, and Y123 condensation energy $U$
 \cite{TallonJL:Cridoh}. $T$-dependence: (d) SPR and peak-width of
 Bi2212 (sample OD84 in Fig.\,\ref{DL_EDC}b); (e)
 $\lambda_{ab}^2(0)/\lambda_{ab}^2(T)$ ($\propto\!n_s$)
 \cite{JacobsT:In-pla,SonierJE:Fieirt,BonnDA:ComtiN}, where
 $\lambda_{ab}$ is the in-plane penetration depth; (f) intensity of
 the neutron ($\pi$,$\pi$) mode \cite{HeH:Resseo}. After
 \textcite{FengDL:Sigsdt}.}\label{DL_SPR}\end{figure}
\textcite{FengDL:Sigsdt} used a phenomenological fitting procedure
to quantify the temperature and doping evolution of the intensity
of the lowest energy peak at ($\pi$,0). In particular, the ratio
between the area of the peak and that of the spectrum integrated
between -0.5 and 0.1\,eV was considered, in the attempt of
extracting more systematic information (i.e., independent of
artifacts due to $k$-dependence of matrix elements and/or
different experimental conditions for the different samples). The
doping and temperature dependence of this quantity, referred to as
the {\it superconducting peak ratio} (SPR), are presented in
Fig.\,\ref{DL_SPR}a and\,\ref{DL_SPR}d, respectively. From the
comparison, in the same figure, with many superfluid-related
quantities measured on Bi2212, Y123, and LSCO,
\textcite{FengDL:Sigsdt} suggested that: (i) the remarkable
similarities between the data presented in Fig.\,\ref{DL_SPR} hint
at a universality in the superconducting properties of the cuprate
HTSCs; (ii) ARPES, which probes single-particle excitations of the
condensate and therefore directly measures the strength of the
pairing (i.e., superconducting gap), may also provide information
on the phase coherence of the superconducting state (note that in
principle this can only be inferred from those techniques which
directly probe the collective motion of the condensate). This
latter point is evidenced by the similar behavior observed, as a
function of hole concentration, for SPR, superfluid density $n_s$
measured by $\mu$SR (Fig.\,\ref{DL_SPR}b), condensation energy $U$
from the specific heat and jump in the specific heat coefficient
(Fig.\,\ref{DL_SPR}c), as well as by the similar temperature
dependence of SPR and $n_s$ measured by microwave or $\mu$SR
spectroscopy (Fig.\,\ref{DL_SPR}d and\,\ref{DL_SPR}e).
\begin{figure}[t!]
\centerline{\epsfig{figure=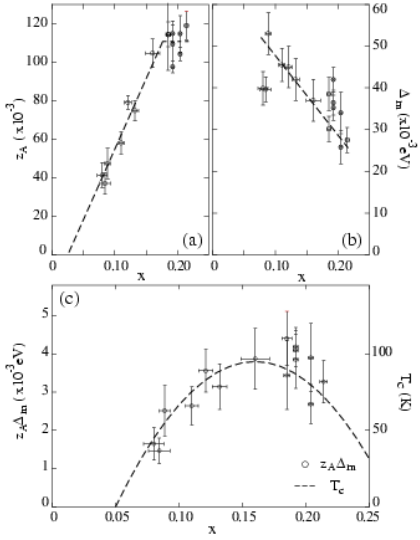,width=6cm,clip=}}
\caption{Doping dependence, as estimated at 14\,K, of: (a)
coherent quasiparticle weight $z_A$, (b) maximum gap size
$\Delta_m$, and (c) $z_A\Delta_m$. In (c) the empirical relation
between $T_c$ and $x$ is also shown \cite{PreslandMR:Gentos}.
After \textcite{DingH:Cohqwi}.}\label{Ding_spr}\end{figure}
Furthermore, contrary to what is expected within the FL-BCS
framework \cite{schrieffer:1}, the SPR and superfluid density
exhibit an abrupt drop near $T_c$ (i.e., disappearance of the
phase coherence) rather than at $T^*$ (i.e., opening of the
pseudo-gap in the underdoped regime), and grow with the hole
concentration $x$ while the gap magnitude, as determined by ARPES
(see Sec.\,\ref{sec:pseudogap} and Fig.\,\ref{pg5}), scales with
($1\!-\!x$).

Similar observations were made by \textcite{DingH:Cohqwi}, who
also found an increase in the peak intensity with doping, as shown
in Fig.\,\ref{Ding_spr}a (here the quantity $z_A$, referred to by
the authors as {\it coherent quasiparticle weight}, is defined
exactly as the above SPR and is therefore a phenomenological
quantity without rigorous theoretical implications). It was
observed that the product $z_A\Delta_m$, where $\Delta_m$ is the
superconducting gap estimated by the peak-position at ($\pi$,0)
for $T\!=\!14$\,K, is directly proportional to the superconducting
transition temperature (Fig.\,\ref{Ding_spr}c). On the basis of
these results, \textcite{DingH:Cohqwi} concluded that
superconductivity is mainly controlled by the quasiparticle
coherence $z_A$ in the underdoped regime and by the
superconducting gap in the overdoped regime, and suggested that a
new quantity, $z_A\Delta_m(T\!=\!0)$, may be the true
superconducting order parameter in the HTSCs.

The emergence of a sharp quasiparticle peak below $T_c$ has been
taken as evidence for some form of coherence transition, which was
also the spirit of the phenomenological analyses discussed above.
While in the BCS framework \cite{schrieffer:1} the superconducting
transition and the opening of a superconducting gap in the single
particle excitation spectrum are due to an effective attractive
interaction between FL quasiparticles which are already
well-defined above $T_c$, in the case of the HTSCs coherent
quasiparticles may be formed only upon entering the
superconducting state [see, e.g., \textcite{ShenZ-X:Novesc}].
Theoretical examples of this kind of transition in the literature
include, e.g., dimensional crossover \cite{CarlsonEW:Dimcq-},
condensation of chargons \cite{senthil:1}, quantum confinement and
Bose condensation \cite{NagaosaN:ConBcg}. Note however that the
above discussed interpretations by \textcite{FengDL:Sigsdt} and
\textcite{DingH:Cohqwi} are not universally accepted. Notably,
\textcite{NormanMR:Temets} suggested that the temperature
dependence of the ARPES spectra is not due to the decrease of
spectral weight of the low-energy peak upon increasing the
temperature above $T_c$, but is instead a reflection of the {\it
quasiparticle lifetime catastrophe}
\cite{KurodaY:SupsC-,NormanMR:Colms-,NormanMR:Temets,AbanovA:relbrn}.
The disappearance of the sharp peak above $T_c$ would be a
consequence of the reduction of the low-energy electron lifetime,
which broadens the quasiparticles out of existence once the
superconducting gap has closed.
\begin{figure}[b!]
\centerline{\epsfig{figure=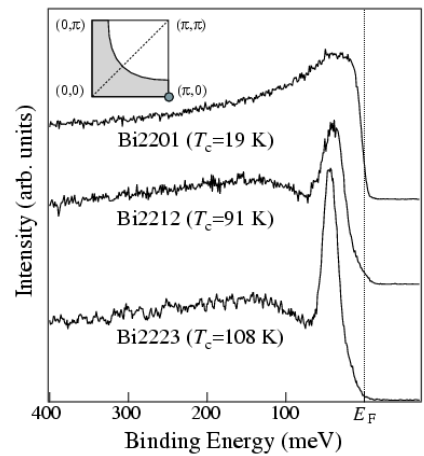,width=6cm,clip=}}
\caption{Superconducting state ($\pi$,0) spectra from Bi2201
($T_c\!=\!19$\,K), Bi2212 ($T_c\!=\!91$\,K), and Bi2223
($T_c\!=\!108$\,K). The data were taken with 21.2\,eV photons at
13.5\,K for Bi2201, and 40\,K for both Bi2212 and Bi2223
\cite{sato:1}.}\label{sato2223}\end{figure}
At this stage, it is hard to judge which interpretation is more
appropriate. On the one hand, the analysis by
\textcite{NormanMR:Temets} is more detailed than the
phenomenological one presented in the two experimental papers
\cite{FengDL:Sigsdt,DingH:Cohqwi}.  On the other hand, this
alternative scenario seems to be in contrast with the experimental
observation that, at least in the optimally-doped/overdoped regime
(see Fig.\,\ref{DL_SPR}d where the FWHM is plotted), the width of
the superconducting peak does not appear to change significantly
across $T_c$ \cite{FedorovAV:Temdps,FengDL:Sigsdt,DingH:Cohqwi}.

\subsection{Bi$_2$Sr$_2$Ca$_2$Cu$_3$O$_{10+\delta}$ and Bi$_2$Sr$_2$CuO$_{6+\delta}$}
\label{sec:speakBi2223}

The temperature dependence seen for the quasiparticle peak in
Bi2212 seems to be confirmed also by the data reported for
optimally doped Bi2223 (Fig.\,\ref{Bi2223_GAP}a
and\,\ref{sato2223_psg}), in which once again the quasiparticle
peak disappears at temperatures slightly larger than
$T_c\!=\!108$\,K \cite{FengDL:2,sato:1}. Furthermore, as shown in
Fig.\,\ref{sato2223} where the superconducting state ($\pi$,0)
ARPES spectra from one, two, and three CuO$_2$ planes Bi-cuprates
are presented, it appears that the superconducting peak intensity
increases with $T_c$, supporting the idea that it is related to
the strength of superconductivity in the system. In particular,
\textcite{FengDL:2} showed that both the peak intensity and the
superconducting gap increase approximately linearly with $T_c$ in
this family of compounds, which seems to indicate that both phase
coherence and pairing strength scale with the number of CuO$_2$
planes. In this regard, it has to be mentioned that for Bi2201,
presumably due to the lower superfluid density of the single layer
compound, only very recently some evidence was collected for the
presence of a sharp peak at low temperatures, which indeed is much
weaker than in the case of Bi2212 and Bi2223  \cite{LanzaraA:1}.

\section{Pseudogap}
\label{sec:pseudogap}

One of the most important contributions of ARPES to the
investigation of the HTSCs, is the discovery of the normal-state
excitation gap or {\it pseudogap}
\cite{LoeserAG:Excgtn,MarshallDS:Uncese,DingH:Speept,levi:2}.
Similar to what we have discussed in Sec.\,\ref{sec_supgap} for
the superconducting gap, the pseudogap can be simply described as
the opening of an energy gap along the underlying FS, but detected
in this case in the normal state.
\begin{figure}[b!]
\centerline{\epsfig{figure=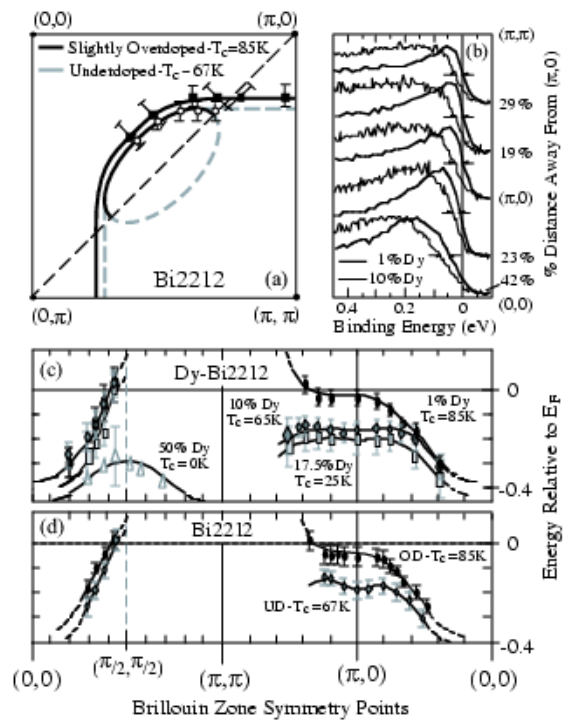,width=8cm,clip=}}
\vspace{0cm} \caption{(a) FS crossings for Bi2212. (b) Leading
edge midpoint shifts for Dy-Bi2212. (c-d) Dispersions determined
from the peak centroids for Dy-Bi2212 and Bi2212, respectively,
for different doping levels. After
\textcite{MarshallDS:Uncese}.}\label{pg1}\end{figure}
In fact, for the underdoped cuprates it was found that the Fermi
level crossings are absent over a large portion of the FS due to
the opening of an excitation gap at temperatures considerably
higher than $T_c$ itself. This phenomenon was first recognized in
the photoemission spectra by \textcite{KingDM:EleseM} in
attempting to connect the Bi2212 data to that from undoped SCOC
\cite{WellsBO:Evkrmb}. The pseudogap has been observed in many of
the cuprate superconductors, but has been most extensively studied
in Bi2212.
\begin{figure}[t!]
\centerline{\epsfig{figure=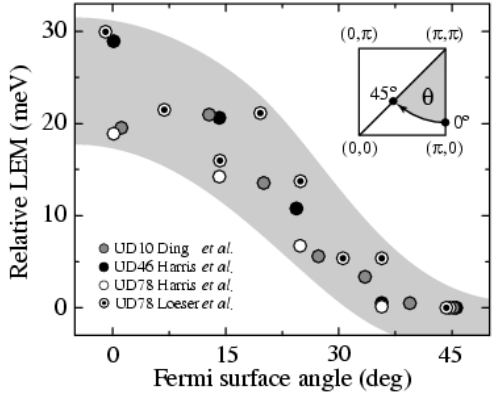,width=7cm,clip=}}
\caption{Momentum dependence of the pseudogap along the expected
FS for underdoped Bi2212 (numbers indicate $T_c$): UD10
\cite{DingH:Speept}; UD46, UD78 \cite{HarrisJM:Anossg}; UD78
\cite{LoeserAG:Excgtn}. Note that here the leading edge midpoint
positions {\it relative} to the value at 45$^\circ$ were
plotted.}\label{pg3b}\end{figure}
As we will discuss in detail later, the main characteristics of
the normal-state pseudogap can be summarized as follows
\cite{ShenZX:Excgtn,randeria:1}: (i) the effect is strong in
underdoped samples, persists to optimal doping, and disappears in
overdoped samples
\cite{LoeserAG:Excgtn,MarshallDS:Uncese,DingH:Speept}. (ii) The
gap has two energy scales, with the low energy one given by the
location of the leading-edge midpoint (which shows a clear gap),
and the higher energy one by the position of the broad peak near
the ($\pi$,0) point (the expected sharp quasiparticle peak at
$E_F$ is converted into a broad feature over $\sim\!100$\,meV and
the low-energy spectral weight is suppressed). The lower energy
scale has a $d$-wave like momentum dependence similar to the one
of the superconducting gap, with a gapless arc near the nodal
region. As a function of doping, the two energy scales track each
other
\cite{MarshallDS:Uncese,NormanMR:DestFs,WhitePJ:Rapsts,HarrisJM:Anossg,CampuzanoJC:Elestr}.
(iii) Upon reducing the hole concentration, the size of the
leading-edge pseudogap increases, in contrast to the decreasing of
$T_c$. This is believed to provide an important piece of evidence
for the non-BCS behavior of the superconducting transition in the
underdoped regime of the HTSCs.

\subsection{Bi$_2$Sr$_2$CaCu$_2$O$_{8+\delta}$}
\label{sec_pgBi2212}

The first specific study of the pseudogap effect in Bi2212 was
performed by \textcite{MarshallDS:Uncese}. Fig.\,\ref{pg1}
reproduces the key data that illustrate the basic phenomenology.
For the optimally doped or overdoped samples, Fermi crossings
around the entire FS are observed (solid squares in
Fig.\,\ref{pg1}a). For the underdoped samples, well defined FS
crossings are detected only within an arc-segment centered on the
(0,0)-($\pi$,$\pi$) line (open circles in Fig.\,\ref{pg1}a), while
they are completely missing near ($\pi$,0). By  comparing the
results from both underdoped and optimally doped Dy-Bi2212
(Fig.\,\ref{pg1}b), it is clear that the spectra from the
underdoped samples pull towards higher binding energy in the
entire region, and in particular at ($\pi$,0). As the FS should be
a continuous surface in momentum space, this behavior was
interpreted as the opening of an anisotropic gap. It has to be
emphasized that the pseudogap phenomenon is relatively insensitive
to the details of the FS topology in this region
(Sec.\,\ref{sec:ns_Bi2212}): in fact, the magnitude of the gap is
rather large, and the spectra from underdoped samples are pulled
back over an extended momentum space region with a weak band
dispersion near the ($\pi$,0) point (see Fig.\,\ref{pg1}c,d, where
the dispersion is summarized for several doping levels). Hence,
the normal state gap and its doping dependence are robust features
in the ARPES spectra.

\textcite{MarshallDS:Uncese} suggested two different ways of
characterizing the normal state pseudogap: by the position of the
leading edge (20-30 meV) or by the position of the broad maximum
of the spectra (100-200 meV), which identify respectively
low-energy and high-energy pseudogap. While the former is well
defined at intermediate low-doping levels but not in the deeply
underdoped regime, the latter is particularly useful in the very
underdoped cases. As we will discuss below, it is generally
believed that low and high-energy scale of the normal-state gap
naturally connect to the gap of the superconductor and the one of
the AF insulator, respectively.
\begin{figure}[t!]
\centerline{\epsfig{figure=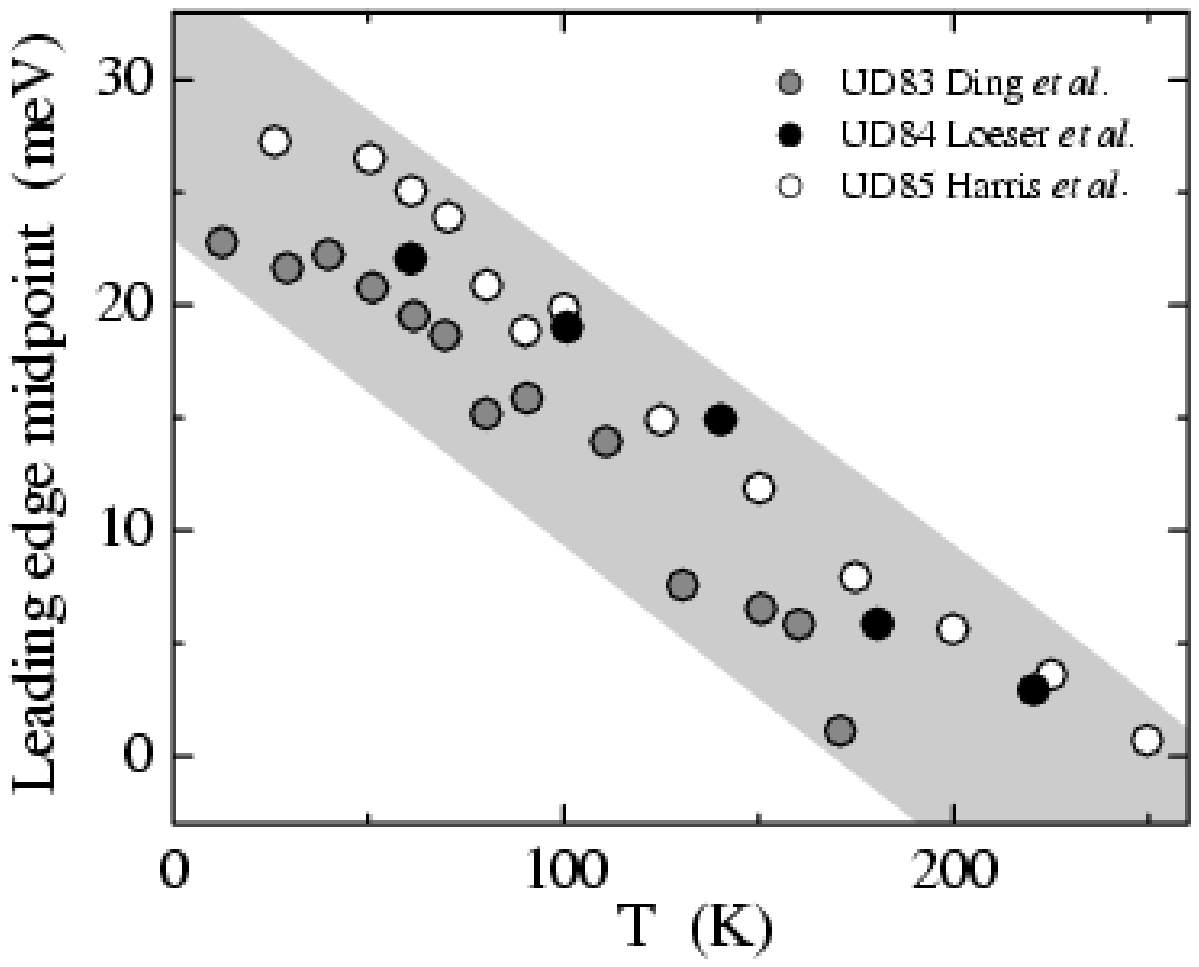,width=7cm,clip=}}
\caption{Temperature dependence of the pseudogap magnitude near
($\pi$,0) for underdoped Bi2212: UD83 \cite{DingH:Speept}; UD84
\cite{LoeserAG:Temddt}; UD85
\cite{HarrisJM:Anossg}.}\label{pg3a}\end{figure}
In this regard, of particular relevance are the momentum
dependence studies of these features. (i) For the low-energy
pseudogap, detailed results were reported by
\textcite{LoeserAG:Excgtn} and \textcite{DingH:Speept}.
\begin{figure}[b!]
\centerline{\epsfig{figure=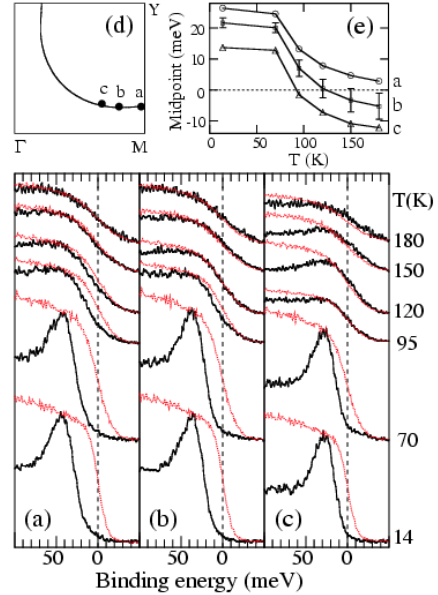,width=6.1cm,clip=}}
\caption{(a-c) ARPES spectra (black lines) from underdoped Bi2212
($T_c\!=\!85$\,K), taken at different momentum space points as
sketched in (d), and reference spectra from polycrystalline Pt
(red lines). (e) Temperature dependence of the Bi2212 leading-edge
midpoints. After \textcite{NormanMR:DestFs}
[Color].}\label{pg2}\end{figure}
Although the spectra are characterized by a peak not as sharp as
the one seen in the superconducting state
(Sec.\,\ref{sec:speak_Bi2212}), the compilation of data presented
in Fig.\,\ref{pg3b} shows that the $k$-dependence of the magnitude
of low-energy pseudogap is similar to the one of the $d$-wave
superconducting gap (Sec.\,\ref{sec:supgapBi2212}), which suggests
the interesting possibility that the two gaps may originate from
the same underlying mechanism. (ii) As for the high-energy
pseudogap, it was suggested that this feature exhibits the same
dispersion along the (0,0)-($\pi$,$0$) and
($\pi$,$\pi$)-($\pi$,$0$) directions \cite{CampuzanoJC:Elestr}. As
these orthogonal directions are equivalent only in the reduced AF
Brillouin zone, it was argued that the high energy pseudogap is a
remnant of the AF undoped insulator [similar point was made by
\textcite{RonningF:PhoerF}, following the proposal of
\textcite{LaughlinRB:Numeed,LaughlinRB:Eviqdp}]. However this
conclusion rests on the assumption that the dispersion of the
high-energy pseudogap is equivalent to the one of the hump
detected in the superconducting state (see
Sec.\,\ref{sec:speak_Bi2212} and, in particular,
Fig.\,\ref{norma2} and\,\ref{pg6}). In light of the recent
detection of bilayer splitting in Bi2212
(Sec.\,\ref{sec:ns_Bi2212}), and of the possible correspondence of
peak and hump features to antibonding and bonding bilayer split
bands, this interpretation of the dispersion of hump and
high-energy pseudogap should be reconsidered (then again, as we
will discussed in Sec\,\ref{sec_pgLSCO} and\,\ref{sec:pseudoNa}, a
similar behavior has been observed also on single-layer underdoped
cuprates).

Important insights also come from the study of the pseudogap as a
function of temperature. Fig.\,\ref{pg3a} summarizes the
temperature dependence of the leading-edge midpoint positions near
($\pi$,0) for underdoped samples with $T_c$ of about 85 K, which
shows that at these dopings the pseudogap opens up around
$T^*\!\sim\!200$\,K
\cite{LoeserAG:Excgtn,DingH:Speept,HarrisJM:Anossg}. More in
detail, as reported by \textcite{NormanMR:DestFs} in a subsequent
paper, we are dealing with a highly anisotropic gap that closes
non-uniformly in momentum space upon increasing the temperature.
\begin{figure}[t!]
\centerline{\epsfig{figure=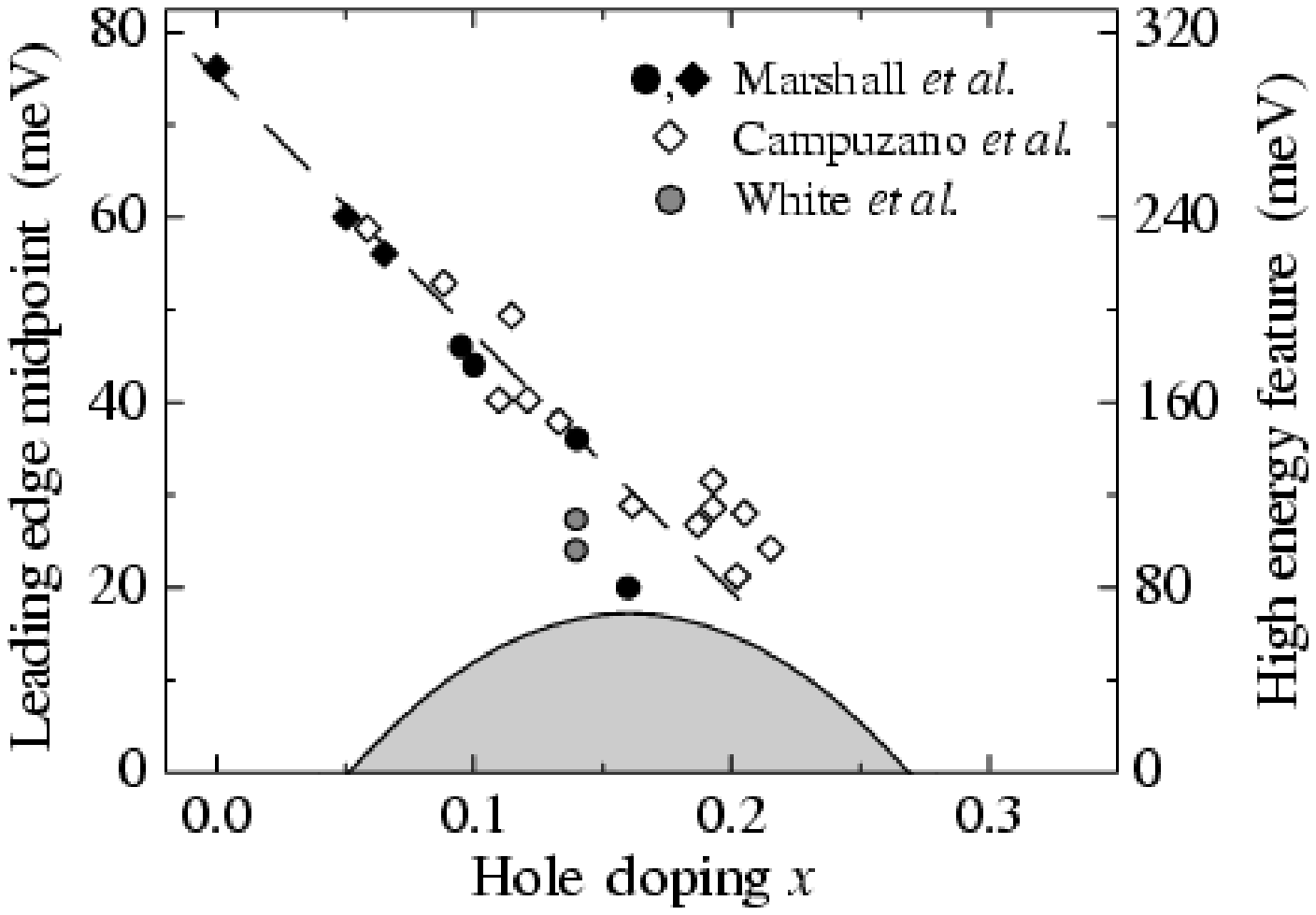,width=8cm,clip=}}
\vspace{0cm} \caption{Doping dependence of the pseudogap magnitude
as determined by the position of leading-edge midpoint (circles,
left axis) and high-energy feature (diamonds, right axis) in the
($\pi$,0) ARPES spectra from Bi2212 [the dome represents the
$d$-wave mean-field approximation $\Delta(x)\!=\!4.3k_BT_c(x)/2$,
left scale \cite{WonH:d-wavs}]. Data from:
\textcite{MarshallDS:Uncese,WhitePJ:Rapsts,CampuzanoJC:Elestr}.}\label{pg5}\end{figure}
In fact, as shown in Fig.\,\ref{pg2}, the temperature at which the
leading-edge midpoint coincides with the reference Fermi energy
obtained from a polycrystalline Pt sample is very different for
the Bi2212 spectra taken at $k$-space points (a), (b), and (c). In
particular, this happens at about 150 K at (a), and 95 K at (c).
Since these results were obtained on a $T_c\!=\!85$\,K underdoped
sample, they indicate that a gap is present at (a) and (b) already
at temperatures well above $T_c$.

The next important issue is the doping dependence of the
normal-state gap magnitude. Fig.\,\ref{pg5} compiles data from
several studies, which indicate that the pseudogap decreases
monotonically upon doping. Here, in order to estimate the
normal-state gap energy, both the position of the leading-edge
midpoint and of the high-energy feature were used. Concerning the
different doping dependence of the experimentally determined
pseudogap and $T_c$ lines in the underdoped regime, two
alternative scenarios are usually discussed: (i) phase
fluctuations and (ii) quantum critical point. The phase
fluctuation picture refers to the idea that superconductivity in
the underdoped regime is determined by the phase stiffness of the
superfluid,\footnote{\textcite{DoniachS:Lon-ra,UemuraYJ:Bassac,ImadaM:TwotMt,EmeryVJ:Imppfs,EmeryVJ:Supbm.,KivelsonSA:Elel-c}.}
which is consistent with the experimental observation that $T_c$
in the underdoped samples is proportional to the superfluid
density \cite{UemuraYJ:Bassac}.
\begin{figure}[b!]
\centerline{\epsfig{figure=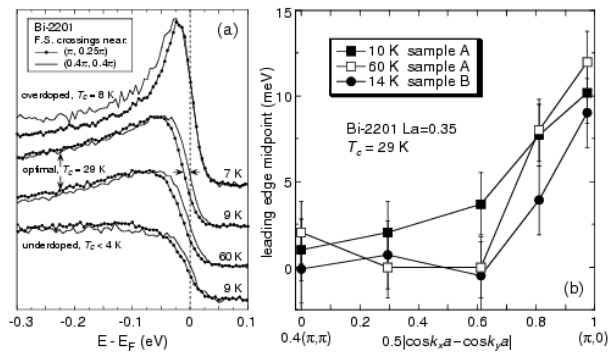,width=1\linewidth,clip=}}
\vspace{0cm} \caption{(a) Bi2201 ARPES spectra measured, for
different doping levels, at the FS crossings in the nodal region
and close to ($\pi$,0). (b) Leading edge midpoint shifts for
slightly underdoped Bi2201 ($T_c\!=\!29$ K). After
\textcite{HarrisJM:Meaaeg}.}\label{BIharris2}\end{figure}
In this view, the pseudogap is a reflection of incoherent
pair-fluctuations above $T_c$,\footnote{Pre-formation of Cooper
pairs at the crossover temperature $T^*\!>\!T_c$ without any phase
coherence, which is established only when they condense into the
coherent superconducting state.} and does not scale with $T_c$
itself. Given the similarity in the momentum dependence between
the pseudogap and the superconducting gap, earlier photoemission
data were mostly interpreted along this line. In the case of the
quantum critical point scenario, superconducting and normal-state
gap have a different
origin,\footnote{\textcite{TallonJL:Cridoh,TallonJL:Facaod,TallonJL:dopdT*}.}
and the suppression of $T_c$ in the underdoped regime is caused by
the development of a competing order responsible for magnetic
correlations independent of the Cu spins \cite{SonierJE:Anowms}.
In this context, possible microscopic descriptions for the
pseudogap phase include the circulating-current phase
\cite{VarmaCM:Non-Fe,VarmaCM:Psepq-}, charge-density wave
\cite{CastellaniC:Sinqst}, and $d$-density wave short-range
fluctuations \cite{ChakravartyS:Hidotc}. It has to be mentioned
that, in addition to scenarios (i) and (ii), there is also a class
of pairing theories that explain the normal state $d$-wave gap as
a signature of pairing in the spin channel. These stem from the
spin-density-wave/spin-bag
approach,\footnote{\textcite{KampfA:Pses-b,KampfAP:Spefps,kusko:1}.}
and from the strong-coupling version of the $t$-$J$ or Hubbard
models,\footnote{\textcite{KotliarG:Supitl,SuzumuraY:MeaftR,TanamotoT:Ferssf,Xiao-GangWen:Theuc.}.}
in which the pseudogap results from AF spin fluctuations and
spinon pairing respectively. Finally, there is the interpretation
of the pseudogap in terms of {\it umklapp} scattering truncation
of the FS near the location of the hot spots, as in
Fig.\,\ref{models}b.\footnote{\textcite{FurukawaN:Trut-d,FurukawaN:InsL-F}.}

\subsection{Bi$_2$Sr$_2$CuO$_{6+\delta}$}
\label{sec_pgBi2201}

Clear evidence for a $d$-wave like pseudogap for underdoped and
optimally doped Bi2201 was reported by \textcite{HarrisJM:Meaaeg},
as shown in Fig.\,\ref{BIharris2} (note that no gap was observed
for the overdoped material). On the basis of the leading-edge
midpoint shift, the gap amplitude was estimated to be
$\sim\!10$\,meV for the optimally doped sample, therefore much
smaller than in Bi2212 ($\sim\!20$-30\,meV, see Fig.\,\ref{pg5}).
As discussed for the case of Bi2212 (Sec.\,\ref{sec_pgBi2212}),
the momentum dependence and magnitude of the normal state
pseudogap are similar to those of the superconducting gap, which
may imply a common origin for the two features. Furthermore, as we
will see in Sec.\,\ref{sec_pgLSCO}, the magnitude of the pseudogap
in Bi2201, in particular as estimated from the position of the
leading-edge midpoint, is comparable to the results obtained for
LSCO. Given that Bi2201 and LSCO are characterized by a similar
value of $T_c^{max}\!\simeq$34-38\,K much lower than the
$\sim\!95$\,K of Bi2212 [see, e.g., \textcite{eisaki:1}], these
findings suggest a direct correlation between $T_c^{max}$ and the
size of the pseudogap (and/or superconducting gap), for the
different families of cuprates.

\subsection{Bi$_2$Sr$_2$Ca$_2$Cu$_3$O$_{10+\delta}$}

\begin{figure}[t!]
\centerline{\epsfig{figure=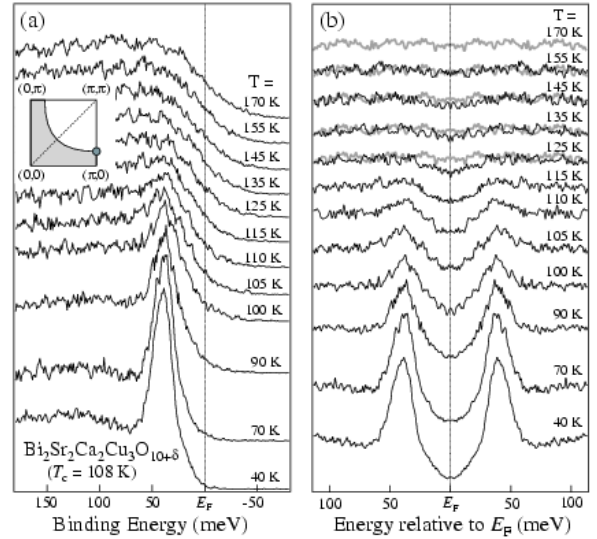,width=8.5cm,clip=}}
\vspace{0cm} \caption{(a) Raw and (b) symmetrized temperature
dependent ARPES spectra measured on Bi2223 in correspondence of
the underlying FS crossing along the ($\pi$,0)-($\pi$,$\pi$)
direction. In (b) the 170\,K spectrum (gray thick line) is
superimposed to those taken at $T\!\geq\!125$\,K for comparison
\cite{sato:1}.}\label{sato2223_psg}\end{figure}

The pseudogap in Bi2223 has been recently studied by
\textcite{sato:1} and \textcite{FengDL:2}, on nearly optimally
doped samples ($T_c\!=\!108$\,K). The ($\pi$,0) ARPES spectra and
the results obtained by symmetrizing the spectra with respect to
$E_F$ [as defined by \textcite{NormanMR:Phetl-}] are presented as
a function of temperature in Fig.\,\ref{sato2223_psg}a,b. At low
temperature, a large superconducting gap is clearly visible, as
discussed in Sec.\,\ref{sec:Bi2223_supgap}.
\begin{figure}[b!]
\centerline{\epsfig{figure=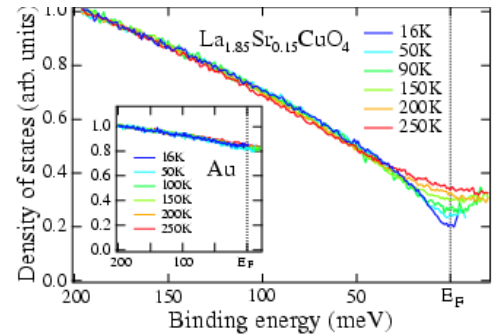,width=7cm,clip=}}
\vspace{0cm} \caption{Density of states for optimally doped LSCO
($T_c\!=\!38$\,K) and polycrystalline Au. After
\textcite{SatoT:PseodL} [Color].}\label{pseudoLSCO1}\end{figure}
However, as emphasized by the direct comparison between the 170\,K
symmetrized spectrum an those taken at $T\!\geq\!125$\,K
(Fig.\,\ref{sato2223_psg}b), the gap is still open at
$T\!\simeq\!T_c$ and it does not close until $T^*\!\simeq\!135$\,K
(as evidenced by the suppression of the low-energy spectral
weight). Note that the absolute difference between $T_c$ and $T^*$
is rather small in this case, presumably because the doping level
is very close to optimal. Similar results were reported by
\textcite{FengDL:2}, who showed that a gap is present at spectra
taken above $T_c$ near ($\pi$,0) but not along the nodal
direction, which suggests a $d$-wave symmetry for the pseudogap in
Bi2223.

\subsection{La$_{2-x}$Sr$_x$CuO$_4$}
\label{sec_pgLSCO}

The investigation of the normal state pseudogap in LSCO by ARPES
has been complicated by the poor stability of the cleaved surface
at temperature higher than $T_c$, even in ultra-high vacuum.
Therefore, the first evidence for this phenomenon in LSCO was
reported by \textcite{InoA:Dopdds} on the basis of
angle-integrated photoemission experiments, in which clean
surfaces were obtained by repeatedly scraping (i.e., every 40
minutes) the sample surface {in situ} with a diamond file (a
procedure which obviously prevents the acquisition of angle
resolved spectra, due to the roughness of the scraped surface).
These data indicated a systematic depression of the density of
states (more pronounced at lower dopings), which has been
considered indicative of a pseudogap with an energy scale of about
100-200 meV at 5-10\% doping.

Using a similar procedure, \textcite{SatoT:PseodL} investigated
the temperature dependence of the pseudogap in optimally doped
LSCO ($x\!=\!0.15$, $T_c\!=\!38$\,K). In order to extract a more
direct representation of the density of states near the Fermi
level, the spectra were divided by the Fermi-Dirac distribution
function (at the corresponding temperature) convoluted with a
Gaussian (to account for the instrumental resolution of 7\,meV).
\begin{figure}[t!]
\centerline{\epsfig{figure=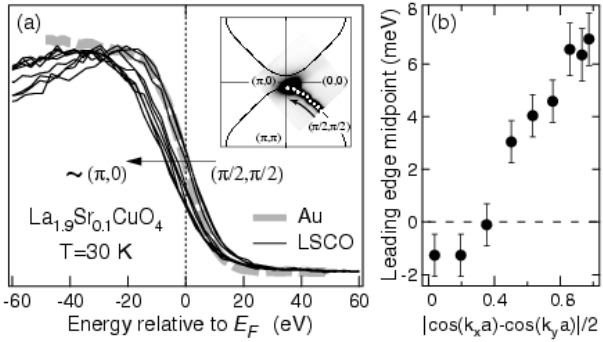,width=1\linewidth,clip=}}
\vspace{0cm} \caption{(a) ARPES spectra from along the FS contour
(see inset) measured at 30\,K with 55.5\,eV photons on 10\% doped
LSCO ($T_c\!=\!29$\,K). A spectrum from polycrystalline Au is
shown for comparison. (b) Corresponding momentum dependence of the
leading-edge gap. After
\textcite{yoshida:1}.}\label{pseudoLSCO2}\end{figure}
In contrast to what is observed in a normal metal like Au (inset
of Fig.\,\ref{pseudoLSCO1}), on LSCO the intensity close to the
Fermi level ($\leq\!30$\,meV) increases smoothly upon raising the
temperature (Fig.\,\ref{pseudoLSCO1}). This effect is observable
over a temperature range much larger than $T_c$, and thus provides
direct evidence for the existence of normal state pseudogap.

\begin{figure}[t!]
\centerline{\epsfig{figure=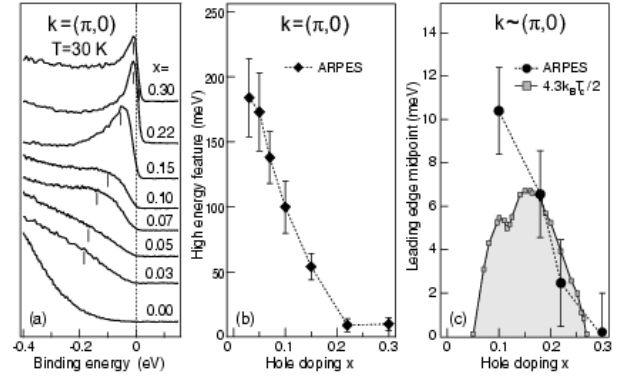,width=1\linewidth,clip=}}
\vspace{0cm} \caption{(a) Doping dependence of the LSCO ARPES
spectra ($h\nu\!=\!55.5$\,eV) and of the (b) high-energy feature
at ($\pi$,0), and (c) of the leading-edge midpoint in
correspondence of the underlying FS {\it near} ($\pi$,0). Note
that the points in (c) and the spectra in (a) are only apparently
at variance (see, e.g., $x\!=\!0.1$): the results are not from the
exact same momenta \cite{yoshida:1}. The dome in (c) is the
$d$-wave mean-field approximation $\Delta(x)\!=\!4.3k_BT_c(x)/2$
\cite{WonH:d-wavs}.}\label{pseudoLSCO3}\end{figure}

Detailed angle-resolved investigations of the pseudogap in LSCO
were recently reported by \textcite{InoA:Dop-de} and
\textcite{yoshida:1}. As shown in Fig.\,\ref{pseudoLSCO2}a where
the shift of the leading-edge midpoint along the underlying FS is
presented for 10\% doped LSCO, a clear gap is observable close to
($\pi$,0) but not at ($\pi$/2,$\pi$/2), with an overall momentum
dependence consistent with the $d$-wave functional form
(Fig.\,\ref{pseudoLSCO2}b). Note that the spectra presented in
Fig.\,\ref{pseudoLSCO2} were taken at the temperature of 30\,K,
therefore not much higher than $T_c\!=\!29$\,K. This was necessary
in order to avoid aging of the sample surface over the time
required to obtain a complete set of data in momentum space. It
has to be emphasized that, however, temperature dependent ARPES
experiments restricted to the ($\pi$,0) region show that the gap
is still open at temperatures as high as 90\,K and therefore well
into the normal state, consistent with the results presented in
Fig.\,\ref{pseudoLSCO1}. Regarding the absolute magnitude of the
pseudogap and its doping evolution, detailed estimates could also
be obtained from angle-resolved experiments
(Fig.\,\ref{pseudoLSCO3}a). The results for the high-energy
pseudogap at ($\pi$,0) and the low-energy pseudogap at the
underlying Fermi vector near ($\pi$,0) are presented in
Fig.\,\ref{pseudoLSCO3}b and\,c, respectively. Similar to the case
of the Bi-based HTSCs, these findings support the picture of a
smooth evolution of the pseudogap into the superconducting gap
upon increasing the doping. Furthermore, as mentioned in
Sec.\,\ref{sec_pgBi2201}, the results from these two different
families indicate a direct scaling of the leading-edge pseudogap
with $T_c^{max}$.

It has been observed that the doping dependence of the high energy
pseudogap (Fig.\,\ref{pseudoLSCO3}b) indicates that this effect is
intimately related to the anomalous behavior seen in
thermodynamics and transport properties. In particular, for
$x\!<\!0.2$ along with the opening of the pseudogap one observes a
suppression of the electronic specific heat and a decrease of the
effective mass \cite{InoA:Dopdds,InoA:Dop-de}.
This suggests that a decrease in the density of states (or
carriers) is responsible for the metal-insulator transition in
LSCO, and not a divergence of the effective mass
\cite{ImadaM:Met-in}. This may also be related to the fact that
the coherent quasiparticle weight decreases upon reducing the
doping level (Fig.\,\ref{pseudoLSCO3}a).

\subsection{Ca$_{2-x}$Na$_x$CuO$_2$Cl$_2$}
\label{sec:pseudoNa}

Many attempts have been made to connect the electronic structure
of the doped HTSC to that of the insulator. This is also the case
in the context of the pseudogap discussion, as many believe that
the pseudogap is a quantity that can already be defined in the
insulator. Since the best insulator data were collected on SCOC
and CCOC, and the best metal data were obtained on Bi2212, it is
natural that the possibility of such a connection has been most
intensively explored for these two families of materials.
\begin{figure}[t!]
\centerline{\epsfig{figure=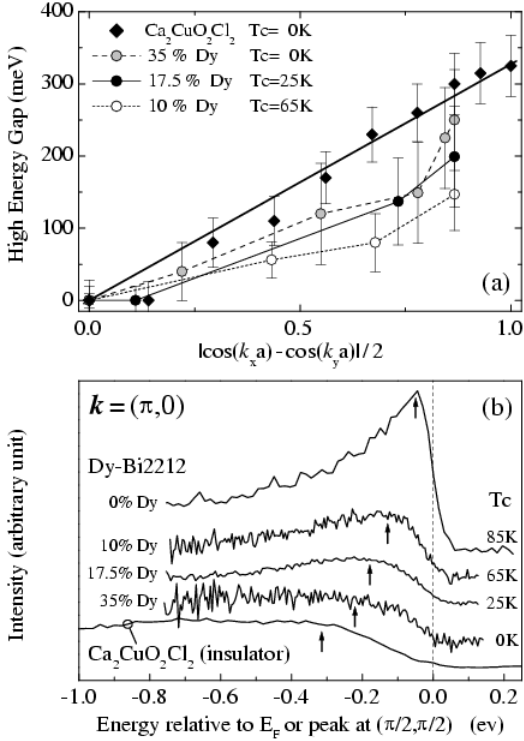,width=7cm,clip=}}
\vspace{0cm} \caption{(a) High energy pseudogap plotted versus
$|\!\cos k_x\!-\! \cos k_y|/2$ for CCOC and Dy-Bi2212. (b) Doping
dependence of the ($\pi$,0) ARPES spectra measured at 100\,K with
25.2\,eV photons on CCOC, and at 110\,K with 22.4\,eV photons on
Dy-Bi2212. After
\textcite{RonningF:PhoerF}.}\label{Filip_DW}\end{figure}
The first explicit attempt in this direction was made by
\textcite{LaughlinRB:Numeed} who showed that the ($\pi$,0) ARPES
spectra evolve continuously from one system to another, making a
strong case for the high-energy broad peak being derived from the
insulator. The idea of remnant FS discussed in
Sec.\,\ref{sec:_ccocRFS} was inspired by this approach, which
suggested to consider the evolution of the ($\pi$,0) spectra in
terms of a gap evolution rather than a trivial band dispersion
even in the insulator.

The ($\pi$,0) spectra from Bi2212, Dy-Bi2212, and CCOC are
compared in Fig.\,\ref{Filip_DW}b \cite{RonningF:PhoerF}, which
shows a smooth evolution of lineshapes and peak positions (see
arrows): upon underdoping the quasiparticle peak at ($\pi$,0)
broadens and shifts to higher binding energies. Note that for CCOC
the zero in energy does not correspond to $E_F$ but to the peak
position at ($\pi$/2,$\pi$/2) which, in turn, corresponds to the
top of the valence band located $\sim\!700$ meV below $E_F$
because of the Mott gap. The high-energy pseudogap
(Fig.\,\ref{Filip_DW}b) is characterized by a $d$-wave like
dispersion not only in the underdoped systems but also in the
undoped insulator. This is shown in Fig.\,\ref{Filip_DW}a where
the dispersion of the high energy pseudogap along the FS
(remnant-FS for CCOC) is plotted against the $d$-wave functional
form (a fit for CCOC is shown). Although their sizes vary, the
superconducting gap, the pseudogap of the underdoped system, and
the gap of the insulator have a similar $d$-wave form, suggesting
a common origin \cite{ZacherMG:Intsag}. This is consistent with
the idea of one underlying symmetry principle that unifies the AF
insulator and the $d$-wave superconductor
\cite{ZhangSC:1,ZacherMG:Intsag}.

Recently, single crystals of Na-doped CCOC have been made
available (as discussed earlier in Sec.\,\ref{sec:ns_ccoc}), which
allowed the study of the doping evolution from insulator to
superconductor within the same material family
\cite{kohsaka:1,ronning:2}. The ($\pi$,0) spectra from 10\% doped
Na-CCOC (Fig.\,\ref{NArawdata}a) and CCOC (Fig.\,\ref{Filip_DW}b)
exhibit the same lineshape, although the insulator data are
shifted below $E_F$ by $\sim\!700$\,meV \cite{ronning:2}.
Furthermore, as shown in Fig.\,\ref{NAcont}b, the 10\% Na-doped
sample is characterized by a Fermi arc centered around the nodal
region, while around ($\pi$,0) the spectra are pushed to much
higher energies. This behavior is almost identical to the one seen
for underdoped Bi2212 in Fig.\,\ref{pg1}, which supports the
earlier conclusions based on the comparison between CCOC and
Bi2212 at various dopings.  Hence, these findings make a strong
case for the high-energy pseudogap to have its origin in the
electronic structure of the insulator. In addition, regarding
possible complications due to bilayer splitting effects in the
($\pi$,0) data from Bi2212, it has to be emphasized that the
similar behavior observed for single-layer Na-CCOC and bilayer
Bi2212 provides further support for the presence of two energy
scales in the pseudogap phenomenon and for their observed doping
evolution.

\subsection{Discussion}
\label{sec_psudo}

The results from Bi2201, Bi2212, and LSCO, with the presence of
two-energy scales for both single and double layer materials,
indicate the generality of the pseudogap phenomenology. The low
energy pseudogap smoothly evolves into the superconducting gap
upon going from underdoped to overdoped regime. Both gaps exhibit
the same overall $d$-wave symmetry. Furthermore, the magnitude of
the leading-edge pseudogap scales with the maximum $T_c$ of the
different HTSC families. The high-energy pseudogap, which allows a
more meaningful description of the data in the very underdoped
regime, seems to originate from the AF insulator, as first
proposed on the basis of CCOC and Bi2212 data. In this regard,
more direct evidence is provided by the results from Na-CCOC. In
particular, we have discussed that the similar $d$-wave form
observed for the superconducting gap, the pseudogap of the
underdoped system, and the angular dependent part of the gap along
the remnant FS of the insulator might suggest a common origin for
these different excitation gaps \cite{ZacherMG:Intsag}, and the
existence of a unifying symmetry principle for the AF insulator
and the $d$-wave superconductor \cite{ZhangSC:1,ZacherMG:Intsag}.
In this regard however, as discussed in detail by
\textcite{ronning:1}, an important caveat is the rounded
electronic dispersion observed near the nodes of the $d$-wave
functional form for both the superconducting and normal state gap.
This finding, which is also consistent with the observations of
Fermi arcs rather than a uniformly gapped FS in the pseudogap
regime (Fig.\,\ref{NAcont},\,\ref{pg1}, and\,\ref{pg2}), suggests
that high order harmonics may have to be included in the expansion
of the gap function \cite{MesotJ:Supgaq,Guinea:1}, as already
discussed in Sec.\,\ref{sec:supgapBi2212}
and\,\ref{sec:sgdiscussion}. This would in fact account for the
lack of a well defined ``V''-like cusp in the gap momentum
dependence in the nodal region. The presence or absence thereof is
a critical information in discriminating between different
microscopic models connecting the AF insulator to the $d$-wave
superconductor.

It has to be noted that the detection of a $d$-wave like pseudogap
and, in particular, the increase of its maximum amplitude at
($\pi$,0) upon reducing doping emphasize the inadequacy of a rigid
band description for the electronic structure of the HTSCs
\cite{KingDM:EleseM}. In other words, the evolution from insulator
to overdoped metal cannot be described in terms of the simple
shift of the chemical potential in a rigid band picture. In fact,
in the insulator the binding energies of the lowest energy states
at ($\pi$/2,$\pi$/2) and ($\pi$,0) differ by $\sim\!300$\,meV as
shown in Fig.\,\ref{Filip_cartoon}, while in the overdoped metal a
Fermi level crossing can be found near both momenta [the rigid
chemical potential shift would instead results in FS pockets
closed around ($\pi$/2,$\pi$/2)].

\section{Self energy and collective modes}
\label{sec:selfmodes}

As discussed in Sec.\,\ref{sec:spectral}, the introduction of the
electron self energy $\Sigma({\bf
k},\omega)\!=\!\Sigma^{\prime}({\bf
k},\omega)\!+\!i\Sigma^{\prime\prime}({\bf k},\omega)$ is a
powerful way to account for many-body correlations in solids. Its
real and imaginary parts correspond, respectively, to the energy
renormalization with respect to the bare band energy
$\epsilon_{\bf k}$ and to the finite lifetime of the
quasiparticles in the interacting system. Owing to the energy and
momentum resolution nowadays achievable
(Sec.\,\ref{sec:stateart}), both components of the self energy can
be in principle estimated very accurately from the analysis of the
ARPES intensity in terms of {\it energy distribution curves}
(EDCs) and/or {\it momentum distribution curves}
(MDCs).\footnote{This is one of the aspects that make ARPES such a
powerful tool for the investigation of complex materials, as
exemplified by the recently reported experimental determinations
of many-body effects in different systems. For a review, see:
\textcite{JohnsonPD:Phoss-,KevanSD:Manbes,GweonG-H:ARPlsF}.} In
this regard, it is important to realize that in some cases the MDC
analysis may be more effective than the analysis of the EDCs in
extracting information on the self energy, as noted by
\textcite{VallaT:Eviqcb} who first used this approach for Bi2212
(see Fig.\,\ref{valla1} and\,\ref{vallamdc}). EDCs are typically
characterized by a complex lineshape (Fig.\,\ref{valla1}) because
of the nontrivial $\omega$ dependence of the self energy (see,
e.g., Eq.\,\ref{eq:seFL} and\,\ref{eq:seMFL}), the presence of
additional background (Sec.\,\ref{sec:matele}), and the low-energy
cutoff due to the Fermi function. Furthermore, as evidenced by the
generic expression for the spectral function $A({\bf k},\omega)$
in Eq.\,\ref{eq:akw1}, the EDC peak position is determined by
$\Sigma^{\prime}({\bf k},\omega)$ as well as
$\Sigma^{\prime\prime}({\bf k},\omega)$, because both terms are
strongly energy dependent. On the other hand, if the self energy
is independent of $k$ normal to the FS (and the matrix elements
are a slowly-varying function of $k$), then the corresponding MDCs
are simple Lorentzians centered at
$k\!=\!k_F\!+\![\omega\!-\!\Sigma^\prime(\omega)]/v_F^0$ with FWHM
given by $2\Sigma^{\prime\prime}(\omega)/v_F^0$, where $v_F^0$ is
the bare Fermi velocity normal to the FS.\footnote{This is
obtained by simply approximating $\epsilon_{\bf
k}\!\simeq\!v_F^0(k\!-\!k_F)$ in the spectral function $A({\bf
k},\omega)$ in Eq.\,\ref{eq:akw1} \cite{KaminskiA:Rensls}.}
Indeed, Lorentzian lineshapes were observed experimentally for the
MDCs (Fig.\,\ref{valla1} and\,\ref{vallamdc}), and this approach
has been extensively used in the literature as we will see
throughout this section. However, it has to be noted that,
although the results of MDC and EDC analyses should coincide,
differences in both dispersions and peak widths can be observed,
in particular at high energies due to the $\omega$ dependence of
$\Sigma({\bf k},\omega)$, or near the band maxima and minima (see,
e.g., the simulations in Fig.\,\ref{mn_mode}).

As we will elaborate in Sec.\,\ref{sec:selfcupr} and
\ref{sec:energyscale}, the peak-dip-hump structure and the
corresponding step-edge in $\Sigma^{\prime\prime}({\bf k},\omega)$
observed for the cuprates are crucial hallmarks for quasiparticles
interacting with a dispersionless collective bosonic mode. This
problem has been studied in great detail for the strong-coupling
BCS superconductors, in which the bosonic mode is an Einstein
phonon \cite{engelsbergs;1,schrieffer:1,scalapino:1}. Therefore,
before discussing the specific case of the cuprate HTSCs, in the
following section we will briefly review the effects of
electron-phonon coupling on the ARPES spectra from simple metallic
surfaces for which, given that the electron-electron interaction
is relatively unimportant, the established theoretical formalism
can be applied very
effectively.\footnote{\textcite{BalasubramanianT:Larvte,HofmannP:Ele-la,HengsbergerM:Phossc,HengsbergerM:Ele-ph,VallaT:Eviqcb,LaShellS:Nonstp,RotenbergE:Coubav}.}

\subsection{Electron-phonon coupling on metallic surfaces}
\label{sec:elphcoupl}

\begin{figure}[b!]
\centerline{\epsfig{figure=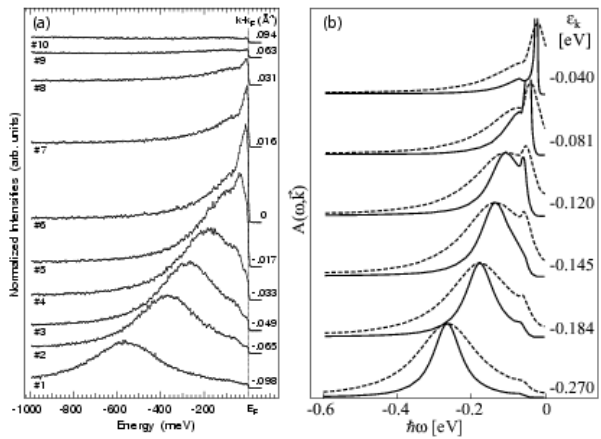,clip=,angle=0,width=1\linewidth}}
\caption{(a) ARPES spectra for the Be(0001) surface state measured
at 12\,K with (HeI) 21.2\,eV photons [after
\textcite{HengsbergerM:Phossc}]. (b) Spectral function within the
Debye model for $\omega_D\!=\!65$\,meV and $\lambda\!=\!0.65$ with
(dashed) and without (solid) impurity scattering [after
\textcite{LaShellS:Nonstp}].} \label{baer}
\end{figure}

\begin{figure}[t!]
\centerline{\epsfig{figure=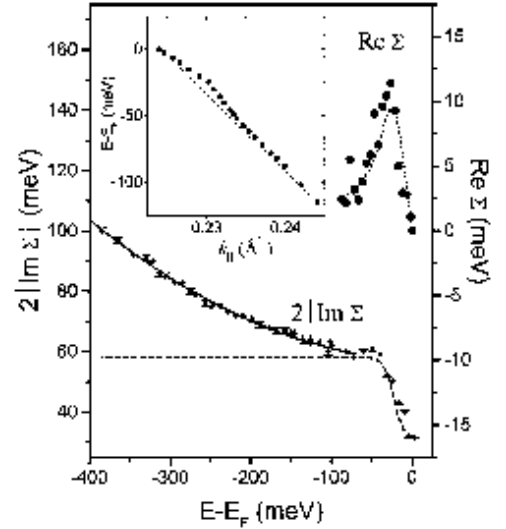,clip=,angle=0,width=7cm}}
\caption{Self energy estimated from the Mo(110) surface state
ARPES spectra, and corresponding quasiparticle dispersion (inset).
Calculated electron-phonon contributions to real and imaginary
part of $\Sigma({\bf k},\omega)$, respectively, are indicated by
dotted and dashed lines (the latter was offset by 26\,meV to
account for impurity scattering). After \textcite{VallaT:Man-bo}.}
 \label{valla0}
\end{figure}

The electron-phonon interaction involving surface phonons and the
$\overline{\Gamma}$-surface state on the Be(0001) surface was
investigated by two groups, and qualitatively similar conclusion
were drawn
\cite{BalasubramanianT:Larvte,HengsbergerM:Phossc,HengsbergerM:Ele-ph,LaShellS:Nonstp}.
Fig.\,\ref{baer}a shows results by \textcite{HengsbergerM:Phossc}
for the Be(0001) surface state along the $\overline{\Gamma M}$
direction of the surface Brillouin zone; a feature is seen
dispersing towards the Fermi level. Close to $E_F$ the spectral
function exhibits a complex structure characterized by a broad
hump and a sharp peak, with the latter being confined to within an
energy range given by the typical bandwidth $\omega_{ph}$ of the
surface phonons.
This behavior corresponds to a two-branch splitting of the
near-$E_F$ dispersion, with a transfer of spectral weight between
the two branches as a function of binding energy (this is more
clearly emphasized by the simulation shown in Fig.\,\ref{baer}b).
While the high-energy dispersion is representative of the bare
quasiparticles, at low energy the dispersion is renormalized by
the electron-phonon interaction (this behavior is shown, for a
similar electron-phonon coupled system, in the inset of
Fig.\,\ref{valla0}).  In other words, the weaker dispersion
observed at energies smaller than $\omega_{ph}$ describes dressed
quasiparticles with an effective mass enhanced by a factor of
$(1\!+\!\lambda)$, where $\lambda$ is the electron-phonon coupling
parameter \cite{Ashcroft:1}. The latter can be estimated from the
ratio of renormalized (${\bf v}_{{\bf k}}^{ }$) and bare (${\bf
v}_{{\bf k}}^{0}$) quasiparticle velocities, according to the
relation ${\bf v}_{\bf k}^{
}\!=\!\hbar^{-1}\partial\varepsilon_{\bf k}/\partial {\bf k}
\!=\!(1\!+\!\lambda)^{-1}{\bf v}_{\bf k}^{0}$. This way, for the
data presented in Fig.\,\ref{baer}a the value $\lambda\!=\!1.18$
was obtained [note that alternatively $\lambda$ could also be
estimated from the temperature dependence of the linewidth near
$E_F$ \cite{BalasubramanianT:Larvte}]. In a follow up paper,
\textcite{HengsbergerM:Phossc} provided a more detailed analysis
of the data, by including also the effect of impurities and
electron-electron interactions, which however led to similar
conclusions.

A slightly different analysis of similar data from Be(0001) was
performed by \textcite{LaShellS:Nonstp}. In this case, the authors
used the isotropic zero temperature Debye model with a constant
electron-phonon interaction matrix element to express the
Eliashberg coupling function (i.e., the phonon density of states
weighted by the electron-phonon coupling strength) as
$\alpha^2F(\omega)\!=\!\lambda(\omega/\omega_D)^2$, for
$\omega\!<\!\omega_D$, and otherwise zero (here $\omega_D$ is the
maximum phonon energy and the dimensionless mass renormalization
parameter $\lambda$ represents the coupling strength).
Within the above approximation the electron-phonon contribution to
the self energy $\Sigma({\bf k},\omega)$, if effectively treated
as momentum independent, can be calculated from
$|\Sigma^{\prime\prime}(\omega)|\!=\!\pi\hbar\int\alpha^2F(\omega^\prime)d\omega^\prime$.
For real ($\Sigma^{\prime}$) and imaginary
($\Sigma^{\prime\prime}$) parts one obtains:
\parbox{7cm}{\begin{eqnarray*}
\Sigma^{\prime}(\omega)&=&
-(\lambda\hbar\omega_D/3)\!\times\![(\omega/\omega_D)^3\ln|(\omega_D^2\!-\!\omega^2)/\omega^2|\\
&\,&+\ln|(\omega_D\!+\!\omega)/(\omega_D\!-\!\omega)|+\omega/\omega_D]
\\
\\
|\Sigma^{\prime\prime}(\omega)|&=&
\hbar\lambda\pi|\omega|^3/(3\,\omega_D^2)\, , \hspace{.66cm}
|\omega|<\omega_D
\\
\\
|\Sigma^{\prime\prime}(\omega)|&=& \hbar\lambda\pi\omega_D/3\, ,
\hspace{1.575cm} |\omega|>\omega_D\,
.
\end{eqnarray*}}\hfill
\parbox{1cm}{\vspace{0.5cm}\begin{eqnarray}\end{eqnarray}}
The spectral function $A({\bf k},\omega)$ for this model is shown
in Fig.\,\ref{baer}b for parameter values $\lambda\!=\!0.65$ and
$\omega_D\!=\!65$\,meV; it qualitatively reproduces the basic
features of the ARPES spectra presented in Fig\,\ref{baer}a, and
in particular the double structure with a dip at approximately
$\omega_D$. By assuming the self energy to be independent of (or
weakly dependent on) momentum, a fact that is qualitatively
consistent with the Lorentzian lineshape of the MDCs (see previous
section), \textcite{LaShellS:Nonstp} extracted the self energy
using only the free parameter $\epsilon_k$, which they took as a
second-order polynomial imposing the additional condition that
$\Sigma^{\prime}(\omega)$ must vanish at $\omega\!=\!0$ and
$\omega\!\rightarrow\!\infty$. These results are in good agreement
with theoretical simulations based on the Debye model.

Another example of an electron-phonon coupled system is the
surface state of Mo(110) studied by \textcite{VallaT:Man-bo}. In
this case, the real and imaginary part of the self energy shown in
Fig.\,\ref{valla0} were obtained directly from the EDC analysis:
$\Sigma^{\prime\prime}$ corresponds to the EDC width and
$\Sigma^{\prime}$ to the difference between the observed
quasiparticle dispersion and a straight line approximating the
dispersion of the non-interacting system (Fig.\,\ref{valla0},
inset). The step-like change at 30 meV in $\Sigma^{\prime\prime}$
is interpreted as the phonon contribution (dashed line) and the
parabolic part at higher energies is attributed to
electron-electron interactions. The phonon contribution to the
real part of the self energy is calculated from the Kramers-Kronig
relations (see Sec.\,\ref{sec:spectral}) and agrees well with the
data (dotted line). As an additional confirmation of the
electron-phonon description, it was noted that also the observed
temperature dependence of the scattering rate is well reproduced
by the calculations \cite{VallaT:Man-bo}. The difference between
the Mo(110) results and those from Be(0001), or the simulations
presented in Fig.\,\ref{baer}b, is that no double structure is
seen directly in the EDCs for the Mo(110) surface state, although
one should expect it for the parameters extracted from the
experiment. The reason for this discrepancy probably lies in the
energy scale of the Debye frequency, which is about 30 and 65\,meV
for Mo and Be, respectively. For $\omega_D\!=\!30$\,meV the
complex lineshape is probably smeared out because of impurities or
finite resolution.

A similar manifestation of two-branch splitting in the ARPES
spectra as a consequence of electron-phonon coupling was observed
for hydrogen and deuterium adsorbed onto the W(110) surface
\cite{RotenbergE:Coubav}. In this case, the direct coupling
between an adsorbate optical phonon and an intrinsic surface
electronic state was evidenced by the hydrogen/deuterium isotope
effect.

\subsection{Self-energy effects in the HTSCs}
\label{sec:selfcupr}

\subsubsection{The ($\pi$,0) region}
\label{sec:pi0reg}

Earlier ARPES investigation of the self-energy corrections in the
cuprates focused on the antinodal ($\pi$,0) region of Bi2212,
where across $T_c$ the lineshape evolves into the well-known
peak-dip-hump structure already discussed in
Sec.\,\ref{sec:speak_Bi2212}. However, as recently recognized and
elaborated in detail in Sec.\,\ref{sec:bilayer}, the ($\pi$,0)
spectra from Bi2212 are severely distorted by the bilayer
splitting
effects.\footnote{\textcite{FengDL:Bilste,FengDL:3,ChuangY-D:Doutbo,ChuangY-D:1,kordyuk:1,gromko:2,BorisenkoSV:1}.}
In particular, the results reported by \textcite{kordyuk:1} and
\textcite{gromko:2} most convincingly suggested that the peak and
hump observed in the superconducting state correspond to
antibonding and bonding bilayer split bands, at {\it any} doping
levels (i.e., even at under and optimal doping). This invalidates
much of the quantitative lineshape analysis of data from this
momentum-space region.
\begin{figure}[t!]
\centerline{\epsfig{figure=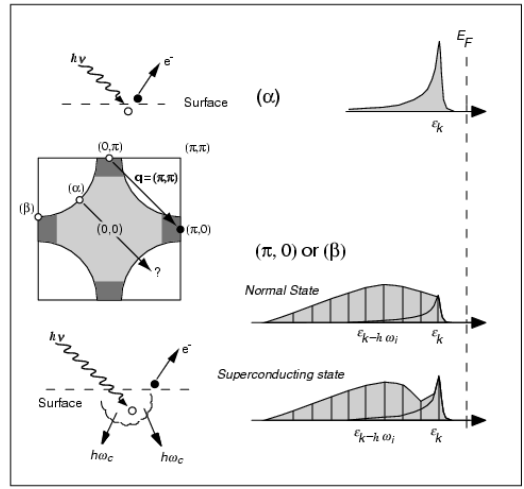,clip=,angle=0,width=7.5cm}}
\caption{Photoemission processes and corresponding lineshapes for
weak ($\alpha$) and strong coupling [$\beta$ or ($\pi$,0)]. Phase
space considerations for quasiparticles coupled to ($\pi$,$\pi$)
collective modes are also sketched \cite{ShenZ-X:Momtdd}.}
\label{zxmode}
\end{figure}
However, as the methodology is still very instructive, in the
following we will nevertheless review the literature devoted to
this issue.

As for the detailed understanding of the peak-dip-hump structure
seen at ($\pi$,0) in the superconducting state, there are two
distinct aspects of this effort:

\noindent (i) The development of a way to extract self-energy
information directly from the ARPES data. The upshot of this
approach is that one does not need input from any microscopic
model, thus it may in principle allow one to analyze the data with
no preconceived judgment. The price to pay is that certain
approximations are necessary and, as a consequence, one has to be
cautious in the quantitative evaluation of the final
results.\footnote{\textcite{NormanMR:Unudls,NormanMR:Colms-,NormanMR:Phetl-,NormanMR:Magcmd,NormanMR:Exttes}.}

\noindent (ii) The analysis of the photoemission data within the
context of a specific microscopic
model.\footnote{\textcite{NormanMR:Unudls,NormanMR:Magcmd,ShenZ-X:Momtdd,NormanMR:Colms-,CampuzanoJC:Elestr,AbanovA:relbrn,EschrigM:Neurmp,EschrigM:Review}.}
Within this approach, in analogy to the case of the
electron-phonon coupled systems discussed in
Sec.\,\ref{sec:elphcoupl}, the peak-dip-hump structure would
result from the coupling between the quasiparticles and a
collective bosonic mode. In this case the latter is assumed to be
electronic in nature and, in particular, the so-called
$Q\!=\!(\pi$,$\pi$) resonant magnetic mode observed in inelastic
neutron scattering experiments in YBCO
\cite{Rossat-MignodJ:Neusst,MookHA:Polndt,HungFaiFong:Phomns} and
Bi2212 \cite{mook:1,FongHF:Neusme}. Related to this scenario is
the approach schematically summarized in Fig.\,\ref{zxmode}, which
emphasizes that the ${\rm Q}\!\simeq\!(\pi,\pi)$ scattering will
severely affect the ($\pi$,0) spectra (i.e., broadening due to the
simultaneous excitation of collective modes). As a consequence of
the sharpening of the quasiparticle pole in the superconducting
state, a sharp peak would eventually emerge below $T_c$, as
observed in the experiments.

\begin{figure}[b!]
\centerline{\epsfig{figure=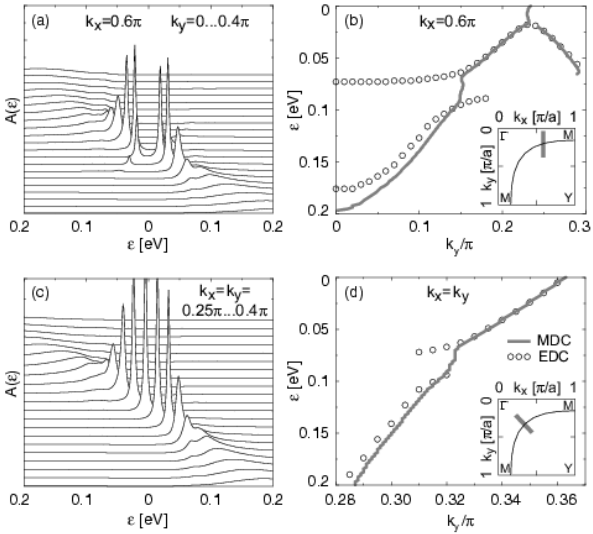,clip=,angle=0,width=8.5cm}}
\caption{Resonant magnetic-mode model calculations: (a,c) electron
removal and addition spectra and (b,d) corresponding MDC and EDC
derived dispersions, for momentum-space cuts parallel to $M$-$Y$
(top panels) and to the nodal direction $\Gamma$-$Y$ (bottom
panels). After \textcite{EschrigM:Review}.} \label{mn_mode}
\end{figure}

It should be noted that the connection between the ($\pi$,0)
peak-dip-hump structure and a sharp electronic resonant mode in
the cuprates was first proposed by \textcite{NormanMR:Unudls},
following earlier work on electron-phonon coupling
\cite{engelsbergs;1,scalapino:1}. Fig.\,\ref{mn_mode} reproduces
the calculations performed at two different momenta in the
superconducting state for electrons coupled to a collective mode
of energy $\Omega_{res}$ as well as a gapped continuum, which
account for the complete spin excitation spectrum
\cite{EschrigM:Review}.
As a result of the interaction, the spectral function is
characterized by a peak-dip-hump-like lineshape as well as a {\it
break} and {\it two-branch} behavior (i.e., for peak and hump) in
the EDC-derived quasiparticle dispersion [the energy of dip and
dispersion kink is comparable to the one of the ($\pi$,0)
superconducting state peak plus $\Omega_{res}$]. This behavior is
particularly evident in the ($\pi$,0) region
(Fig.\,\ref{mn_mode}a-b), but similar although weaker effects can
also be seen along the nodal direction (Fig.\,\ref{mn_mode}c-d).
An obvious difference between the two momentum space regions in
Fig.\,\ref{mn_mode} comes from the presence of the $d$-wave
superconducting gap: while the quasiparticle peak crosses $E_F$
with a linear dispersion along (0,0)-($\pi$,$\pi$), it disperses
backward losing spectral weight for $k\!>\!k_F$ around ($\pi$,0),
where the gap along the FS is maximum. In Fig.\,\ref{mn_mode} it
should also be noted that the two-branch behavior seen in the EDCs
corresponds to a {\it kink} in the MDC-derived quasiparticle
dispersion, which emphasizes the crossover between high and
low-energy branches.

\begin{figure}[t!]
\centerline{\epsfig{figure=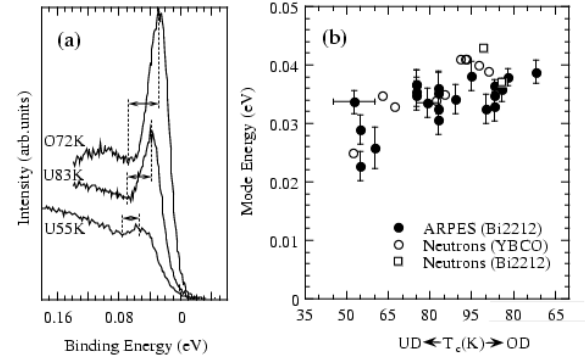,width=7.75cm,clip=}}
\vspace{0cm} \caption{Doping dependence of: (a) the ($\pi$,0)
ARPES spectra from Bi2212; (b) the collective mode energy as
inferred from the ARPES spectra (i.e., difference between peak and
dip positions) and neutron data. After
\textcite{CampuzanoJC:Elestr}.} \label{campu2}
\end{figure}

The agreement between the above calculations and ARPES data such
as those presented in Fig.\,\ref{norma2},\,\ref{pg6},
and\,\ref{kami2} was taken as evidence for the validity of the
approach.  Furthermore, as mentioned above, according to the
magnetic-mode scenario the energy separation between peak and dip
in the ARPES spectra (Fig.\,\ref{campu2}a) should correspond to
the energy $\Omega_{res}$ of the mode itself
\cite{CampuzanoJC:Elestr}.
Indeed, some correlation between these characteristic energies was
experimentally observed over a broad doping range, as shown in
Fig.\,\ref{campu2}b where ARPES and inelastic neutron scattering
data are compared. \textcite{CampuzanoJC:Elestr} argued that these
results, which have also been confirmed by more recent
photoemission and neutron data reported for the very same sample
by \textcite{mesot:1}, provide direct evidence for the bosonic
mode being the ($\pi$,$\pi$) resonant magnetic mode. However, in
light of the recent detection of bilayer splitting effects that
dominate the ($\pi$,0) spectra (with an energy scale coinciding
with that of peak and hump), also this quantitative comparison of
neutrons and Bi2212 ARPES data needs to be carefully reevaluated
[see, e.g., \textcite{EschrigM:Review}].

\subsubsection{The nodal direction}
\label{sec:selfnodal}

\begin{figure}[t!]
\centerline{\epsfig{figure=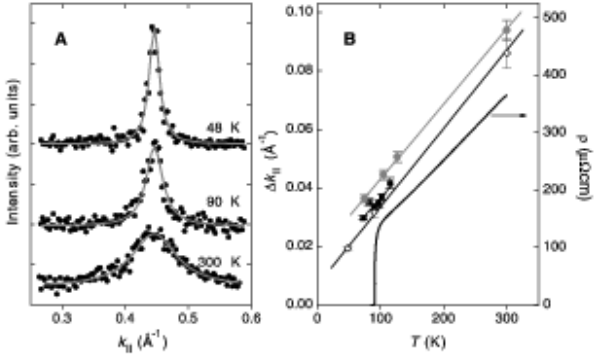,clip=,angle=0,width=8.5cm}}
\caption{(a) Temperature dependent MDCs at $E_F$ measured along
the nodal direction on optimally doped Bi2212 ($T_c\!=\!91$\,K),
with 21.2\,eV photons. (b) Corresponding MDC widths versus
temperature for three different samples (symbols), together with
DC resistivity data \cite{VallaT:Eviqcb}.} \label{vallamdc}
\end{figure}

As the data from Bi2212 along the (0,0)-($\pi$,$\pi$) direction
are not complicated by superstructure contaminations (at least
along $\Gamma Y$, as shown in Fig.\,\ref{ding2}
and\,\ref{fretwell1}a,b) or bilayer splitting effects (see
Fig.\,\ref{pasha_bilayer} and\,\ref{bilayerfs}), in principle the
quasiparticle lineshape analysis in this region of momentum space
is more reliable. In this context, detailed ARPES studies of the
single-particle self energy have been very recently reported by
several
groups.\footnote{\textcite{VallaT:Eviqcb,VallaT:Temdsr,BogdanovPV:Eviesq,KaminskiA:Rensls,LanzaraA:Eviuse,yusof:1,KaminskiA:Quatss,JohnsonPD:Doptdt}.}
Interestingly, from a temperature dependence study on optimally
doped Bi2212 ($T\!=\!48$\,K ARPES data are shown in
Fig.\,\ref{valla1}), \textcite{VallaT:Eviqcb} reported that the
width of the Lorentzian-like MDCs at $E_F$ decreases linearly as a
function of temperature (Fig.\,\ref{vallamdc}), similar to what is
observed for the scattering rate in normal state resistivity data
and consistent with the marginal Fermi liquid (MFL) description
[\textcite{VarmaCM:Phetns}; see Sec.\,\ref{sec:spectral}].
Furthermore, this behavior seemed to smoothly persist into the
superconducting state in contrast to what is seen by transport and
optical measurements. The linear dependence of the scattering rate
on temperature (at small binding energies) and on energy (at large
binding energies), with no indications for an extraneous energy
scale set by conventional sources of quasiparticle scattering, was
considered by \textcite{VallaT:Eviqcb} a signature of a nearby
quantum critical point.\footnote{See for example:
\textcite{ChakravartyS:Two-di,VarmaCM:Phetns,VarmaCM:Non-Fe,LittlewoodPB:Phetns,SachdevS:Uniq-c,SokolA:Towump,EmeryVJ:SoloKa,CastellaniC:Sinqst,ChakravartyS:Hidotc}.}

\begin{figure}[b!]
\centerline{\epsfig{figure=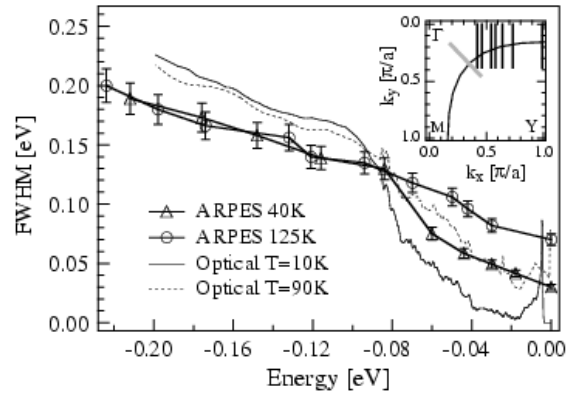,clip=,angle=0,width=8cm}}
\caption{EDC width versus the quasiparticle binding energy for
optimally doped Bi2212 ($T_c\!=\!89$\,K), from along the nodal
direction (thick gray line in the Brillouin zone sketch), and
frequency dependent scattering rate from infrared reflectivity
data \cite{PuchkovAV:psesh-}. After \textcite{KaminskiA:Quatss}.}
\label{kaminski}
\end{figure}

However, evidences for an additional energy scale in the
quasiparticle self energy, at least below $T_c$, were later
reported by two other groups
\cite{KaminskiA:Quatss,BogdanovPV:Eviesq,LanzaraA:Eviuse}.
\textcite{KaminskiA:Quatss} observed a clear drop in the
low-energy scattering rate in particular below $T_c$, as estimated
from the width of the EDCs measured along the nodal direction on
optimally doped Bi2212. This behavior is shown in
Fig.\,\ref{kaminski} where the scattering rate obtained from ARPES
and optical experiments are compared for two different
temperatures above and below $T_c$.  In addition, from the
analysis of both EDCs and MDCs taken on Bi2212 for several dopings
along the (0,0)-($\pi$,$\pi$) direction,
\textcite{BogdanovPV:Eviesq} found a kink at approximately
$50\pm15$ meV in the quasiparticle dispersion (see
Fig.\,\ref{ale_1} which shows, in addition to the Bi2212 results,
similar data reported by \textcite{LanzaraA:Eviuse} for a wide
range of HTSCs). This is counter to the linear dispersion
predicted by band-structure calculations in this energy range
\cite{MassiddaS:ElespB,KrakauerH:Effbh-}. The kink appeared to be
more pronounced at low temperatures and in underdoped samples;
below $T_c$, a drop in the EDC-derived quasiparticle scattering
rate was observed at the kink energy, similar to that shown in
Fig.\,\ref{kaminski}.

It should be noted that a kink in the dispersion is also present
in the superconducting state results by \textcite{VallaT:Eviqcb},
although this aspect of the data was not considered by those
authors (see white circles in Fig.\,\ref{valla1}).
However, as real and imaginary part of the self energy are related
by Kramers-Kronig relations, the presence of a dispersion
renormalization below a certain binding energy corresponds
necessarily to a well defined structure in the quasiparticle
scattering rate, contrary to the initial claim of the linear
temperature dependence of the scattering rate and absence of any
energy scale [as later recognized by those same authors
\cite{JohnsonPD:Doptdt}].

\subsection{Origin of the energy scale in the HTSCs}
\label{sec:energyscale}

While there has been an initial disagreement with one group
arguing for the absence of any energy scale at all temperatures
\cite{VallaT:Eviqcb}, the field eventually converged to the
conclusion that the kink in the dispersion and the change in the
EDC width observed along the nodal direction indicate the presence
of a well defined energy scale in the electron self energy for
Bi2212, at least in the superconducting state.
\begin{figure}[b!]
\centerline{\epsfig{figure=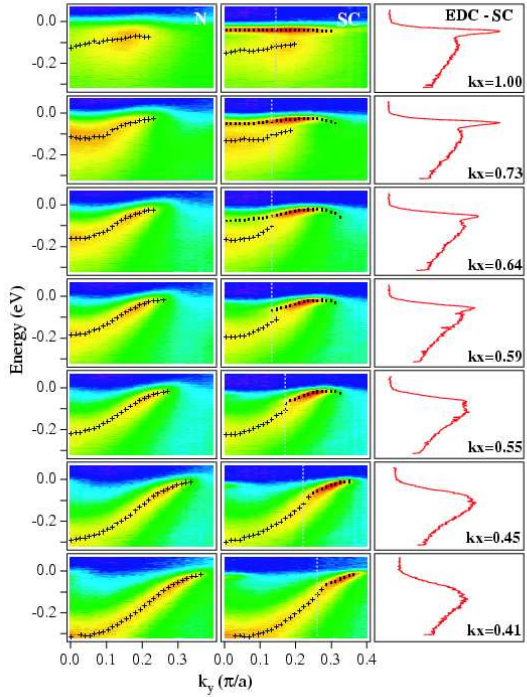,clip=,angle=0,width=7cm}}
\caption{(left) Normal ($T\!=\!140$\,K) and (middle)
superconducting state ($T\!=\!40$\,K) ARPES intensity measured on
optimally doped Bi2212 ($T_c\!=\!89$\,K) with 22\,eV photons (the
location of the cuts in the Brillouin zone is indicated by the
black lines in the sketch of Fig.\,\ref{kaminski}). Dots and
crosses mark the position of peak and hump, respectively. (right)
Superconducting state EDCs from the momenta indicated by the
dashed lines in the middle panels. After
\textcite{KaminskiA:Rensls} [Color].} \label{kami2}
\end{figure}
Furthermore, the coupling between the quasiparticles and some
collective mode is considered to be the most likely origin of this
phenomenon
\cite{BogdanovPV:Eviesq,KaminskiA:Quatss,KaminskiA:Rensls,LanzaraA:Eviuse,JohnsonPD:Doptdt}.
In fact, other possible causes, such as the opening of the
superconducting gap, have been ruled out as will be elaborated
later. However, there is currently no consensus on whether or not
these effects are limited to temperatures lower than $T_c$ and, in
turn, on the precise nature of the collective mode. As for the
latter, in addition to the ($\pi$,$\pi$) resonant magnetic-mode
scenario already discussed in relation to the self-energy effects
in the ($\pi$,0) region, also lattice vibrations have been
proposed as a possible candidate. A direct comparison of these two
competing scenarios will be presented in the following.

\subsubsection{Resonant magnetic-mode scenario}

In the context of a coupling between quasiparticles and a ${\rm
Q}\!=\!(\pi,\pi)$ mode, the strongest effects on the self energy
are expected around ($\pi$,0), as emphasized in the inset of
Fig.\,\ref{zxmode}. In addition, as already discussed in relation
to Fig.\,\ref{mn_mode}, \textcite{KaminskiA:Rensls} argued that
the coupling between quasiparticles and spin fluctuations could
also be responsible for the drop in the scattering rate and the
dispersion renormalization observed in the superconducting state
of Bi2212 along the (0,0)-($\pi$,0) direction (Fig\,\ref{kaminski}
and \ref{kami2} for $k_x\!=\!0.41$).
\begin{figure}[t!]
\centerline{\epsfig{figure=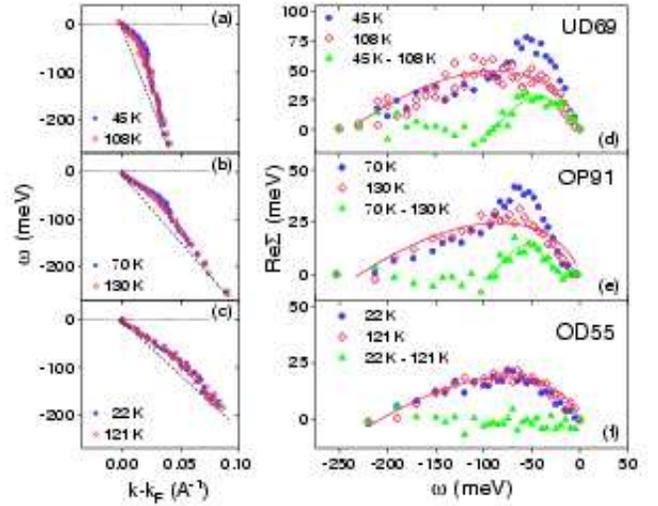,width=0.99\linewidth,clip=}}
\vspace{0cm} \caption{(a-c) MDC-derived quasiparticle dispersion
along the (0,0)-($\pi$,$\pi$) direction for underdoped, optimally
and overdoped Bi2212 above (red) and below (blue) $T_c$. (d-f)
Corresponding $\Sigma^{\prime}$ (green symbols: difference between
superconducting and normal state results). Red lines: marginal FL
fits to the data. After \textcite{JohnsonPD:Doptdt}
[Color].}\label{johnson1}\end{figure}
Within this approach
\cite{KaminskiA:Rensls,EschrigM:Neurmp,EschrigM:Review} a unified
view is given for the peak-dip-hump structure and two-branch
behavior detected in the ($\pi$,0) region (Fig\,\ref{kami2}, top
panels), and the quasiparticle renormalization observed along the
nodal direction (Fig\,\ref{kami2}, bottom panels). However, as
already discussed in Sec.\,\ref{sec:pi0reg}, an important caveat
is that the effects identified as due to self-energy corrections
in the ($\pi$,0) region could mostly be a direct manifestation of
bonding and antibonding bilayer split bands.

Similar results and conclusions were later reported by
\textcite{JohnsonPD:Doptdt}, who investigated the self-energy
corrections along the nodal direction of Bi2212 for different
doping levels, above and below $T_c$. Fig.\,\ref{johnson1} shows
the MDC-derived quasiparticle dispersion and an approximate
estimate of the real part of the self energy. This was obtained
from the difference between the MDC-derived dispersion and the
dashed straight lines that in Fig.\,\ref{johnson1}a-c intersect
the dispersion at $k_F$ and at 200-250\,meV, where
$\Sigma^{\prime}$ is then set to zero (strictly speaking, however,
the true noninteracting dispersion should be used as a reference,
which would not cross the interacting one). Although these results
are not characterized by a strong temperature dependence (i.e.,
the main deviation from the dashed lines in
Fig.\,\ref{johnson1}a-c is already present in the normal state),
\textcite{JohnsonPD:Doptdt} argued that, while the normal state
dispersion shows only a smooth MFL-like renormalization (as
defined in Sec.\,\ref{sec:spectral}), in the superconducting state
the self energy is characterized by the emergence of a sharp
structure for under and optimal doping. On the basis of this
temperature and doping dependence, \textcite{JohnsonPD:Doptdt}
attributed the self-energy change across $T_c$ (green symbols in
Fig.\,\ref{johnson1}) to the magnetic-mode contribution.
Interestingly, it was also noted that for $\omega\!\geq\!50$\,meV
the band velocity decreases with doping and that in the underdoped
regime both EDCs and MDCs widths are larger than the corresponding
binding energy. These results were discussed by
\textcite{JohnsonPD:Doptdt} as indicative of electron (hole)
fractionalization, following an earlier work by
\textcite{OrgadD:Eviefp}.

\subsubsection{Electron-phonon coupling scenario}

\begin{figure}[t!]
\centerline{\epsfig{figure=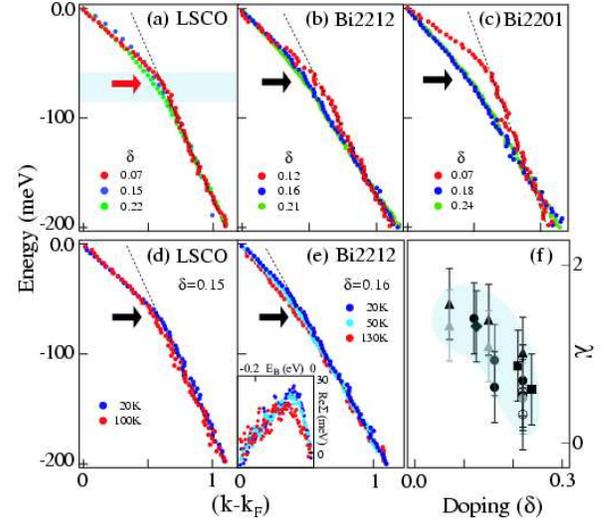,width=8.3cm,clip=}}
\caption{Quasiparticle dispersion of Bi2212, Bi2201 and LSCO along
the nodal direction, plotted versus the rescaled momentum for
different dopings (a-c) and temperatures (d,e). Black arrows
indicate the position of the kink and the red arrow the energy of
the ${\rm q}\!=\!(\pi,0)$ oxygen stretching phonon mode. Inset of
(e): temperature dependent $\Sigma^{\prime}$ for optimally doped
Bi2212. (f) Doping dependence of $\lambda^\prime$ along
(0,0)-($\pi$,$\pi$) for all the different cuprates
\cite{LanzaraA:Eviuse} [Color].}\label{ale_1}\end{figure}

Alternative to the magnetic-mode scenario discussed in the
previous section, \textcite{LanzaraA:Eviuse} and
\textcite{ShenZ-X:1} argued that the self-energy effects in the
ARPES data from Bi2212 along the nodal direction can be best
described in terms of electron-phonon coupling. This
interpretation is based on the following observations. First, the
kink in the quasiparticle dispersion as well as the drop in the
scattering rate are seen in all the $p$-type cuprates studied and
the behavior is almost identical. This is shown in
Fig\,\ref{ale_1}a-c that present the dispersion along
(0,0)-($\pi$,$\pi$) for LSCO, Bi2212, and Bi2201. In all cases one
can observe a clear change in band velocity, which becomes more
pronounced in the underdoped samples.\footnote{For the $n$-type
cuprates no detailed study is available; at this stage, as shown
in Fig.\,\ref{Sato_EDC},  no effect is detected on NCCO along the
nodal direction \cite{SatoT:Obsdx,ShenZ-X:1}, while a break in the
dispersion at about 50\,meV was observed along the
($\pi$,0)-($\pi$,$\pi$) direction \cite{SatoT:Obsdx}.} Note that
the observation of a similar energy scale (50-80\,meV) in systems
characterized by very different gap energies (10-20\,meV for LSCO
and Bi2201, 30-50\,meV for Bi2212) rules out the superconducting
gap as a possible origin. Furthermore, this finding was taken by
\textcite{LanzaraA:Eviuse} also as a strong evidence against the
magnetic-mode interpretation and in favor of the phonon scenario.
In fact, while the kink in the dispersion appears to be similar in
the different families of compounds, the magnetic resonant mode
behaves very differently: the latter is not seen in LSCO, and is
expected to have an energy three times lower in Bi2201 as compared
to Bi2212 or YBCO, because its characteristic energy scales
approximately with $T_c$.
On the contrary, the similar energy scale observed for the
different compounds is consistent with the coupling between
quasiparticles and oxygen phonons: as evidenced by neutron
scattering experiments \cite{McQueeneyRJ:AnodLp,petrov:1}, these
lattice vibrational modes are strongly coupled to the charge
carriers and their characteristic frequencies fall in the same
energy range of the kink observed in the quasiparticle dispersion
(black arrows in Fig\,\ref{ale_1}b-e). For instance, for LSCO the
($\pi$,0) oxygen stretching mode has an energy of 70 meV as
indicated by the red arrow in Fig\,\ref{ale_1}a
\cite{McQueeneyRJ:AnodLp}, and coincides with the energy of the
kink detected by ARPES.

A second important piece of evidence in favor of the phonon
scenario comes from the temperature dependence of the kink in the
quasiparticle dispersion. The data in Fig\,\ref{ale_1}d-e indicate
that the effect smoothly evolves upon raising the temperature and
persists well above $T_c$. This is particulary obvious in the
100\,K data from LSCO, for which $T_c\!\simeq\!38$\,K. The inset
in panel (e) presents the real part of the self energy for
optimally doped Bi2212 extracted with the same procedure used by
\textcite{JohnsonPD:Doptdt} for the data presented in
Fig.\,\ref{johnson1}. The results from the two groups are very
similar except for a subtle difference: while both data sets show
a slightly more pronounced structure at low-temperature, the
superconducting state data by \textcite{JohnsonPD:Doptdt} hint at
a sharp peak superimposed on the normal state result
(Fig.\,\ref{johnson1}), which could be taken as evidence for a
qualitatively different effect below and above $T_c$. On the other
hand, the data by \textcite{LanzaraA:Eviuse} show a more
continuous evolution, with a persistence of the effect into the
normal state. Therefore the temperature dependence of the ARPES
data from Bi2212 and especially LSCO was taken by
\textcite{LanzaraA:Eviuse} as a direct and conclusive evidence
against the magnetic-mode interpretation because the latter, as
indicated by elastic neutron scattering experiments, turns on
below $T_c$
\cite{DaiP:MagduY,DaiPC:magest,FongHF:Neusme,HeH:Resseo}.

\begin{figure}[t!]
\centerline{\epsfig{figure=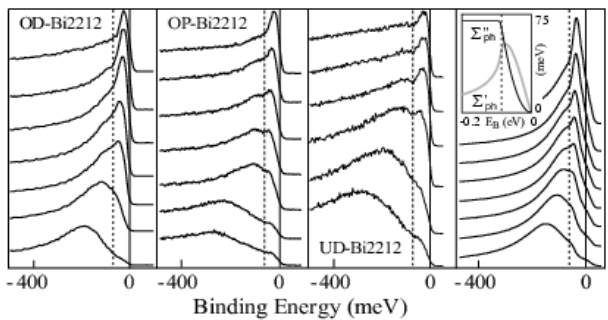,width=1\linewidth,clip=}}
\vspace{0cm} \caption{Low-temperature spectra from overdoped (OD),
optimally doped (OP), and underdoped (UD) Bi2212 along
(0,0)-($\pi$,$\pi$). Right panel: spectral function for an
electron-phonon coupled system in the Debye model at $T\!=\!0$;
$\Sigma^{\prime}(\omega)$ and $\Sigma^{\prime\prime}(\omega)$ are
shown in the inset, where the dashed line indicates the maximum
phonon energy \cite{LanzaraA:Eviuse}.}\label{ale_2}\end{figure}

Fig.\,\ref{ale_2} directly compares the low-temperature ARPES
spectra taken with 55\,eV photons on overdoped (OD), optimally
doped (OP) and underdoped (UD) Bi2212. It should be emphasized
that these data were recorded along the nodal direction, where the
two bilayer split bands are degenerate in energy
(Sec.\,\ref{sec:bilayer}).
The most important aspect is that these spectra are characterized
by a clear peak-dip-hump structure, in strong resemblance with
both the simulated spectral function for an electron-phonon
coupled systems (Fig.\,\ref{ale_2}, right panel), and the results
for the Be(0001) surface state presented in Fig.\,\ref{baer}a.
Note that the peak-dip-hump structure is better resolved in these
data than in those reported in an earlier publication
\cite{BogdanovPV:Eviesq} because of the improved signal-to-noise
ratio and possibly the higher photon energy used in the more
recent experiments. Furthermore, the dip energy marked by dashed
lines in Fig.\,\ref{ale_2} corresponds to that of the kink in the
dispersion (Fig\,\ref{ale_1}). However, while the kink is still
recognizable above $T_c$, the peak-dip-hump structure is smeared
out at high temperature. Similar results were also obtained along
the nodal direction for  LSCO \cite{zhou:1} and Bi2201
\cite{LanzaraA:1}. In particular, for Bi2201 the peak-dip-hump
structure could be clearly observed also above $T_c$, as this is
low enough that the system can be studied in the normal state
avoiding, at the same time, excessive thermal broadening.

From the ratio of the quasiparticle velocities at energies above and
below the kink, \textcite{LanzaraA:Eviuse} estimated the quantity
$\lambda^\prime$ which is proportional to the electron-phonon coupling
parameter $\lambda$ [note that this is the same procedure already
discussed in Sec.\,\ref{sec:elphcoupl} for Be(0001); however, in the
present case because of the underlying electron-electron correlations
the quasiparticle velocity is already strongly renormalized over a
large energy scale]. The doping dependence of $\lambda^\prime$ is
presented in Fig.\,\ref{ale_1}f and it clearly shows that the
electron-phonon coupling increases in all the systems (i.e., LSCO,
Bi2201, and Bi2212) upon reducing the doping level.

The ARPES results presented in this section indicate that in
addition to the electron self-energy corrections due to
electron-electron correlations, which are responsible for the
renormalization of the electronic structure of the cuprates over a
large energy scale (see, e.g., Sec.\,\ref{sec:ns_ccoc}), also the
contribution from electron-phonon interaction must be taken into
account, as the latter appears to have a direct influence on the
quasiparticle dynamics. In this regard \textcite{ShenZ-X:1} went a
step further and, by considering both the diagonal and
off-diagonal channels of electron-phonon coupling for $p$ and
$n$-type HTSC superconductors, asserted that electron-phonon
interaction is also an essential ingredient to pairing (a more
detailed discussion of this issue is however beyond the scope of
our review).

\subsubsection{Discussion}
\label{sec:selfdiscu}

As we have seen throughout this section, the investigation of the
self-energy corrections in Bi2212 is an extremely complex issue.
There are now strong evidences that the normal and superconducting
state low-energy dispersion along the nodal direction is
characterized by a sharp break accompanied by a double peak
structure in the EDCs (as indicated not only by the data from
Bi2212 but also from Bi2201 and LSCO). This identifies an energy
scale in the quasiparticle dynamics which is naturally explained
as a consequence of relatively strong electron-phonon coupling. On
the other hand, the results we discussed for the antinodal region
are much less certain because of the complications due to the
superstructure contaminations and bilayer splitting effects. Most
recently, however, new and more convincing evidence for the
presence of a kink in the Bi2212 quasiparticle dispersion near
($\pi$,0) below $T_c$ was reported by
\textcite{gromko:2,gromko:3}.
\begin{figure}[t!]
\centerline{\epsfig{figure=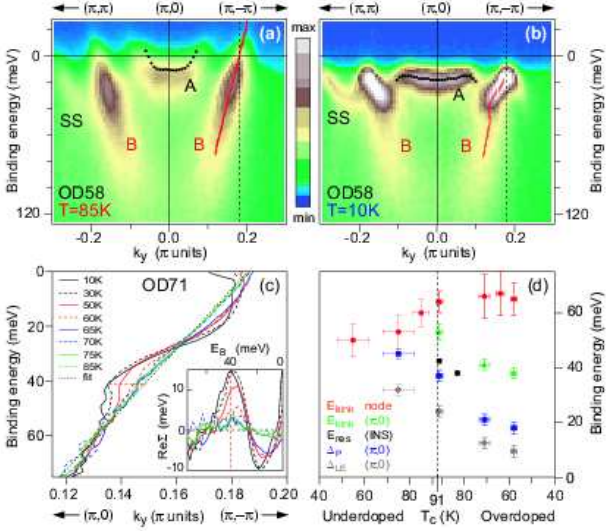,width=1\linewidth,clip=}}
\vspace{0cm} \caption{(a) Normal and (b) superconducting state
ARPES intensity measured on overdoped Bi2212 ($T_c\!=\!58$\,K)
with 20\,eV photons [red and black lines: bonding (B) and
antibonding (A) band dispersion; SS: superstructure replica]. (c)
MDC-derived dispersion along ($\pi$,0)-($\pi$,-$\pi$). Inset:
temperature dependence of $\Sigma^{\prime}(\omega)$ defined as the
difference between the data and the linear fit to the 85\,K
result. (d) Doping dependence of kink energy, gap magnitude (from
peak and leading-edge midpoint positions), and resonant mode
frequency (neutron scattering data). After \textcite{gromko:2}
[Color].}\label{gromko1}\end{figure}
As shown in Fig.\,\ref{gromko1}a where the normal state ARPES
intensity from overdoped Bi2212 along the
($\pi$,$\pi$)-($\pi$,0)-($\pi$,$-\pi$) direction is presented,
these authors could clearly resolved the bonding and antibonding
split bands as well as their superstructure replicas. Below
$T_c\!=\!58$\,K (Fig.\,\ref{gromko1}b), one can observe the
opening of the superconducting gap and the sharpening and/or
intensity increase of the low energy peaks, which give rise to the
traditional ($\pi$,0) peak-dip-hump structure. In addition, the
bonding band MDC-derived dispersion, which was linear above $T_c$,
is now characterized by a sharp kink at 40\,meV (this is further
substantiated by the detailed study on a $T_c\!=\!71$\,K overdoped
sample presented in Fig.\,\ref{gromko1}c). It has to be emphasized
that, on the one hand, this work confirms prior reports [see,
e.g., \textcite{NormanMR:Momdct}]. On the other hand, while the
kink in the ($\pi$,0) region was believed to be connected to the
traditional peak-dip-hump structure, these results indicate the
independency of the two phenomena, with the traditional
peak-dip-hump structure being simply a bilayer splitting effect.
Furthermore, \textcite{gromko:3} reported that the emergence of
the ($\pi$,0) kink at 40\,meV in the bonding band is accompanied
by the development of a separate antinodal peak-dip-hump structure
much weaker than the traditional one, consistent with an earlier
report [\textcite{FengDL:Bilste}; see crosses and bars in
Fig.\,\ref{bilayerpi0edc}b].

From the comparison between the results above and those  presented
in Sec.\,\ref{sec:energyscale}, it is clear that the kinks in the
antinodal and nodal dispersions have a completely different
origins: (i) while the former has a dramatic temperature
dependence and is observed only below $T_c$
(Fig.\,\ref{gromko1}c), the latter is seen already above $T_c$ and
is essentially independent of temperature (Fig.\,\ref{ale_1}).
(ii) As summarized by \textcite{gromko:2} in Fig.\,\ref{gromko1}d
where several energy scales are plotted versus doping, the
antinodal kink has a characteristic energy significantly smaller
than the nodal one. (iii) Different is also the doping dependence
of the energy scale (Fig.\,\ref{gromko1}d), and of the magnitude
of the velocity renormalization; as for the latter, which is
proportional to the coupling strength, while the nodal kink gets
weaker upon increasing doping, the antinodal one remains very
strong in the overdoped regime.\footnote{It should be noted that
it is hard to judge whether in the antinodal region there is an
additional kink at higher binding energies, similar to the one
observed along the nodal direction. This is due to the very
limited energy range of the data at ($\pi$,0), which is defined by
the bottom of the bonding band (see Fig.\,\ref{bilayerpi0edc}
and\,\ref{gromko1}).}

Because the antinodal kink is clearly associated with the onset of
superconductivity and it is limited to a small momentum space
region around ($\pi$,0) where the superconducting gap is maximum,
\textcite{gromko:2} suggested that this effect is due to the
coupling between the electrons and the $Q\!=\!(\pi,\pi)$ magnetic
resonant mode. This, in turn, could be the source of pairing for
high-$T_c$ superconductivity. However, in this regard a few
caveats have to be mentioned. First of all, there are alternative
explanations for the observed effect that have to be ruled out,
such as the very same opening of the superconducting gap and the
dramatic change in quasiparticle dynamics associated with it (see
Sec.\,\ref{sec_suppeak}). Furthermore, at this stage there is no
experimental evidence from inelastic neutron scattering
experiments for the existence of the resonant mode in the
extremely overdoped regime ($T_c\!\sim\!58$\,K), where the
antinodal kink remain very strong. For a more conclusive
assignment, a careful and more detailed investigation of these
issues is required.

\section{Concluding remarks}

The results discussed in the course of this review article
demonstrate that the HTSCs are complex materials characterized by
many competing degrees of freedom. As a consequence, both the
experimental investigation and the theoretical understanding are
extremely difficult. However, the remarkable improvement of the
ARPES technique has allowed the community to considerably deepen
our knowledge of the electronic structure of cuprates, a
fundamental step towards the development of a comprehensive
theoretical description of these systems.

Looking back at the last decade, there are many topics for which
the ARPES results have obtained a general consensus and have
profoundly impacted our understanding of the physics of the copper
oxide superconductors. (i) The importance of electronic
correlations (Mott physics), and the renormalization of the
bandwidth from $t$ to $J$ scale for the undoped parent compounds
of the HTSCs. (ii) The existence of a well defined FS for the
overdoped metal in all the different families of cuprates; in this
context, the observation of two bilayer split sheets of FS in
Bi2212 is the most direct evidence for the remarkable precision
achieved by these measurements. (iii) The qualitative doping
evolution of the electronic structure in the different HTSC
families, and the influence of AF correlations in the p-type
underdoped cuprates and especially in the n-type ones, in which
the hot-spot physics is still observed at optimal doping. (iv) The
overall $d$-wave symmetry of the superconducting gap detected
below $T_c$ in both hole and electron doped systems, which
supports the universality of the pairing nature in the HTSCs. (v)
The opening of the normal-state pseudogap at a temperature
$T^*\!>\!T_c$ in the underdoped regime, with a $d$-wave form
similar to the one of the superconducting gap. (vi) The emergence
of a coherent quasiparticle peak below $T_c$ near ($\pi$,0) in
Bi2212, whose spectral weight scales with the doping level $x$ in
the underdoped regime. (vii) The presence of an energy scale of
about 40-80\,meV in the quasiparticle dynamics, as evidenced by
the sharp dispersion renormalization and drop in the scattering
rate observed at those energies at different momenta.
Independently of its final interpretation, this latter effect is
likely of great significance to pairing because its energy scale
falls in between the values of $J$ and $\Delta$.

Looking towards the future, ARPES as a technique continues to
experience a rapid development. In gas-phase photoemission
experiments, which however are performed in a very different
configuration, the energy resolution has now reached exceptional
values of better than 0.01\,meV \cite{HollensteinU:Selfih}. One
would expect further progress also in solid state experiments. In
addition to that, the most significant technological advances
include the implementation of circularly polarized photon sources,
spin-polarized detectors, and time-resolved systems. In this
regard, it should be mentioned that most recently
\textcite{KaminskiA:Spobts}, by detecting a difference in the
ARPES intensity measured with left and right {\it circularly}
polarized photons on Bi2212 at ($\pi$,0), concluded that the
opening of the normal-state pseudogap below $T^*$ in the
underdoped regime corresponds to a phase transition into a state
characterized by a spontaneously broken time-reversal symmetry [as
suggested by \textcite{varma:2}]. In turn, this would imply that
$T^*$ identifies a new phase transition rather than the
fluctuation regime for the superconducting order parameter.
Although the validity of these conclusions remains to be tested
further, this and other new experiments, not only limited to the
specific field of HTSCs [for a general overview see, e.g.,
\textcite{Grioni:1}], indicate that the study of the electronic
structure of complex materials by ARPES will remain a vibrant and
rapidly evolving field.

\section*{Acknowledgments}

We would first like to express our gratitude to the current and
former members of the Stanford photoemission group for their
cooperation, daily discussions, and many useful comments: N.P.
Armitage, P. Bogdanov, Y. Chen, T. Cuk, D.S. Dessau, H. Eisaki,
D.L. Feng, S.L. Friedmann, M.Z. Hasan, J.M. Harris, N.J.C. Ingle,
S.A. Kellar, C. Kim, D.M. King, A. Lanzara, W.S. Lee, A.G. Loeser,
D.H. Lu, D.S. Marshall, A.Y. Matsuura, T.-U. Nahm, C.-H. Park,
A.V. Puchkov, F. Ronning, M.C. Schabel, K.M. Shen, Z.Y. Wang, B.O.
Wells, P.J. White, W.L.Yang, and X.J. Zhou. We also would like to
take this opportunity to acknowledge our collaborators in the
field: A.V. Balatsky, R.J. Birgeneau, D.A. Bonn, E.W. Carlson,
J.N. Eckstein, V.J. Emery, A. Fujimori, M. Greven, W.N. Hardy, A.
Ino, A. Kapitulnik, M.A. Kastner, K. Kishio, S.A. Kivelson, R.
Liang, S. Maekawa, Y. Maeno, L.L. Miller, T. Mizokawa, N. Nagaosa,
D. Orgad, F. Parmigiani, G.A. Sawatzky, J.R. Schrieffer, W.E.
Spicer, H. Takagi, S. Tajima, T. Timusk, T. Tohyama, Y. Tokura, S.
Uchida, D. van der Marel, T. Yoshida, R. Yosjizaki, and Z.-X.
Zhao. Over the years, we have also benefited tremendously from
scientific discussions with S. Doniach, R.B. Laughlin, and S.-C.
Zhang, and many other colleagues: E. Abrahams, P. Aebi, O.K.
Andersen, P.W. Anderson, E. Arrigoni, A. Bansil, J.C. Campuzano,
S. Chakravarty, A.V. Chubukov, E. Dagotto, J.C. Davis, H. Ding, Y.
Endoh, M.E. Flatt$\acute{{\rm e}}$, H. Fukuyama, D.-H. Lee, W.
Hanke, W.A. Harrison, M. Imada, P.D. Johnson, B. Keimer, P.A. Lee,
R.S. Markiewicz, J. Mesot, A. J. Millis, K.A. Moler, H.A. Mook,
M.R. Norman, N.P. Ong, J. Orenstein, J. Osterwalder, S.H. Pan, M.
Randeria, S. Sachdev, D.J. Scalapino, N.V. Smith, M. Tachiki, J.M.
Tranquada, Y.J. Uemura, C.M. Varma, M.G. Zacher, and  J. Zaanen.
Last, it is a pleasure to thank E. Abrahams, P. Aebi, N.P.
Armitage, E. Arrigoni, T. Cuk, N.J.C. Ingle, D.H. Lu, D.W. Lynch,
R.S. Markiewicz, N. Nagaosa, M.R. Norman, F. Ronning, K.M. Shen,
J. van den Brink, for the critical reading of this review article.
This project was carried out at the Stanford Synchrotron Radiation
Laboratory (SSRL) which is operated by the DOE Office of Basic
Energy Sciences, Division of Chemical Sciences. The Office's
Division of Materials Science has provided funding for this
project. The Stanford work was also supported by the NFS Grant No.
DMR0071897 and ONR Grant No. N00014-98-1-0195.

\bibliographystyle{apsrmp}
\bibliography{RMP_ARPES_final}

\end{document}